\documentclass[12pt]{article}
\usepackage[latin1]{inputenc}
\usepackage[T1]{fontenc}
\usepackage{pslatex}
\usepackage{txfonts}
\usepackage{color}
\usepackage{cite}
\usepackage{axodraw}
\usepackage{epsfig}
\usepackage{latexsym,feynarts,exscale}
\usepackage[dvips]{rotating}
\def\skstreichen#1{{\color{green}\ifmmode\scriptstyle\else\tiny\fi #1 \color{black}}}

\makeatletter

\def\paragraph{\@startsection{paragraph}{4}{\z@}{+2.00ex plus
 +1ex minus +.2ex}{1.5ex plus .2ex}{\it\normalsize}}

\def\section{\@startsection {section}{1}{\z@}{+3.0ex plus +1ex minus
  +.2ex}{2.3ex plus .2ex}{\normalsize\bf\boldmath}}
\def\subsection{\@startsection{subsection}{2}{\z@}{+2.5ex plus +1ex
minus +.2ex}{1.5ex plus .2ex}{\normalsize\bf\boldmath}}
\def\subsubsection{\@startsection{subsubsection}{3}{\z@}{+3.25ex plus
 +1ex minus +.2ex}{1.5ex plus .2ex}{\normalsize\it}}

 \expandafter\ifx\csname mathrm\endcsname\relax\def\mathrm#1{{\rm #1}}\fi

\renewcommand{\theequation}{\thesection.\arabic{equation}}
\newcounter{saveeqn}

\@addtoreset{equation}{section}

\def\asymp#1%
{\mathrel{\raisebox{-.4em}{$\widetilde{\scriptstyle #1}$}}}

\def\Nequal#1%
{\mathrel{\raisebox{-.5em}{$\stackrel{=}{\scriptstyle\rm#1}$}}}
\newcommand{\dsl}[1]{\not \hspace{-0.7mm}#1}
\def\dsl{\mathpalette\make@slash}
\def\make@slash#1#2{\setbox\z@\hbox{$#1#2$}%
  \hbox to 0pt{\hss$#1/$\hss\kern-\wd0}\box0}

\def\beq{\begin{equation}}
\def\eeq{\end{equation}}
\def\beqar{\begin{eqnarray}}
\def\eeqar{\end{eqnarray}}
\def\barr#1{\begin{array}{#1}}
\def\earr{\end{array}}
\def\bfi{\begin{figure}}
\def\efi{\end{figure}}
\def\btab{\begin{table}}
\def\etab{\end{table}}
\def\bce{\begin{center}}
\def\ece{\end{center}}
\def\nn{\nonumber}

\def\text{\textstyle}

\def\al{\alpha}

\def\Ga{\Gamma}
\def\ga{\gamma}
\def\de{\delta}
\def\De{\Delta}

\def\veps{\varepsilon}

\def\La{\Lambda}
\def\la{\lambda}

\def\si{\sigma}
\def\Si{\Sigma}

\def\refeq#1{\mbox{(\ref{#1})}}

\def\reffi#1{\mbox{Figure~\ref{#1}}}

\def\refta#1{\mbox{Table~\ref{#1}}}

\def\refse#1{\mbox{Section~\ref{#1}}}

\def\refapp#1{\mbox{App.~\ref{#1}}}
\def\citere#1{\mbox{Ref.~\cite{#1}}}
\def\citeres#1{\mbox{Refs.~\cite{#1}}}

\newcommand\reftasx[2]{\mbox{Tables~\ref{#1}~--~\ref{#2}}}

\newcommand\reffisb[3]{Figures~\ref{#1}, \ref{#2}, and \ref{#3}}
\newcommand\reffisa[2]{Figures~\ref{#1} and \ref{#2}}
\newcommand\reffisx[2]{\mbox{Figures~\ref{#1}~--~\ref{#2}}}

\newcommand\refeqsa[2]{Eqs.~(\ref{#1}) and (\ref{#2})}
\newcommand\refeqsx[2]{\mbox{Eqs.~(\ref{#1})~--~(\ref{#2})}}

\newcommand{\TeV}{\unskip\,\mathrm{TeV}}
\newcommand{\GeV}{\unskip\,\mathrm{GeV}}
\newcommand{\MeV}{\unskip\,\mathrm{MeV}}
\newcommand{\pba}{\unskip\,\mathrm{pb}}
\newcommand{\fb}{\unskip\,\mathrm{fb}}

\newcommand{\ri}{{\mathrm{i}}}
\newcommand{\rd}{{\mathrm{d}}}

\newcommand{\M}{{\cal{M}}}

\def\mathswitchr#1{\relax\ifmmode{\mathrm{#1}}\else$\mathrm{#1}$\fi}
\newcommand{\Pf}{\mathswitch  f}

\newcommand{\PV}{\mathswitchr V}
\newcommand{\PW}{\mathswitchr W}

\newcommand{\PZ}{\mathswitchr Z}
\newcommand{\PA}{\mathswitchr A}
\newcommand{\Pg}{\mathswitchr g}

\newcommand{\PH}{\mathswitchr H}
\newcommand{\Pe}{\mathswitchr e}
\newcommand{\Pne}{\mathswitch \nu_{\mathrm{e}}}

\newcommand{\Pnmu}{\mathswitch \nu_{\mu}}
\newcommand{\Pnmubar}{\mathswitch \bar\nu_{\mu}}
\newcommand{\Pd}{\mathswitchr d}
\newcommand{\Pdbar}{\bar{\mathswitchr d}}
\newcommand{\Pu}{\mathswitchr u}
\newcommand{\Pubar}{\bar{\mathswitchr u}}
\newcommand{\Ps}{\mathswitchr s}

\newcommand{\Pc}{\mathswitchr c}

\newcommand{\Pb}{\mathswitchr b}
\newcommand{\Pbbar}{\mathswitchr{\bar b}}
\newcommand{\Pp}{\mathswitchr p}
\newcommand{\Pt}{\mathswitchr t}

\newcommand{\Pep}{\mathswitchr {e^+}}

\newcommand{\Pmum}{\mathswitchr {\mu^-}}

\newcommand{\PWp}{\mathswitchr {W^+}}
\newcommand{\PWm}{\mathswitchr {W^-}}

\newcommand{\PQ}{\mathswitchr Q}

\newcommand{\PD}{\mathswitchr D}
\newcommand{\PDbar}{\bar{\mathswitchr D}}
\newcommand{\PU}{\mathswitchr U}

\newcommand{\Pmu}{\mathswitchr\mu}
\newcommand{\Pl}{\mathswitchr l}

\newcommand{\Pnl}{\mathswitchr \nu_{\mathrm{l}}}

\newcommand{\Pnlbar}{\mathswitchr {\bar\nu}_{\mathrm{l}}}

\newcommand{\Pqbar}{\bar{\mathswitchr q}}

\newcommand{\Pq}{\mathswitchr q}
\newcommand{\bPp}{\bar{\mathswitchr p}}
\newcommand{\bPq}{\bar{\mathswitchr q}}
\newcommand{\bPd}{\bar{\mathswitchr d}}
\newcommand{\bPu}{\bar{\mathswitchr u}}

\newcommand{\bPb}{\bar{\mathswitchr b}}
\newcommand{\bPt}{\bar{\mathswitchr t}}

\def\mathswitch#1{\relax\ifmmode#1\else$#1$\fi}

\newcommand{\MV}{\mathswitch {M_\PV}}
\newcommand{\MW}{\mathswitch {M_\PW}}

\newcommand{\MZ}{\mathswitch {M_\PZ}}
\newcommand{\MH}{\mathswitch {M_\PH}}

\newcommand{\GW}{\Gamma_{\PW}}

\newcommand{\GZ}{\Gamma_{\PZ}}
\newcommand{\GV}{\Gamma_{\PV}}
\newcommand{\GH}{\Gamma_{\PH}}

\newcommand{\GF}{\mathswitch {G_\mu}}

\newcommand{\gs}{g_{\mathrm{s}}}
\newcommand{\als}{\al_{\mathrm{s}}}

\def\solid{\raise.9mm\hbox{\protect\rule{1.1cm}{.2mm}}}
\def\dash{\raise.9mm\hbox{\protect\rule{2mm}{.2mm}}\hspace*{1mm}}

\def\lra{\mathop{\mathrm{\leftrightarrow}}\nolimits}

\def\draftdate{\relax}
\def\mda{\relax}
\def\mua{\relax}
\def\mla{\relax}
\def\Mda{\relax}
\def\Mua{\relax}
\def\Mla{\relax}
\def\draft{
\def\thtystars{******************************}
\def\sixtystars{\thtystars\thtystars}
\typeout{}
\typeout{\sixtystars**}
\typeout{* Draft mode!
         For final version remove \protect\draft\space in source file *}
\typeout{\sixtystars**}
\typeout{}
\def\draftdate{\today}
\def\mua{\marginpar[\boldmath\hfil$\uparrow$]%
                   {\boldmath$\uparrow$\hfil}%
                    \typeout{marginpar: $\uparrow$}\ignorespaces}
\def\mda{\marginpar[\boldmath\hfil$\downarrow$]%
                   {\boldmath$\downarrow$\hfil}%
                    \typeout{marginpar: $\downarrow$}\ignorespaces}
\def\mla{\marginpar[\boldmath\hfil$\rightarrow$]%
                   {\boldmath$\leftarrow $\hfil}%
                    \typeout{marginpar: $\lra$}\ignorespaces}
\def\Mua{\marginpar[\boldmath\hfil$\Uparrow$]%
                   {\boldmath$\Uparrow$\hfil}%
                    \typeout{marginpar: $\uparrow$}\ignorespaces}
\def\Mda{\marginpar[\boldmath\hfil$\Downarrow$]%
                   {\boldmath$\Downarrow$\hfil}%
                    \typeout{marginpar: $\downarrow$}\ignorespaces}
\def\Mla{\marginpar[\boldmath\hfil$\Rightarrow$]%
                   {\boldmath$\Leftarrow $\hfil}%
                    \typeout{marginpar: $\lra$}\ignorespaces}
\overfullrule 5pt
\oddsidemargin -15mm
\marginparwidth 29mm
}

\def\stars{\strut\leaders\hbox{*}\hfill\strut}
\def\starline{\hfil\strut\hfil\hbox to \textwidth {\stars}\hfil}

\newcommand{\antiquarks}{(anti-)quarks}
\newcommand{\antiquark}{(anti-)quark}

\newcommand{\antineutrinos}{(anti-)neutrinos}
\newcommand{\tppWWj}{$\Pp\Pp/\Pp\bPp\to\PW\PW+\mathrm{jet+X}$}
\newcommand{\tppllj}{$\Pp\Pp/\Pp\bPp\to\ell\bar\ell' \nu_{\ell'}\bar\nu_{\ell}+\mathrm{jet+X}$}

\newenvironment{diagram}{\begin{feynartspicture}(120,120)(1,1)}{\end{feynartspicture}}

\newcommand{\muren}{\mu_{\mathrm{ren}}}
\newcommand{\mufact}{\mu_{\mathrm{fact}}}
\newcommand{\Nf}{N_\mathrm{f}}

\newcommand{\thetac}{\theta_{\mathrm{C}}}
\newcommand{\tppbWWj}{$\Pp\bPp\to\PW\PW+\mathrm{jet+X}$}
\newcommand{\tppbWWjj}{$\Pp\bPp\to\PW\PW+2\mathrm{jets+X}$}
\newcommand{\DIRa}{\Delta^{\mathrm{IR}}_1(\mu)}
\newcommand{\DIRb}{\Delta^{\mathrm{IR}}_2(\mu)}
\newcommand{\DUV}{\Delta^{\mathrm{UV}}_1(\mu)}

\def\ssize{\scriptsize}

\begin{document}
\enlargethispage{2cm}
\thispagestyle{empty}
\def\thefootnote{\fnsymbol{footnote}}
\setcounter{footnote}{1}
\null
\draftdate
\hfill HU-EP-09/25\\
\strut\hfill MPP-2009-153\\
\strut\hfill PSI-PR-09-11\\
\strut\hfill SFB/CPP-09-45\\
\vspace{1.5cm}
\begin{center}
  {\Large \bf\boldmath
    NLO QCD corrections to $\Pp\Pp/\Pp\bar\Pp\to\PW\PW+\mathrm{jet}+X$ 
    \\[.3em]
    including leptonic W-boson decays
    \par} 
  
\vspace{1.5cm}
{\large\sc 
  S.\ Dittmaier$^{1,2}$, S.\ Kallweit$^{2,3}$ and P.\ Uwer$^4$ } \\[.5cm]
$^1$ \it Albert-Ludwigs-Universit\"at Freiburg, Physikalisches Institut, \\
D-79104 Freiburg, Germany
\\[.5cm]
$^2$ {\it Max-Planck-Institut f\"ur Physik
  (Werner-Heisenberg-Institut), \\
  D-80805 M\"unchen, Germany}
\\[.5cm]
$^3$ {\it Paul Scherrer Institut, W\"urenlingen und Villigen,\\
  CH-5232 Villigen PSI, Switzerland} 
\\[0.5cm]
$^4$ {\it Institut f\"ur Physik, Humboldt-Universit\"at zu Berlin,\\
D-10099 Berlin, Germany}
\par \vskip 1em
\end{center}\par
\vfill {\bf Abstract:} 
We report on the calculation of the next-to-leading order QCD corrections
to the production of W-boson pairs in association with a hard
jet at the Tevatron and the LHC, which is an important source 
of background for Higgs and new-physics searches. 
Leptonic decays of the $\PW$ bosons are included by applying an improved 
version of the narrow-width approximation that treats the $\PW$ bosons 
as on-shell particles, but keeps the information on the $\PW$~spin.
Contributions from external bottom quarks are neglected, because they
are either numerically suppressed or should be attributed
to different processes such as $\PW\Pt$ or $\Pt\bar\Pt$ production.
A survey of differential NLO QCD cross sections 
is provided both for the LHC and the Tevatron. 
The QCD corrections stabilize the leading-order prediction for 
the cross section with respect to scale variations. 
However, the scale dependence of the next-to-leading order results for
the LHC is only reduced considerably 
if a veto against the emission of a second hard jet is applied.
In general, the corrections do not simply rescale the differential
leading-order cross sections.
In particular, their shapes are distorted if an additional energy 
scale is involved.

\vfill
\noindent
August 2009 
\null
\setcounter{page}{0}

\clearpage

\def\thefootnote{\arabic{footnote}}
\setcounter{footnote}{0}

\section{Introduction}
The search for new-physics particles---including the Standard Model
Higgs boson---will be the primary task in high-energy physics in the 
era of the LHC. The extremely complicated
hadron collider environment does not only require
sufficiently precise predictions for new-physics signals, but also
for many complicated background reactions that cannot entirely be
measured from data. Among such background processes, several involve
three, four, or even more particles in the final state, rendering
the necessary next-to-leading-order (NLO) calculations in QCD very
complicated. This problem lead to the creation of an
``experimenters' wishlist for NLO calculations''
\cite{Buttar:2006zd,Campbell:2006wx,Bern:2008ef} that 
were still missing at that time, but are required
for successful LHC analyses. The process
$\Pp\Pp\to\PW^+\PW^-{+} \mathrm{jet+X}$ made it to the top of this list.
Meanwhile the $2\to3$ particle processes on the list have been
evaluated at NLO QCD, and we further contribute to and extend that part 
in this paper; most notably also some of the $2\to4$ processes have been
calculated to NLO recently, viz.\ for the production of
$\Pt\bar\Pt\Pb\bar\Pb$~\cite{Bredenstein:2008zb}
and $\PW+3\mathrm{jets}$ final states~\cite{Ellis:2008qc}.
Moreover, benchmark results for the virtual corrections
have been presented for a specific phase-space point for all $2\to4$ processes
on the list in \citere{vanHameren:2009dr}.

The process of $\PW\PW$+jet production
is an important source for background to the
production of a Higgs boson that subsequently decays into a W-boson
pair, where additional jet activity might arise from the 
production. In particular, it has been shown recently that 
the sensitivity of the Tevatron experiments CDF and D0 concerning the
discovery of a Standard Model Higgs boson in the mass range 
$135{-}190\GeV$ could improve significantly by studying
$\PH\to \PW^+\PW^-\to \Pl^+\Pl^- /\!\!\!\!\!p_T$ together 
with an additional jet
\cite{Mellado:2007fb}.

$\PW\PW$+jet production
delivers also potential background to new-physics searches, such as
supersymmetric particles, because of leptons and missing transverse
momentum from the W~decays. Besides the process is 
interesting in its own right, since W-pair production processes
enable a direct analysis of the non-abelian
gauge-boson self-interactions, and a large fraction of W~pairs
will show up with additional jet activity at the LHC.
Last but not least $\PW\PW$+jet at NLO
also delivers the real--virtual contributions 
to the next-to-next-to-leading-order (NNLO) calculation of $\PW$-pair 
production, 
for which further building blocks are presented in \citere{Chachamis:2008yb}.

Here we report on the calculation of the process 
$\Pp\Pp/\Pp\bPp\to\PW^+\PW^-{+}\mathrm{jet+X}$ in NLO QCD including 
leptonic $\PW$-boson decays. 
Results of this calculation on integrated cross sections, which are not 
sensitive to the W~decays, have already been published in
\citere{Dittmaier:2007th}.
Parallel to our work, another NLO study~\cite{Campbell:2007ev} of 
$\Pp\Pp\to\PW^+\PW^-{+} \mathrm{jet+X}$ at the LHC appeared, where also
the W~decays have been taken into account.
Moreover, a third calculation focusing on WW+jet production is still
in progress~\cite{Sanguinetti:2008xt}.
In \citere{Bern:2008ef} the three different approaches are briefly
described, and a detailed comparison of the virtual corrections at a single
phase-space point is presented, revealing mutual agreement.

The paper is organized as follows.
In \refse{se:LO} we describe the general setup of the calculation,
paying particular attention to the treatment of the various quark
flavours. Section~\ref{se:NLO} provides details of the NLO
calculation, and Section~\ref{se:decays} describes
the methods to include the W~decays.
Our detailed discussion of numerical results is presented in
\refse{se:numres}, which covers both integrated and differential
cross sections. In \refse{se:concl} we give our conclusions, and
the Appendix contains a derivation of the improved narrow-width 
approximation.

\section{Leading-order calculation and calculational framework}
\label{se:LO}

\begin{figure}
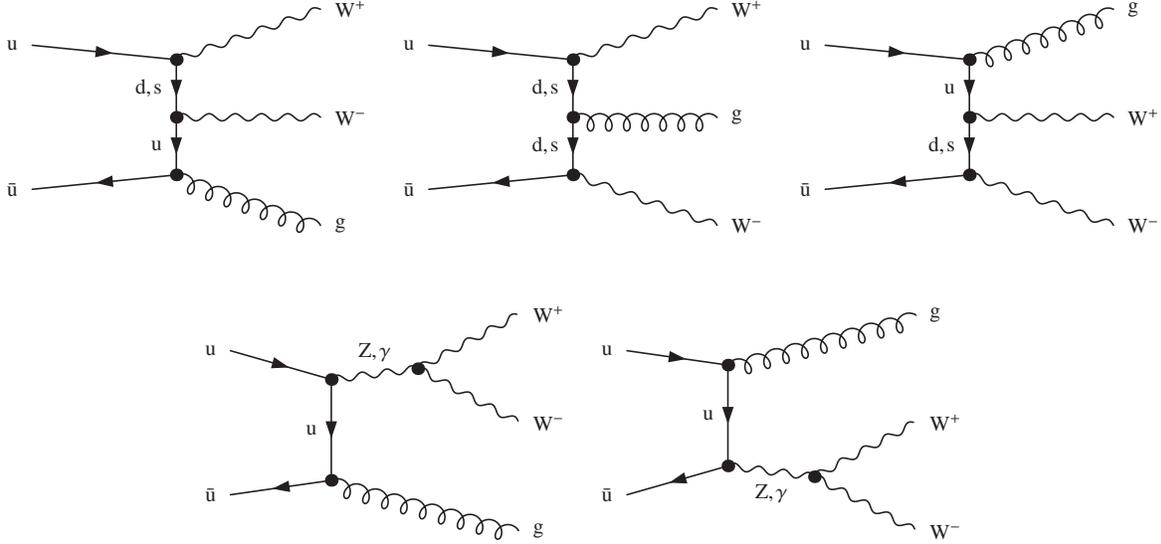

\centering{
\begin{diagram}
\FADiagram{}
\FALabel(-1,15)[r]{\ssize$\Pu$}
\FAProp(0.,15.)(10.,14.)(0.,){/Straight}{1}
\FALabel(-1,5)[r]{\ssize$\bPu$}
\FAProp(0.,5.)(10.,6.)(0.,){/Straight}{-1}
\FALabel(9.,12.)[r]{\ssize$\Pd,\Ps$}
\FAProp(10.,14.)(10.,10.)(0.,){/Straight}{1}
\FALabel(9.,8.)[r]{\ssize$\Pu$}
\FAProp(10.,10.)(10.,6.)(0.,){/Straight}{1}
\FALabel(21,17.5)[l]{\ssize$\PWp$}
\FAProp(10.,14.)(20.,17.5)(0.,){/Sine}{0}
\FALabel(21,10.)[l]{\ssize$\PWm$}
\FAProp(10.,10.)(20.,10.)(0.,){/Sine}{0}
\FALabel(21,2.5)[l]{\ssize$\Pg$}
\FAProp(20.,2.5)(10.,6.)(0.,){/Cycles}{0}
\FAVert(10.,14.){0}
\FAVert(10.,6.){0}
\FAVert(10.,10.){0}
\end{diagram}
\hspace*{2em}
\begin{diagram}
\FADiagram{}
\FALabel(-1,15)[r]{\ssize$\Pu$}
\FAProp(0.,15.)(10.,14.)(0.,){/Straight}{1}
\FALabel(-1,5)[r]{\ssize$\bPu$}
\FAProp(0.,5.)(10.,6.)(0.,){/Straight}{-1}
\FALabel(9.,12.)[r]{\ssize$\Pd,\Ps$}
\FAProp(10.,14.)(10.,10.)(0.,){/Straight}{1}
\FALabel(9.,8.)[r]{\ssize$\Pd,\Ps$}
\FAProp(10.,10.)(10.,6.)(0.,){/Straight}{1}
\FALabel(21,17.5)[l]{\ssize$\PWp$}
\FAProp(10.,14.)(20.,17.5)(0.,){/Sine}{0}
\FALabel(21,10.)[l]{\ssize$\Pg$}
\FAProp(20.,10.)(10.,10.)(0.,){/Cycles}{0}
\FALabel(21,2.5)[l]{\ssize$\PWm$}
\FAProp(10.,6.)(20.,2.5)(0.,){/Sine}{0}
\FAVert(10.,14.){0}
\FAVert(10.,6.){0}
\FAVert(10.,10.){0}
\end{diagram}
\hspace*{2em}
\begin{diagram}
\FADiagram{}
\FALabel(-1,15)[r]{\ssize$\Pu$}
\FAProp(0.,15.)(10.,14.)(0.,){/Straight}{1}
\FALabel(-1,5)[r]{\ssize$\bPu$}
\FAProp(0.,5.)(10.,6.)(0.,){/Straight}{-1}
\FALabel(9.,8.)[r]{\ssize$\Pd,\Ps$}
\FAProp(10.,14.)(10.,10.)(0.,){/Straight}{1}
\FALabel(9.,12.)[r]{\ssize$\Pu$}
\FAProp(10.,10.)(10.,6.)(0.,){/Straight}{1}
\FALabel(21,17.5)[l]{\ssize$\Pg$}
\FAProp(20.,17.5)(10.,14.)(0.,){/Cycles}{0}
\FALabel(21,10.)[l]{\ssize$\PWp$}
\FAProp(10.,10.)(20.,10.)(0.,){/Sine}{0}
\FALabel(21,2.5)[l]{\ssize$\PWm$}
\FAProp(10.,6.)(20.,2.5)(0.,){/Sine}{0}
\FAVert(10.,14.){0}
\FAVert(10.,6.){0}
\FAVert(10.,10.){0}
\end{diagram}
}
\vspace*{-1ex}
\centering{
\begin{diagram}
\FADiagram{}
\FALabel(-1,15)[r]{\ssize$\Pu$}
\FAProp(0.,15.)(7.,13.)(0.,){/Straight}{1}
\FALabel(-1,5)[r]{\ssize$\bPu$}
\FAProp(0.,5.)(7.,6.)(0.,){/Straight}{-1}
\FALabel(6.,9.5)[r]{\ssize$\Pu$}
\FAProp(7.,13.)(7.,6.)(0.,){/Straight}{1}
\FALabel(21,17.5)[l]{\ssize$\PWp$}
\FAProp(13.,13.75)(20.,17.5)(0.,){/Sine}{0}
\FALabel(21,10.)[l]{\ssize$\PWm$}
\FAProp(13.,13.75)(20.,10.)(0.,){/Sine}{0}
\FALabel(10.,14.25)[b]{\ssize$\PZ,\gamma$}
\FAProp(7.,13.)(13.,13.75)(0.,){/Sine}{0}
\FALabel(21,2.5)[l]{\ssize$\Pg$}
\FAProp(20.,2.5)(7.,6.)(0.,){/Cycles}{0}
\FAVert(7.,13){0}
\FAVert(7.,6.){0}
\FAVert(13.,13.75){0}
\end{diagram}
\hspace*{2em}
\begin{diagram}
\FADiagram{}
\FALabel(-1,15)[r]{\ssize$\Pu$}
\FAProp(0.,15.)(7.,14.)(0.,){/Straight}{1}
\FALabel(-1,5)[r]{\ssize$\bPu$}
\FAProp(0.,5.)(7.,7.)(0.,){/Straight}{-1}
\FALabel(6.,10.5)[r]{\ssize$\Pu$}
\FAProp(7.,14.)(7.,7.)(0.,){/Straight}{1}
\FALabel(21,10.)[l]{\ssize$\PWp$}
\FAProp(13.,6.25)(20.,10.)(0.,){/Sine}{0}
\FALabel(21,2.5)[l]{\ssize$\PWm$}
\FAProp(13.,6.25)(20.,2.5)(0.,){/Sine}{0}
\FALabel(10.,5.75)[t]{\ssize$\PZ,\gamma$}
\FAProp(7.,7.)(13.,6.25)(0.,){/Sine}{0}
\FALabel(21,17.5)[l]{\ssize$\Pg$}
\FAProp(20.,17.5)(7.,14.)(0.,){/Cycles}{0}
\FAVert(7.,14){0}
\FAVert(7.,7.){0}
\FAVert(13.,6.25){0}
\end{diagram}
}
\vspace*{-1.5ex}
\caption{Diagrams contributing to one specific LO subprocess: 
The two $\PW$ bosons couple directly to the fermion chain (upper diagrams) 
or by means of an intermediate neutral gauge boson (lower diagrams).}
\label{fi:LOdiagrams}
\end{figure}

At leading order (LO), hadronic $\PW\PW{+}$jet production receives 
contributions from the partonic processes 
$\Pq\Pqbar\to\PW^+\PW^- \Pg$, $\Pq\Pg\to\PW^+\PW^- \Pq$, and 
$\Pg\Pqbar\to\PW^+\PW^- \Pqbar$, where $\Pq$ stands for up- or down-type 
quarks. Note that the amplitudes for $\Pq=\Pu,\Pd$ are not the same, 
even for vanishing light-quark masses. All three channels are related 
by crossing symmetry.  
The LO diagrams for the sample process $\Pu\bar\Pu\to\PW^+\PW^- \Pg$ 
are shown in \reffi{fi:LOdiagrams}.

\subsection{Quark-mixing effects}

To very good approximation the Cabibbo--Kobayashi--Maskawa (CKM)
matrix can be assumed to be of block-diagonal form allowing mixing
only between 
the two light generations, namely
\begin{eqnarray}
\label{eq:pr:CKMmatrix}
V_{\rm{CKM}} =\left(\begin{array}{ccc}
V_{\Pu\Pd} & V_{\Pu\Ps} & 0\\
V_{\Pc\Pd} & V_{\Pc\Ps} & 0\\
0 & 0 & 1\\
\end{array}\right)=
\left(\begin{array}{ccc}
\cos\theta_{\rm{C}} & \sin\theta_{\rm{C}} & 0\\
-\sin\theta_{\rm{C}} & \cos\theta_{\rm{C}} & 0\\
0 & 0 & 1\\
\end{array}\right),
\end{eqnarray}
where $\theta_{\rm{C}}$ denotes the Cabibbo angle. 
The approximation is justified by the fact that the neglected off-diagonal 
matrix elements are very small. 
Moreover, ignoring the masses of the light quarks, 
the dependence on the CKM matrix 
drops out for a remarkable set of subprocesses.
\begin{figure}
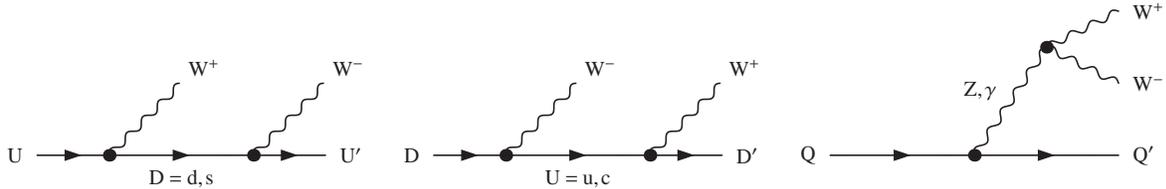

\centering{
\begin{diagram}
\FADiagram{}
\FALabel(10.5,15.5)[lb]{\ssize$\PWp$}
\FALabel(20.5,15.5)[lb]{\ssize$\PWm$}
\FALabel(-1,10)[r]{\ssize$\PU$}
\FALabel(21,10)[l]{\ssize$\PU'$}
\FAProp(0.,10.)(5.,10.)(0.,){/Straight}{1}
\FAProp(5.,10.)(15.,10.)(0.,){/Straight}{1}
\FAProp(15.,10.)(20.,10.)(0.,){/Straight}{1}
\FAProp(5.,10.)(10.,15.)(0.,){/Sine}{0}
\FAProp(15.,10.)(20.,15.)(0.,){/Sine}{0}
\FALabel(10,9)[t]{\ssize$\PD=\Pd,\Ps$}
\FAVert(5.,10.){0}
\FAVert(15.,10.){0}
\end{diagram}
\hspace*{2em}
\begin{diagram}
\FADiagram{}
\FALabel(10.5,15.5)[lb]{\ssize$\PWm$}
\FALabel(20.5,15.5)[lb]{\ssize$\PWp$}
\FALabel(-1,10)[r]{\ssize$\PD$}
\FALabel(21,10)[l]{\ssize$\PD'$}
\FAProp(0.,10.)(5.,10.)(0.,){/Straight}{1}
\FAProp(5.,10.)(15.,10.)(0.,){/Straight}{1}
\FAProp(15.,10.)(20.,10.)(0.,){/Straight}{1}
\FAProp(5.,10.)(10.,15.)(0.,){/Sine}{0}
\FAProp(15.,10.)(20.,15.)(0.,){/Sine}{0}
\FALabel(10,9)[t]{\ssize$\PU=\Pu,\Pc$}
\FAVert(5.,10.){0}
\FAVert(15.,10.){0}
\end{diagram}
\hspace*{2em}
\begin{diagram}
\FADiagram{}
\FALabel(21,20.)[l]{\ssize$\PWp$}
\FALabel(21,15.)[l]{\ssize$\PWm$}
\FALabel(11.5,14.5)[r]{\ssize$\PZ,\gamma$}
\FALabel(-1,10)[r]{\ssize$\PQ$}
\FALabel(21,10)[l]{\ssize$\PQ'$}
\FAProp(0.,10.)(10.,10.)(0.,){/Straight}{1}
\FAProp(10.,10.)(20.,10.)(0.,){/Straight}{1}
\FAProp(10.,10.)(15.,17.5)(0.,){/Sine}{0}
\FAProp(15.,17.5)(20.,20.)(0.,){/Sine}{0}
\FAProp(15.,17.5)(20.,15.)(0.,){/Sine}{0}
\FAVert(10.,10.){0}
\FAVert(15.,17.5){0}
\end{diagram}
}
\vspace*{-8ex}
\caption{Electroweak part of diagrams with two W bosons coupling directly 
  to the same fermion chain (left and central diagram): With light-quark 
  masses neglected, all contributions with \mbox{$\PU\neq \PU'$} or 
  \mbox{$\PD\neq \PD'$} vanish in the block-diagonal approximation of the 
  CKM matrix due to cancellations between diagrams with different intermediate 
  states. The contributions with \mbox{$\PU=\PU'$} or \mbox{$\PD=\PD'$} 
  behave as in the case of a trivial (unit) CKM matrix. Diagrams with an 
  intermediate neutral electroweak vector boson (right diagram) 
  do not depend on the CKM matrix, and thus \mbox{$\PQ= \PQ'$} holds.}
\label{fi:CKMdependence}
\end{figure}

For \tppWWj, this happens if both $\PW$ bosons couple to the same fermion
chain, as illustrated in \reffi{fi:CKMdependence}. 
Independent of gluonic couplings to this fermion chain, which do not 
affect the electroweak structure, the unitarity of the CKM matrix leads to
\begin{eqnarray}
  \label{eq:pr:unitarity}
  \sum_{\PD=\Pd,\Ps}V_{\PU'\PD}V_{\PU\PD}^\ast
  &=&\sum_{\PD=\Pd,\Ps}V_{\PU'\PD}V_{\PD\PU}^\dagger
  =\delta_{\PU\PU'}\;,\hspace*{.2cm}
  \sum_{\PU=\Pu,\Pc}V_{\PU\PD'}^\ast V_{\PU\PD}
  =\sum_{\PU=\Pu,\Pc}V_{\PD'\PU}^\dagger V_{\PU\PD}=\delta_{\PD\PD'}\;,
  \hspace{2em}
\end{eqnarray}
with the nomenclature of \reffi{fi:CKMdependence}, when the intermediate quark
state is summed over. Diagrams with the $\PW$-boson pair coupling to 
the fermion chain by means of an intermediate vector boson 
are independent of the CKM matrix anyway.
Therefore, a remarkable set of subprocesses of \tppWWj{ }is not
influenced by the explicit entries of the CKM matrix due to its unitarity.

\begin{figure}
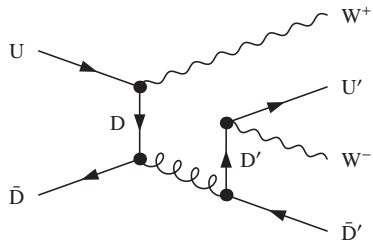

\centering{
\begin{diagram}
\FADiagram{}
\FALabel(-1,15)[r]{\ssize$\PU$}
\FAProp(0.,15.)(7.,12.5)(0.,){/Straight}{1}
\FALabel(-1,5)[r]{\ssize$\PDbar$}
\FAProp(7.,7.5)(0.,5.)(0.,){/Straight}{1}
\FALabel(6.,10.)[r]{\ssize$\PD$}
\FAProp(7.,12.5)(7.,7.5)(0.,){/Straight}{1}
\FALabel(21,17.5)[l]{\ssize$\PWp$}
\FAProp(7.,12.5)(20.,17.5)(0.,){/Sine}{0}
\FAProp(7.,7.5)(13.,5.)(0.,){/Cycles}{0}
\FALabel(21.,12.5)[l]{\ssize$\PU'$}
\FAProp(13.,10.)(20.,12.5)(0.,){/Straight}{1}
\FALabel(21,2.5)[l]{\ssize$\PDbar'$}
\FAProp(13.,5.)(20.,2.5)(0.,){/Straight}{-1}
\FALabel(21,7.5)[l]{\ssize$\PWm$}
\FAProp(13.,10.)(20.,7.5)(0.,){/Sine}{0}
\FALabel(14,7.5)[l]{\ssize$\PD'$}
\FAProp(13.,5.)(13.,10.)(0.,){/Straight}{1}
\FAVert(7.,12.5){0}
\FAVert(7.,7.5){0}
\FAVert(13.,10.){0}
\FAVert(13.,5.){0}
\end{diagram}
}
\vspace*{-1.5ex}
\caption[A sample diagram with two W bosons coupling to different %
fermion chains.]{A sample diagram with two W bosons coupling to
  different fermion chains: Contributions of this kind yield a 
  non-trivial dependence on the CKM matrix elements.}
\label{fi:CKMtwofermionchains}
\end{figure}

The only subprocesses contributing to \tppWWj{ }that depend on the 
explicit entries of the CKM matrix are those containing two fermion 
chains with the two W bosons coupling to different fermion chains at 
least in some diagrams. These are the real-emission subprocesses including 
both two external up-type 
and two down-type \mbox{\antiquarks}. A sample diagram is shown 
in \reffi{fi:CKMtwofermionchains}.


\subsection{Treatment of bottom \antiquarks}
\label{se:pr:bquarks}

The situation of a hard jet resulting from an outgoing bottom 
\mbox{\antiquark} has to be considered with care: The long bottom 
lifetime results in a resolvable second vertex that is  
displaced from the primary interaction point, allowing in principle to
tag the bottom flavour.
This, however, does not imply that all events involving outgoing bottom 
quarks can be separated. For instance, in real-emission subprocesses
only one of the two outgoing jets must be detected:
This can be the light jet as well, 
while the b-jet leaves the detector unseen in direction of the beam axis. 
Moreover, the b-tagging efficiency is below 100\%, so that not 
all ``b-jet events'' can be isolated.
The influence of incoming bottom \mbox{\antiquarks} on the hadronic 
cross section is suppressed with respect to other incoming \antiquarks\
by the small bottom PDFs in the colliding hadrons. 

In the process class \tppWWj, 
however, the suppression of subprocesses 
with external bottom \mbox{\antiquarks} may be
overcompensated by top 
resonances, always showing up along with final-state bottom 
\mbox{\antiquarks}. Moreover, outgoing $\Pb\bPb$ pairs appearing in 
real-emission subprocesses even contain two potentially
resonant top-quark propagators in some diagrams without PDF suppression. These 
subprocesses should, however, in general not be assigned to WW+jet 
production, since they are actually off-shell continuations of 
$\Pt\bPt$ production. From this point of view, the subprocesses with
only one outgoing bottom \mbox{\antiquark} can be seen as
contributions to $\PWm\Pt$ production or $\PWp\bPt$ production,
with the off-shell decay of the top \mbox{\antiquark} included. 
In our approach, precisely described in the following, we exclude these
resonance reactions upon omitting contributions from channels involving
external bottom \antiquarks. Thus, if desired, contributions from these 
channels could be easily added, but their calculation requires a
consistent treatment of top \antiquarks\ as unstable particles, as for
instance provided by the complex-mass 
scheme~\cite{Denner:1999gp,Denner:2005fg}.

For the calculation of cross sections, we follow two different strategies regarding the treatment of bottom \mbox{\antiquarks}:

\paragraph{Five-flavour scheme} 
In the first approach, bottom \mbox{\antiquarks} are treated as
massless particles, so five-flavour PDFs and a five-flavour running 
of $\als$ are used. In this framework, the strategy is to neglect all 
contributions containing external bottom \mbox{\antiquarks}. For the 
initial state, this approach is justified by the smallness of their
PDFs. Final-state bottom \mbox{\antiquarks} are excluded by the
assumption that their signal can be distinguished from that of a 
light-quark jet by means of b-tagging. 

An advantage of this procedure can be seen in the fact that the 
influence of the off-shell continuations of $\Pt\bPt$, $\PWm\Pt$, 
and $\PWp\bPt$ production, which is explained in the foregoing
passage, is simply left away. The respective contributions should 
be added to dedicated calculations for these processes. 
Note, however, that the naive application of this procedure would
lead to an ill-defined cross section containing mass singularities. 
The problem is  due to the $g\to \Pb\Pbbar$ splitting which by  
anti-b-tagging would be removed
from the real corrections. This contribution is required to cancel the
mass singularities from the bottom-quark loop in the gluon self-energy 
contributing to the virtual corrections.
We treat this case similar to what has been done in 
Ref.~\cite{Brandenburg:1997pu} and is also
discussed in Ref.~\cite{Banfi:2006hf}. 
If the two bottom quarks are
combined according to the applied jet algorithm the resulting jet 
carries no (net) b-charge and is thus counted as a ``light'' jet. On the
other hand if the two bottom quarks form two individual jets we assume
them as tagged and ignore this configuration. Applying this procedure
restricts the phase-space integration to the collinear configurations which
are needed to cancel the mass singularities in the virtual corrections. 
Within the dipole subtraction formalism we combine the corresponding 
integrated dipoles
with the virtual corrections to obtain a finite result. In addition
one obtains a contribution from the real matrix element with
the unintegrated dipoles subtracted.
However this contribution is only integrated
over the collinear phase space region. Since in this region the unintegrated
dipoles approximate the real matrix elements we can ignore this
contribution to very good approximation. 
The case in which the bottom-quark pair is not produced via
gluon splitting is highly suppressed by anti-b-tagging.
The neglected contribution only delivers sizeable 
contributions from diagrams including resonant top quarks. Exactly 
these contributions, however, are meant to be left out, because they 
should actually be assigned to different process classes.
\\

\paragraph{``Four-flavour scheme''} 
In the second approach, the bottom \mbox{\antiquarks} are understood
as massive particles. We use four-flavour PDF's to describe the
parton contents of the proton.  
Here, no top resonances show up, since no bottom-\mbox{\antiquark}
densities are taken into account and no mixing between the 
two light and the third generation takes place in the chosen 
approximation of the CKM matrix. As a consequence no single
outgoing bottom \mbox{\antiquarks} appear.
No $\Pb\bPb$ pairs are taken into account in final 
states which is justified by assuming 
anti-b-tagging.  The running of $\als$ is driven only by the four
remaining light quarks with both bottom- and top-quark loops in the 
gluon self-energies subtracted at zero momentum. 
As explained in \refse{se:NLO}, 
no large corrections arise from terms proportional to $\als\ln m_\Pb$, because counterterms always contribute in the combination \mbox{$\frac12\delta Z_\PA+\delta Z_{\gs}$}. 
Thus, the \mbox{$\als\ln m_\Pb$} term from the renormalization of the
strong coupling cancels against the corresponding term from the
wave-function renormalization of the gluon. In this
scheme, such a cancellation always 
takes place if the number of external gluons is equal to the number of 
strong couplings in the considered LO process.

\section{Details of the NLO calculation}
\label{se:NLO}

In order to prove the correctness of our results we have evaluated each ingredient twice using independent calculations based---as far as possible---on different methods, yielding results in mutual agreement.

\subsection{Virtual corrections}
\label{se:virtual}
The virtual corrections modify the partonic processes that are already present at LO. At NLO these corrections are induced by self-energy, vertex, box (4-point), and pentagon (5-point) corrections. For illustration the pentagon graphs, which are the most complicated diagrams, are shown in \reffi{fi:bospentagondirect} for one partonic channel.
\begin{figure}
\centering{
\begin{diagram}
\FADiagram{}
\FALabel(-1,15)[r]{\ssize$\Pu$}
\FALabel(-1,5)[r]{\ssize$\bPu$}
\FALabel(21,17.5)[l]{\ssize$\PWp$}
\FALabel(21,10.)[l]{\ssize$\PWm$}
\FALabel(21,2.5)[l]{\ssize$\Pg$}
\FAProp(0.,15.)(5.,13.)(0.,){/Straight}{1}
\FAProp(0.,5.)(5.,7.)(0.,){/Straight}{-1}
\FAProp(20.,17.5)(10.5,14.5)(0.,){/Sine}{0}
\FAProp(20.,2.5)(10.5,5.5)(0.,){/Cycles}{0}
\FAProp(20.,10.)(14.,10.)(0.,){/Sine}{0}
\FAProp(5.,13.)(5.,7.)(0.,){/Cycles}{0}
\FAProp(5.,13.)(10.5,14.5)(0.,){/Straight}{1}
\FALabel(7.25,14.75)[b]{\ssize$\Pu$}
\FAProp(5.,7.)(10.5,5.5)(0.,){/Straight}{-1}
\FALabel(7.25,5.25)[t]{\ssize$\Pu$}
\FAProp(10.5,14.5)(14.,10.)(0.,){/Straight}{1}
\FALabel(13.5,12.)[bl]{\ssize$\Pd,\Ps$}
\FAProp(10.5,5.5)(14.,10.)(0.,){/Straight}{-1}
\FALabel(13.5,8.)[tl]{\ssize$\Pu$}
\FAVert(5.,13.){0}
\FAVert(5.,7.){0}
\FAVert(10.5,14.5){0}
\FAVert(10.5,5.5){0}
\FAVert(14.,10.){0}
\end{diagram}
\hspace*{2em}
\begin{diagram}
\FADiagram{}
\FALabel(-1,15)[r]{\ssize$\Pu$}
\FALabel(-1,5)[r]{\ssize$\bPu$}
\FALabel(21,17.5)[l]{\ssize$\Pg$}
\FALabel(21,10.)[l]{\ssize$\PWp$}
\FALabel(21,2.5)[l]{\ssize$\PWm$}
\FAProp(0.,15.)(5.,13.)(0.,){/Straight}{1}
\FAProp(0.,5.)(5.,7.)(0.,){/Straight}{-1}
\FAProp(20.,17.5)(10.5,14.5)(0.,){/Cycles}{0}
\FAProp(20.,2.5)(10.5,5.5)(0.,){/Sine}{0}
\FAProp(20.,10.)(14.,10.)(0.,){/Sine}{0}
\FAProp(5.,13.)(5.,7.)(0.,){/Cycles}{0}
\FAProp(5.,13.)(10.5,14.5)(0.,){/Straight}{1}
\FALabel(7.25,14.75)[b]{\ssize$\Pu$}
\FAProp(5.,7.)(10.5,5.5)(0.,){/Straight}{-1}
\FALabel(7.25,5.25)[t]{\ssize$\Pu$}
\FAProp(10.5,14.5)(14.,10.)(0.,){/Straight}{1}
\FALabel(13.5,12.)[bl]{\ssize$\Pu$}
\FAProp(10.5,5.5)(14.,10.)(0.,){/Straight}{-1}
\FALabel(13.5,8.)[tl]{\ssize$\Pd,\Ps$}
\FAVert(5.,13.){0}
\FAVert(5.,7.){0}
\FAVert(10.5,14.5){0}
\FAVert(10.5,5.5){0}
\FAVert(14.,10.){0}
\end{diagram}
\hspace*{2em}
\begin{diagram}
\FADiagram{}
\FALabel(-1,15)[r]{\ssize$\Pu$}
\FALabel(-1,5)[r]{\ssize$\bPu$}
\FALabel(21,17.5)[l]{\ssize$\PWp$}
\FALabel(21,10.)[l]{\ssize$\Pg$}
\FALabel(21,2.5)[l]{\ssize$\PWm$}
\FAProp(0.,15.)(5.,13.)(0.,){/Straight}{1}
\FAProp(0.,5.)(5.,7.)(0.,){/Straight}{-1}
\FAProp(20.,17.5)(10.5,14.5)(0.,){/Sine}{0}
\FAProp(20.,2.5)(10.5,5.5)(0.,){/Sine}{0}
\FAProp(20.,10.)(14.,10.)(0.,){/Cycles}{0}
\FAProp(5.,13.)(5.,7.)(0.,){/Cycles}{0}
\FAProp(5.,13.)(10.5,14.5)(0.,){/Straight}{1}
\FALabel(7.25,14.75)[b]{\ssize$\Pu$}
\FAProp(5.,7.)(10.5,5.5)(0.,){/Straight}{-1}
\FALabel(7.25,5.25)[t]{\ssize$\Pu$}
\FAProp(10.5,14.5)(14.,10.)(0.,){/Straight}{1}
\FALabel(13.5,12.)[bl]{\ssize$\Pd,\Ps$}
\FAProp(10.5,5.5)(14.,10.)(0.,){/Straight}{-1}
\FALabel(13.5,8.)[tl]{\ssize$\Pd,\Ps$}
\FAVert(5.,13.){0}
\FAVert(5.,7.){0}
\FAVert(10.5,14.5){0}
\FAVert(10.5,5.5){0}
\FAVert(14.,10.){0}
\end{diagram}
}
\\[-.5em]
\centering{
\begin{diagram}
\FADiagram{}
\FALabel(-1,15)[r]{\ssize$\Pu$}
\FALabel(-1,5)[r]{\ssize$\bPu$}
\FALabel(21,17.5)[l]{\ssize$\PWp$}
\FALabel(21,10.)[l]{\ssize$\Pg$}
\FALabel(21,2.5)[l]{\ssize$\PWm$}
\FAProp(0.,15.)(8.5,14.5)(0.,){/Straight}{1}
\FAProp(0.,5.)(8.5,5.5)(0.,){/Straight}{-1}
\FAProp(20.,17.5)(14.,13.)(0.,){/Sine}{0}
\FAProp(20.,2.5)(14.,7.)(0.,){/Sine}{0}
\FAProp(20.,10.)(5.,10.)(0.,){/Cycles}{0}
\FAProp(14.,13.)(14.,7.)(0.,){/Straight}{1}
\FAProp(14.,13.)(8.5,14.5)(0.,){/Straight}{-1}
\FALabel(12.,14.75)[b]{\ssize$\Pu$}
\FAProp(14.,7.)(8.5,5.5)(0.,){/Straight}{1}
\FALabel(12.,5.25)[t]{\ssize$\Pu$}
\FAProp(8.5,14.5)(5.,10.)(0.,){/Cycles}{0}
\FALabel(15.,12.)[l]{\ssize$\Pd,\Ps$}
\FAProp(5.,10.)(8.5,5.5)(0.,){/Cycles}{0}
\FAVert(14.,13.){0}
\FAVert(14.,7.){0}
\FAVert(8.5,14.5){0}
\FAVert(8.5,5.5){0}
\FAVert(5.,10.){0}
\end{diagram}
}
\vspace*{-1.5ex}
\caption{Pentagon diagrams for \mbox{$\Pu\bPu\to\PWp\PWm\Pg$}.}
\label{fi:bospentagondirect}
\end{figure}

\begin{figure}
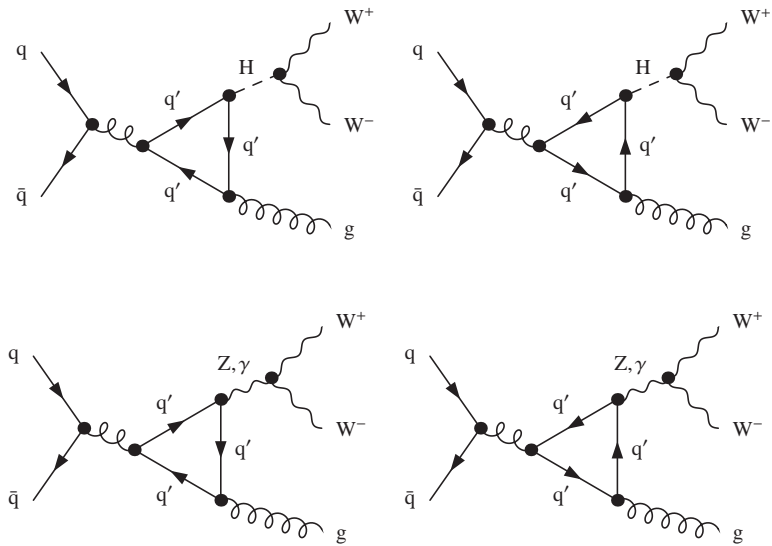

\centering{
\begin{diagram}
\FADiagram{}
\FALabel(-1,15)[r]{\ssize$\Pq$}
\FALabel(-1,5)[r]{\ssize$\bPq$}
\FALabel(21,17.5)[l]{\ssize$\PWp$}
\FALabel(21,10.)[l]{\ssize$\PWm$}
\FALabel(21,2.5)[l]{\ssize$\Pg$}
\FAProp(0.,15.)(3.5,10.)(0.,){/Straight}{1}
\FAProp(0.,5.)(3.5,10.)(0.,){/Straight}{-1}
\FAProp(20.,17.)(16.5,13.5)(0.,){/Sine}{0}
\FAProp(20.,10.)(16.5,13.5)(0.,){/Sine}{0}
\FAProp(20.,3.)(13.,5.)(0.,){/Cycles}{0}
\FAProp(3.5,10.)(7.,8.5)(0.,){/Cycles}{0}
\FAProp(16.5,13.5)(13.,12.)(0.,){/ScalarDash}{0}
\FALabel(14.75,13.5)[br]{\ssize$\PH$}
\FAProp(13.,5.)(7.,8.5)(0.,){/Straight}{1}
\FALabel(9.75,6.)[tr]{\ssize$\Pq'$}
\FAProp(13.,5.)(13.,12.)(0.,){/Straight}{-1}
\FALabel(14.,8.5)[l]{\ssize$\Pq'$}
\FAProp(7.,8.5)(13.,12.)(0.,){/Straight}{1}
\FALabel(9.75,11.)[br]{\ssize$\Pq'$}
\FAVert(3.5,10.){0}
\FAVert(16.5,13.5){0}
\FAVert(13.,5.){0}
\FAVert(7.,8.5){0}
\FAVert(13.,12.){0}
\end{diagram}
\hspace*{2em}
\begin{diagram}
\FADiagram{}
\FALabel(-1,15)[r]{\ssize$\Pq$}
\FALabel(-1,5)[r]{\ssize$\bPq$}
\FALabel(21,17.5)[l]{\ssize$\PWp$}
\FALabel(21,10.)[l]{\ssize$\PWm$}
\FALabel(21,2.5)[l]{\ssize$\Pg$}
\FAProp(0.,15.)(3.5,10.)(0.,){/Straight}{1}
\FAProp(0.,5.)(3.5,10.)(0.,){/Straight}{-1}
\FAProp(20.,17.)(16.5,13.5)(0.,){/Sine}{0}
\FAProp(20.,10.)(16.5,13.5)(0.,){/Sine}{0}
\FAProp(20.,3.)(13.,5.)(0.,){/Cycles}{0}
\FAProp(3.5,10.)(7.,8.5)(0.,){/Cycles}{0}
\FAProp(16.5,13.5)(13.,12.)(0.,){/ScalarDash}{0}
\FALabel(14.75,13.5)[br]{\ssize$\PH$}
\FAProp(13.,5.)(7.,8.5)(0.,){/Straight}{-1}
\FALabel(9.75,6.)[tr]{\ssize$\Pq'$}
\FAProp(13.,5.)(13.,12.)(0.,){/Straight}{1}
\FALabel(14.,8.5)[l]{\ssize$\Pq'$}
\FAProp(7.,8.5)(13.,12.)(0.,){/Straight}{-1}
\FALabel(9.75,11.)[br]{\ssize$\Pq'$}
\FAVert(3.5,10.){0}
\FAVert(16.5,13.5){0}
\FAVert(13.,5.){0}
\FAVert(7.,8.5){0}
\FAVert(13.,12.){0}
\end{diagram}
}
\\[-.5em]
\centering{
\begin{diagram}
\FADiagram{}
\FALabel(-1,15)[r]{\ssize$\Pq$}
\FALabel(-1,5)[r]{\ssize$\bPq$}
\FALabel(21,17.5)[l]{\ssize$\PWp$}
\FALabel(21,10.)[l]{\ssize$\PWm$}
\FALabel(21,2.5)[l]{\ssize$\Pg$}
\FAProp(0.,15.)(3.5,10.)(0.,){/Straight}{1}
\FAProp(0.,5.)(3.5,10.)(0.,){/Straight}{-1}
\FAProp(20.,17.)(16.5,13.5)(0.,){/Sine}{0}
\FAProp(20.,10.)(16.5,13.5)(0.,){/Sine}{0}
\FAProp(20.,3.)(13.,5.)(0.,){/Cycles}{0}
\FAProp(3.5,10.)(7.,8.5)(0.,){/Cycles}{0}
\FAProp(16.5,13.5)(13.,12.)(0.,){/Sine}{0}
\FALabel(15.,13.75)[br]{\ssize$\PZ,\gamma$}
\FAProp(13.,5.)(7.,8.5)(0.,){/Straight}{1}
\FALabel(9.75,6.)[tr]{\ssize$\Pq'$}
\FAProp(13.,5.)(13.,12.)(0.,){/Straight}{-1}
\FALabel(14.,8.5)[l]{\ssize$\Pq'$}
\FAProp(7.,8.5)(13.,12.)(0.,){/Straight}{1}
\FALabel(9.75,11.)[br]{\ssize$\Pq'$}
\FAVert(3.5,10.){0}
\FAVert(16.5,13.5){0}
\FAVert(13.,5.){0}
\FAVert(7.,8.5){0}
\FAVert(13.,12.){0}
\end{diagram}
\hspace*{2em}
\begin{diagram}
\FADiagram{}
\FALabel(-1,15)[r]{\ssize$\Pq$}
\FALabel(-1,5)[r]{\ssize$\bPq$}
\FALabel(21,17.5)[l]{\ssize$\PWp$}
\FALabel(21,10.)[l]{\ssize$\PWm$}
\FALabel(21,2.5)[l]{\ssize$\Pg$}
\FAProp(0.,15.)(3.5,10.)(0.,){/Straight}{1}
\FAProp(0.,5.)(3.5,10.)(0.,){/Straight}{-1}
\FAProp(20.,17.)(16.5,13.5)(0.,){/Sine}{0}
\FAProp(20.,10.)(16.5,13.5)(0.,){/Sine}{0}
\FAProp(20.,3.)(13.,5.)(0.,){/Cycles}{0}
\FAProp(3.5,10.)(7.,8.5)(0.,){/Cycles}{0}
\FAProp(16.5,13.5)(13.,12.)(0.,){/Sine}{0}
\FALabel(15.,13.75)[br]{\ssize$\PZ,\gamma$}
\FAProp(13.,5.)(7.,8.5)(0.,){/Straight}{-1}
\FALabel(9.75,6.)[tr]{\ssize$\Pq'$}
\FAProp(13.,5.)(13.,12.)(0.,){/Straight}{1}
\FALabel(14.,8.5)[l]{\ssize$\Pq'$}
\FAProp(7.,8.5)(13.,12.)(0.,){/Straight}{-1}
\FALabel(9.75,11.)[br]{\ssize$\Pq'$}
\FAVert(3.5,10.){0}
\FAVert(16.5,13.5){0}
\FAVert(13.,5.){0}
\FAVert(7.,8.5){0}
\FAVert(13.,12.){0}
\end{diagram}
}
\vspace*{-1.5ex}
\caption{Fermionic vertex-correction diagrams with intermediate electroweak bosons for \mbox{$\Pq\bPq\to\PWp\PWm\Pg$}.}
\label{fi:fermWWvertexdirect}
\end{figure}

At one-loop level $\PW\PW$+jet production also serves as an off-shell 
continuation of the loop-induced process of Higgs+jet production with 
the Higgs boson decaying into a W-boson pair. 
In these diagrams the off-shell Higgs boson is coupled via a heavy-quark 
loop to two gluons; the graphs for this mechanism are shown in 
\reffi{fi:fermWWvertexdirect} together with vertex-correction diagrams 
with an intermediate electroweak vector boson. 
The Higgs resonance is included in our calculation upon employing
a fixed decay width $\GH$, i.e.\ by replacing the propagator denominator
$(p_\PH^2-\MH^2)$ by $(p_\PH^2-\MH^2+\ri\MH\GH)$, where $p_\PH$ and $\MH$
denote the momentum and the mass of the Higgs boson.
The box contributions to the fermionic corrections are depicted 
in \reffi{fi:fermWWboxdirect}.

\begin{figure}
\centering{
\begin{diagram}
\FADiagram{}
\FALabel(-1,15)[r]{\ssize$\Pq$}
\FALabel(-1,5)[r]{\ssize$\bPq$}
\FALabel(21,17.5)[l]{\ssize$\PWp$}
\FALabel(21,10.)[l]{\ssize$\Pg$}
\FALabel(21,2.5)[l]{\ssize$\PWm$}
\FAProp(0.,15.)(2.5,10.)(0.,){/Straight}{1}
\FAProp(0.,5.)(2.5,10.)(0.,){/Straight}{-1}
\FAProp(20.,17.5)(13.5,13.5)(0.,){/Sine}{0}
\FAProp(20.,10.)(13.5,6.5)(0.,){/Cycles}{0}
\FAProp(20.,2.5)(6.5,6.5)(0.,){/Sine}{0}
\FAProp(2.5,10.)(6.5,13.5)(0.,){/Cycles}{0}
\FAProp(13.5,13.5)(13.5,6.5)(0.,){/Straight}{1}
\FALabel(14.75,10.)[bl]{\ssize$\Pd_j$}
\FAProp(13.5,13.5)(6.5,13.5)(0.,){/Straight}{-1}
\FALabel(10.,14.5)[b]{\ssize$\Pu_i$}
\FAProp(13.5,6.5)(6.5,6.5)(0.,){/Straight}{1}
\FALabel(10.,7.5)[b]{\ssize$\Pd_j$}
\FAProp(6.5,6.5)(6.5,13.5)(0.,){/Straight}{1}
\FALabel(5.5,9.)[tr]{\ssize$\Pu_i$}
\FAVert(2.5,10.){0}
\FAVert(13.5,13.5){0}
\FAVert(13.5,6.5){0}
\FAVert(6.5,6.5){0}
\FAVert(6.5,13.5){0}
\end{diagram}
\hspace*{2em}
\begin{diagram}
\FADiagram{}
\FALabel(-1,15)[r]{\ssize$\Pq$}
\FALabel(-1,5)[r]{\ssize$\bPq$}
\FALabel(21,10.)[l]{\ssize$\PWp$}
\FALabel(21,17.5)[l]{\ssize$\Pg$}
\FALabel(21,2.5)[l]{\ssize$\PWm$}
\FAProp(0.,15.)(2.5,10.)(0.,){/Straight}{1}
\FAProp(0.,5.)(2.5,10.)(0.,){/Straight}{-1}
\FAProp(20.,17.5)(13.5,13.5)(0.,){/Cycles}{0}
\FAProp(20.,10.)(13.5,6.5)(0.,){/Sine}{0}
\FAProp(20.,2.5)(6.5,6.5)(0.,){/Sine}{0}
\FAProp(2.5,10.)(6.5,13.5)(0.,){/Cycles}{0}
\FAProp(13.5,13.5)(13.5,6.5)(0.,){/Straight}{1}
\FALabel(14.75,10.)[bl]{\ssize$\Pu_i$}
\FAProp(13.5,13.5)(6.5,13.5)(0.,){/Straight}{-1}
\FALabel(10.,14.5)[b]{\ssize$\Pu_i$}
\FAProp(13.5,6.5)(6.5,6.5)(0.,){/Straight}{1}
\FALabel(10.,7.5)[b]{\ssize$\Pd_j$}
\FAProp(6.5,6.5)(6.5,13.5)(0.,){/Straight}{1}
\FALabel(5.5,9.)[tr]{\ssize$\Pu_i$}
\FAVert(2.5,10.){0}
\FAVert(13.5,13.5){0}
\FAVert(13.5,6.5){0}
\FAVert(6.5,6.5){0}
\FAVert(6.5,13.5){0}
\end{diagram}
\hspace*{2em}
\begin{diagram}
\FADiagram{}
\FALabel(-1,15)[r]{\ssize$\Pq$}
\FALabel(-1,5)[r]{\ssize$\bPq$}
\FALabel(21,17.5)[l]{\ssize$\PWp$}
\FALabel(21,10.)[l]{\ssize$\PWm$}
\FALabel(21,2.5)[l]{\ssize$\Pg$}
\FAProp(0.,15.)(2.5,10.)(0.,){/Straight}{1}
\FAProp(0.,5.)(2.5,10.)(0.,){/Straight}{-1}
\FAProp(20.,17.)(13.5,13.5)(0.,){/Sine}{0}
\FAProp(20.,10.)(13.5,6.5)(0.,){/Sine}{0}
\FAProp(20.,3.)(6.5,6.5)(0.,){/Cycles}{0}
\FAProp(2.5,10.)(6.5,13.5)(0.,){/Cycles}{0}
\FAProp(13.5,13.5)(13.5,6.5)(0.,){/Straight}{1}
\FALabel(14.75,10.)[bl]{\ssize$\Pd_j$}
\FAProp(13.5,13.5)(6.5,13.5)(0.,){/Straight}{-1}
\FALabel(10.,14.5)[b]{\ssize$\Pu_i$}
\FAProp(13.5,6.5)(6.5,6.5)(0.,){/Straight}{1}
\FALabel(10.,7.5)[b]{\ssize$\Pu_i$}
\FAProp(6.5,6.5)(6.5,13.5)(0.,){/Straight}{1}
\FALabel(5.5,9.)[tr]{\ssize$\Pu_i$}
\FAVert(2.5,10.){0}
\FAVert(13.5,13.5){0}
\FAVert(13.5,6.5){0}
\FAVert(6.5,6.5){0}
\FAVert(6.5,13.5){0}
\end{diagram}
}
\\[-.5em]
\centering{
\begin{diagram}
\FADiagram{}
\FALabel(-1,15)[r]{\ssize$\Pq$}
\FALabel(-1,5)[r]{\ssize$\bPq$}
\FALabel(21,17.5)[l]{\ssize$\PWp$}
\FALabel(21,10.)[l]{\ssize$\Pg$}
\FALabel(21,2.5)[l]{\ssize$\PWm$}
\FAProp(0.,15.)(2.5,10.)(0.,){/Straight}{1}
\FAProp(0.,5.)(2.5,10.)(0.,){/Straight}{-1}
\FAProp(20.,17.5)(13.5,13.5)(0.,){/Sine}{0}
\FAProp(20.,10.)(13.5,6.5)(0.,){/Cycles}{0}
\FAProp(20.,2.5)(6.5,6.5)(0.,){/Sine}{0}
\FAProp(2.5,10.)(6.5,13.5)(0.,){/Cycles}{0}
\FAProp(13.5,13.5)(13.5,6.5)(0.,){/Straight}{-1}
\FALabel(14.75,10.)[bl]{\ssize$\Pu_i$}
\FAProp(13.5,13.5)(6.5,13.5)(0.,){/Straight}{1}
\FALabel(10.,14.5)[b]{\ssize$\Pd_j$}
\FAProp(13.5,6.5)(6.5,6.5)(0.,){/Straight}{-1}
\FALabel(10.,7.5)[b]{\ssize$\Pu_i$}
\FAProp(6.5,6.5)(6.5,13.5)(0.,){/Straight}{-1}
\FALabel(5.5,9.)[tr]{\ssize$\Pd_j$}
\FAVert(2.5,10.){0}
\FAVert(13.5,13.5){0}
\FAVert(13.5,6.5){0}
\FAVert(6.5,6.5){0}
\FAVert(6.5,13.5){0}
\end{diagram}
\hspace*{2em}
\begin{diagram}
\FADiagram{}
\FALabel(-1,15)[r]{\ssize$\Pq$}
\FALabel(-1,5)[r]{\ssize$\bPq$}
\FALabel(21,10.)[l]{\ssize$\PWp$}
\FALabel(21,17.5)[l]{\ssize$\Pg$}
\FALabel(21,2.5)[l]{\ssize$\PWm$}
\FAProp(0.,15.)(2.5,10.)(0.,){/Straight}{1}
\FAProp(0.,5.)(2.5,10.)(0.,){/Straight}{-1}
\FAProp(20.,17.5)(13.5,13.5)(0.,){/Cycles}{0}
\FAProp(20.,10.)(13.5,6.5)(0.,){/Sine}{0}
\FAProp(20.,2.5)(6.5,6.5)(0.,){/Sine}{0}
\FAProp(2.5,10.)(6.5,13.5)(0.,){/Cycles}{0}
\FAProp(13.5,13.5)(13.5,6.5)(0.,){/Straight}{-1}
\FALabel(14.75,10.)[bl]{\ssize$\Pd_j$}
\FAProp(13.5,13.5)(6.5,13.5)(0.,){/Straight}{1}
\FALabel(10.,14.5)[b]{\ssize$\Pd_j$}
\FAProp(13.5,6.5)(6.5,6.5)(0.,){/Straight}{-1}
\FALabel(10.,7.5)[b]{\ssize$\Pu_i$}
\FAProp(6.5,6.5)(6.5,13.5)(0.,){/Straight}{-1}
\FALabel(5.5,9.)[tr]{\ssize$\Pd_j$}
\FAVert(2.5,10.){0}
\FAVert(13.5,13.5){0}
\FAVert(13.5,6.5){0}
\FAVert(6.5,6.5){0}
\FAVert(6.5,13.5){0}
\end{diagram}
\hspace*{2em}
\begin{diagram}
\FADiagram{}
\FALabel(-1,15)[r]{\ssize$\Pq$}
\FALabel(-1,5)[r]{\ssize$\bPq$}
\FALabel(21,17.5)[l]{\ssize$\PWp$}
\FALabel(21,10.)[l]{\ssize$\PWm$}
\FALabel(21,2.5)[l]{\ssize$\Pg$}
\FAProp(0.,15.)(2.5,10.)(0.,){/Straight}{1}
\FAProp(0.,5.)(2.5,10.)(0.,){/Straight}{-1}
\FAProp(20.,17.)(13.5,13.5)(0.,){/Sine}{0}
\FAProp(20.,10.)(13.5,6.5)(0.,){/Sine}{0}
\FAProp(20.,3.)(6.5,6.5)(0.,){/Cycles}{0}
\FAProp(2.5,10.)(6.5,13.5)(0.,){/Cycles}{0}
\FAProp(13.5,13.5)(13.5,6.5)(0.,){/Straight}{-1}
\FALabel(14.75,10.)[bl]{\ssize$\Pu_i$}
\FAProp(13.5,13.5)(6.5,13.5)(0.,){/Straight}{1}
\FALabel(10.,14.5)[b]{\ssize$\Pd_j$}
\FAProp(13.5,6.5)(6.5,6.5)(0.,){/Straight}{-1}
\FALabel(10.,7.5)[b]{\ssize$\Pd_j$}
\FAProp(6.5,6.5)(6.5,13.5)(0.,){/Straight}{-1}
\FALabel(5.5,9.)[tr]{\ssize$\Pd_j$}
\FAVert(2.5,10.){0}
\FAVert(13.5,13.5){0}
\FAVert(13.5,6.5){0}
\FAVert(6.5,6.5){0}
\FAVert(6.5,13.5){0}
\end{diagram}
}
\vspace*{-1ex}
\caption{Fermionic box diagrams for \mbox{$\Pq\bPq\to\PWp\PWm\Pg$}.}
\label{fi:fermWWboxdirect}
\end{figure}

An on-shell renormalization is performed for the wave functions of the external QCD partons and an $\overline{\mathrm{MS}}$ renormalization for the strong coupling with the massive-quark loops in the gluon self-energy subtracted at zero momentum. 
Since only massless external
quarks are involved, the whole counterterm amplitude can be written as
\begin{eqnarray}
\mathcal{M}_{\mathrm{ct}}=\left(\delta Z_\Pq+\frac12\delta Z_\PA+\delta Z_{\gs}\right)\mathcal{M}_{\mathrm{LO}}\;,
\end{eqnarray}
where the renormalization constants are calculated as follows,
\begin{eqnarray}
\label{eq:Zquark}
\delta Z_\Pq&=&-\frac{\als}{3\pi}\left[\DUV-\DIRa\right]\;,\\
\delta Z_\PA&=&-\frac{\als}{2\pi}
\Biggl[\left(\frac{\Nf}{3}-\frac{5}{2}\right)\left(\DUV-\DIRa\right)
+\frac13\sum_{\Pq\atop m_\Pq\neq0}\Biggl(\DUV+\ln\frac{\MW^2}{m_\Pq^2}\Biggr)
\Biggr]\;,\\
\delta Z_{\gs}&=&\frac{\als}{4\pi}\Biggl[\left(\frac{\Nf}{3}-\frac{11}{2}\right)
\left(\DUV-\ln\frac{\muren^2}{\MW^2}\right)
+\frac13\sum_{\Pq\atop m_\Pq\neq0}\Biggl(\DUV+\ln\frac{\MW^2}{m_\Pq^2}\Biggr)
\Biggr]\;.
\end{eqnarray}
The sum over $\Pq$ runs over all massive quarks, namely $\Pq=\Pt$ in the 
five-flavour scheme $(\Nf=5)$
and $\Pq=\Pb,\Pt$ in the four-flavour scheme $(\Nf=4)$.
Both UV and IR (soft and collinear) divergences are regularized in
$D=4-2\epsilon$ dimensions, and the divergences arising as poles in $\epsilon$
are quantified by 
\beq
\Delta_k(\mu) = \left(\frac{4\pi\mu^2}{\MW^2}\right)^\epsilon
\frac{\Gamma(1+\epsilon)}{\epsilon^k}, \qquad k=1,2,
\eeq
where $\mu$ is the arbitrary reference scale of dimensional regularization.
Superscripts ``UV'' or ``IR'' on $\Delta_k(\mu)$
indicate the origin of the divergences.

As is well known, the use of the Dirac matrix $\gamma_5$ deserves
some care within dimensional regularization, because the Dirac algebra
with $\gamma_5$ does not admit a straightforward analytical continuation
to $D\ne4$ dimensions. In both calculations
of the virtual corrections we use a simple recipe with 
a $\gamma_5$ that anticommutes with all other Dirac matrices but
nevertheless obeys the usual trace relations from four dimensions.%
\footnote{The authors of \citere{Sanguinetti:2008xt} adopted the
more rigorous, but cumbersome approach of `t~Hooft and 
Veltman~\cite{'tHooft:1972fi},
where $\gamma_5$ is split into an anticommuting four-dimensional part 
and a commuting remainder, and find full agreement with our result
(see also \citere{Bern:2008ef} for some details).}
Although this approach is, of course, not fully consistent from the
mathematical point of view, it is well known that it delivers the
correct results at NLO as long as closed fermions loops are calculated
``generically'', i.e.\ they are calculated for general fermion flavours
followed by a subsequent summation over all fermions by only changing
the quantum numbers. 
The arguments supporting this scheme in the presence of UV divergences
can be found in \citere{Jegerlehner:2000dz} (see also references therein).
In view of dimensionally regularized IR divergences, the situation
is much simpler because of their factorization off
tree-level structures. As long as it is guaranteed that these residual
tree structures are treated in the same way in the virtual and real
corrections, between which the cancellations of IR
divergences take place, the
IR-finite sum is well defined. In fact for the compensation of 
IR divergences no \mbox{(anti-)c}om\-mu\-ta\-tion rules involving $\gamma_5$ 
are needed at all. Since the naive scheme described above does not induce
any spurios terms from the regularization, it automatically delivers
the correct IR-finite sum of virtual and real corrections.

\textit{Version 1} of the virtual corrections is essentially obtained as for the related processes of $\Pt\bar\Pt\PH$
\cite{Beenakker:2002nc} and $\Pt\bar\Pt{+}$jet \cite{Dittmaier:2007wz}
production.
Feynman diagrams and amplitudes are generated with 
{\sl Feyn\-Arts}~1.0 \cite{Kublbeck:1990xc}
and further processed with in-house {\sl Mathematica} routines,
which automatically create an output in {\sl Fortran}. 
The IR divergences (soft and collinear) are analytically separated
from the finite remainder in terms of triangle subdiagrams,
as described in \citeres{Beenakker:2002nc,Dittmaier:2003bc}.
This separation, in particular, allows for a transparent evaluation
of so-called rational terms that originate from $D$-dependent terms
multiplying IR divergences, which appear as single or double poles in 
$\epsilon$.
As generally shown in \citere{Bredenstein:2008zb}, after properly
separating IR from UV divergences such rational terms
originating from IR
divergences completely cancel; this general result is confirmed in our
explicit calculation.
For the results presented in \citere{Dittmaier:2007th},
the pentagon tensor integrals were directly reduced to box
integrals following \citere{Denner:2002ii}, while box and lower-point 
integrals were reduced \`a la Passarino--Veltman \cite{Passarino:1978jh} 
to scalar integrals. This procedure completely avoids inverse Gram 
determinants of external momenta in the reduction step from 5-point to 
4-point integrals, but the reduction of box and lower-point tensor
integrals involves such inverse determinants via the 
Passarino--Veltman algorithm. Although these inverse determinants
jeopardize the numerical stability in regions where such determinants
are small, sufficient numerical stability was already achieved.
Meanwhile the tensor reduction has been further improved using
the methods of \citere{Denner:2005nn}. In detail the reduction
of pentagons is performed by the more recent procedure of
\citere{Denner:2005nn} (similar to a method proposed in
\citere{Binoth:2005ff}), and the Passarino--Veltman reduction for
4-point integrals, etc., is supplemented by the dedicated 
expansions for small Gram and kinematical determinants in the
regions where these determinants become small.%
\footnote{Similar procedures based on expansions in small determinants 
  have also been proposed in \citere{Giele:2004ub}.}
The scalar one-loop integrals
are either calculated analytically or using the results of
\citeres{'tHooft:1978xw,Beenakker:1988jr,Denner:1991qq}.

\textit{Version 2} of the evaluation of loop diagrams starts
with the generation of diagrams and amplitudes via 
{\sl Feyn\-Arts}~3.4 \cite{Hahn:2000kx}
which are then further manipulated with {\sl FormCalc}~6.0
\cite{Hahn:1998yk} and eventually
automatically translated into {\sl Fortran} code.
The whole reduction of tensor to scalar integrals is done with the
help of the {\sl LoopTools} library \cite{Hahn:1998yk},
which employs the method of \citere{Denner:2002ii} for the
5-point tensor integrals, Passarino--Veltman \cite{Passarino:1978jh}
reduction for the lower-point tensors, and the {\sl FF} package 
\cite{vanOldenborgh:1989wn,vanOldenborgh:1991yc} for the evaluation 
of regular scalar integrals.
The dimensionally regularized soft or collinear singular 3- and 4-point
integrals had to be added to this library. To this end, the
explicit results of \citere{Dittmaier:2003bc} for the vertex and of 
\citere{Bern:1993kr}
for the box integrals (with appropriate analytical continuations)
are taken.
Actually the {\sl FormCalc} package assumes a four-dimensional
regularization scheme for IR divergences (such as the concept of
an infinitesimal photon mass in QED), i.e.\ rational terms of IR
origin are neglected by {\sl FormCalc}. However, as mentioned above,
in \citere{Bredenstein:2008zb} it was generally shown that such
rational terms consistently cancel if UV and IR divergences are
properly separated. Owing to this property the algebraic result
of {\sl FormCalc} for the unrenormalized amplitudes could be used 
without any modification, apart from supplementing the needed
IR-singular scalar integrals.

\subsection{Real corrections}
\label{se:real}
The matrix elements for the real corrections are given by the processes
$0 \to \PW^+\PW^-   \Pq \bar \Pq \Pg \Pg$ and
$0 \to \PW^+\PW^-   \Pq \bar \Pq \Pq' \bar \Pq'$
with a large variety of flavour insertions for the light quarks
$\Pq$ and $\Pq'$.
The partonic processes are obtained from these matrix elements 
by all possible crossings of quarks and gluons into the initial state.
The evaluation of the real-emission amplitudes is
performed in two independent ways. Both evaluations employ 
(independent implementations of) the dipole subtraction formalism 
\cite{Catani:1996vz}
for the extraction of IR singularities and for their
combination with the virtual corrections. 

\textit{Version 1} employs the Weyl--van-der-Waerden formalism (as described in
\citere{Dittmaier:1998nn}) for the calculation of the helicity amplitudes.
The phase-space integration is performed by a 
multi-channel Monte Carlo integrator~\cite{Berends:1994pv} 
with weight optimization~\cite{Kleiss:1994qy} 
written in {\sl C++}, which is constructed similar to {\sl RacoonWW}
\cite{Denner:1999gp,Roth:1999kk}.
The results for cross sections with two resolved hard jets
have been checked against results obtained with
{\sl Whizard}~1.50~\cite{Kilian:2007gr} 
and {\sl Sherpa}~1.0.8~\cite{Gleisberg:2003xi}. 
Details on this part of the calculation can be found in
\citere{Kallweit:2006dt}.
In order to improve the integration, additional channels are 
included for the integration of the
difference of the real-emission matrix elements and
the subtraction terms.
\looseness-1

\textit{Version 2} is based on scattering amplitudes calculated 
with {\sl Madgraph} \cite{Stelzer:1994ta} generated code.
The code has been modified to allow for a non-diagonal 
quark mixing matrix and the extraction of the required colour and 
spin structures. The latter enter the evaluation of the dipoles in the
Catani--Seymour subtraction method. The evaluation of the individual dipoles 
was performed using a {\sl C++} library developed during the calculation of 
the NLO corrections for $\Pt\bar\Pt{+}$jet \cite{Dittmaier:2007wz}.
For the phase-space integration a
simple mapping has been used where the phase space is generated from 
a sequential splitting.\\

In the Catani--Seymour dipole subtraction formalism all the IR
divergent pieces with LO kinematics are collected in the so-called 
$\mathcal{I}$-operator---for details we refer to \citere{Catani:1996vz}. Since
the IR finiteness of the virtual corrections combined with the IR
singularities obtained from the real corrections provides an important
check of the calculation, we reproduce here the explicit form of the 
$\mathcal{I}$-operator,
\begin{eqnarray}
\label{eq:cs:Ioperator}
&&\mathcal{I}(p_a,p_b;\ldots,p_i) = 
\int_1 d\sigma^{\mathrm{A}}_{ab}(p_a,p_b;\ldots p_i)
=-\frac{\alpha_{\mathrm{s}}}{2\pi}\,
\overline{|\mathcal{A}_{\mathrm{LO},ab}(p_a,p_b;\ldots,p_i)|^2}\nn\\
&&\hspace*{0pt}\times\left\{-4T_{\mathrm{R}}C_{\mathrm{A}}\left[4\DIRb+2\DIRa\left(\frac{10}3-\ln\frac{2p_ap_i}{\MW^2}-\ln\frac{2p_bp_i}{\MW^2}\right)\right.\right.\nn\\
&&\hspace*{20pt}\left.\left.+\ln^2\frac{2p_ap_i}{\MW^2}+\ln^2\frac{2p_bp_i}{\MW^2}-\frac{10}3\left(\ln\frac{2p_ap_i}{\MW^2}+\ln\frac{2p_bp_i}{\MW^2}\right)+\frac{190}9-\frac{8\pi^2}3\right]\right.\nn\\
&&\hspace*{4pt}\left.+4T_{\mathrm{R}}(C_{\mathrm{A}}-2C_{\mathrm{F}})\left[2\DIRb+\DIRa\left(3-2\ln\frac{2p_ap_b}{\MW^2}\right)+\ln^2\frac{2p_ap_b}{\MW^2}-3\ln\frac{2p_ap_b}{\MW^2}+10-\frac{4\pi^2}3\right]\right.\nn\\
&&\hspace*{4pt}\left.-4T_{\mathrm{R}}^2N_\Pf\left[-\frac43\DIRa+\frac23\left(\ln\frac{2p_ap_i}{\MW^2}+\ln\frac{2p_bp_i}{\MW^2}\right)-\frac{32}9\right]\right\}+\mathcal{O}(\epsilon),
\end{eqnarray}
where $\mathcal{A}_{\mathrm{LO},ab}$ denotes the colour-stripped LO amplitude for the incoming quark--antiquark pair $ab$ and the outgoing gluon $i$. 
Due to the simple colour structure the $\mathcal{I}$-operator does not
involve non-trivial colour correlations. The $\mathcal{I}$-operators for the remaining subprocesses are obtained by exchanging the respective momenta and using the crossed matrix elements.

\section{Inclusion of gauge-boson decays}
\label{se:decays}
Since the produced $\PW$ bosons are unstable particles, their decays
should be included into the analysis. In this context, especially the
case of both gauge bosons decaying leptonically is of interest due to
its clean signature in the detector. Therefore, only leptonic
gauge-boson decays are considered here. The inclusion of the decays is
performed following three different strategies for the LO processes,
which are presented in the following paragraphs. The improved 
narrow-width approximation---the on-shell approximation
  together with the exact treatment of the W-polarization---delivers an 
appropriate compromise between 
the complexity of the calculations and the accuracy of the results, 
which is demonstrated in a comparison of LO results presented in \refse{se:nu:approximations}. 
Therefore, the NLO calculation is performed by means of the improved narrow-width approximation.

\subsection{Full calculation with off-shell $\PW$ bosons}
\label{se:dy:fullcalculation}

The diagrams contributing to \mbox{\tppllj}
can be subdivided into two classes, namely diagrams showing two gauge-boson propagators that can become resonant and other diagrams containing only one $\PW$ resonance. 
The doubly-resonant diagrams comprise all the WW+jet-production diagrams of \reffi{fi:dy:WWdoubleresonant} with the two massive gauge bosons decaying 
into leptons.%
\def\PWplus{\PW}
\def\PWminus{\PW}
\begin{figure}
\centering{
\begin{diagram}
\FADiagram{}
\FAProp(0.,15.)(6.,14.5)(0.,){/Straight}{1}
\FALabel(-1.,15.)[r]{\ssize$\Pu$}
\FAProp(0.,5.)(6.,5.5)(0.,){/Straight}{-1}
\FALabel(-1.,5.)[r]{\ssize$\bar\Pu$}
\FAProp(14.,17.5)(6.,14.5)(0.,){/Sine}{0}
\FALabel(10,17.5)[b]{\ssize$\PWplus$}
\FAProp(14.,17.5)(20.,20)(0.,){/Straight}{1}
\FALabel(21,20)[l]{\ssize$\Pne$}
\FAProp(14.,17.5)(20.,15)(0.,){/Straight}{-1}
\FALabel(21,15)[l]{\ssize$\Pe^+$}
\FAProp(14.,2.5)(6.,5.5)(0.,){/Sine}{0}
\FALabel(10,2.5)[t]{\ssize$\PWminus$}
\FAProp(14.,2.5)(20.,0)(0.,){/Straight}{-1}
\FALabel(21,0)[l]{\ssize$\Pnmubar$}
\FAProp(14.,2.5)(20.,5)(0.,){/Straight}{1}
\FALabel(21,5)[l]{\ssize$\Pmu^-$}
\FAProp(20.,10.)(6.,10.)(0.,){/Cycles}{0}
\FALabel(21,10.)[l]{\ssize$\Pg$}
\FAProp(6.,14.5)(6.,10.)(0.,){/Straight}{1}
\FALabel(4.93,12.25)[r]{\ssize$\Pd,\Ps$}
\FAProp(6.,5.5)(6.,10.)(0.,){/Straight}{-1}
\FALabel(4.93,7.75)[r]{\ssize$\Pd,\Ps$}
\FAVert(14.,17.5){0}
\FAVert(14.,2.5){0}
\FAVert(6.,14.5){0}
\FAVert(6.,5.5){0}
\FAVert(6.,10.){0}
\end{diagram}
\hspace*{2em}
\begin{diagram}
\FADiagram{}
\FAProp(0.,15.)(6.,14.5)(0.,){/Straight}{1}
\FALabel(-1.,15.)[r]{\ssize$\Pu$}
\FAProp(0.,5.)(6.,5.5)(0.,){/Straight}{-1}
\FALabel(-1.,5.)[r]{\ssize$\bar\Pu$}
\FAProp(14.,12.5)(6.,10.)(0.,){/Sine}{0}
\FALabel(10,9.5)[t]{\ssize$\PWplus$}
\FAProp(14.,12.5)(20.,15)(0.,){/Straight}{1}
\FALabel(21,15)[l]{\ssize$\Pne$}
\FAProp(14.,12.5)(20.,10)(0.,){/Straight}{-1}
\FALabel(21,10)[l]{\ssize$\Pe^+$}
\FAProp(14.,2.5)(6.,5.5)(0.,){/Sine}{0}
\FALabel(10,2.5)[t]{\ssize$\PWminus$}
\FAProp(14.,2.5)(20.,0)(0.,){/Straight}{-1}
\FALabel(21,0)[l]{\ssize$\Pnmubar$}
\FAProp(14.,2.5)(20.,5)(0.,){/Straight}{1}
\FALabel(21,5)[l]{\ssize$\Pmu^-$}
\FAProp(20.,20.)(6.,14.5)(0.,){/Cycles}{0}
\FALabel(21,20.)[l]{\ssize$\Pg$}
\FAProp(6.,14.5)(6.,10.)(0.,){/Straight}{1}
\FALabel(4.93,12.25)[r]{\ssize$\Pu$}
\FAProp(6.,5.5)(6.,10.)(0.,){/Straight}{-1}
\FALabel(4.93,7.75)[r]{\ssize$\Pd,\Ps$}
\FAVert(14.,12.5){0}
\FAVert(14.,2.5){0}
\FAVert(6.,10.){0}
\FAVert(6.,5.5){0}
\FAVert(6.,14.5){0}
\end{diagram}
\hspace*{2em}
\begin{diagram}
\FADiagram{}
\FAProp(0.,15.)(6.,14.5)(0.,){/Straight}{1}
\FALabel(-1.,15.)[r]{\ssize$\Pu$}
\FAProp(0.,5.)(6.,5.5)(0.,){/Straight}{-1}
\FALabel(-1.,5.)[r]{\ssize$\bar\Pu$}
\FAProp(14.,17.5)(6.,14.5)(0.,){/Sine}{0}
\FALabel(10.,17.5)[b]{\ssize$\PWplus$}
\FAProp(14.,17.5)(20.,20)(0.,){/Straight}{1}
\FALabel(21.,20)[l]{\ssize$\Pne$}
\FAProp(14.,17.5)(20.,15)(0.,){/Straight}{-1}
\FALabel(21.,15)[l]{\ssize$\Pe^+$}
\FAProp(14.,7.5)(6.,10.)(0.,){/Sine}{0}
\FALabel(10.,10.5)[b]{\ssize$\PWminus$}
\FAProp(14.,7.5)(20.,5.)(0.,){/Straight}{-1}
\FALabel(21.,5.)[l]{\ssize$\Pnmubar$}
\FAProp(14.,7.5)(20.,10.)(0.,){/Straight}{1}
\FALabel(21.,10.)[l]{\ssize$\Pmu^-$}
\FAProp(20.,0.)(6.,5.5)(0.,){/Cycles}{0}
\FALabel(21.,0.)[l]{\ssize$\Pg$}
\FAProp(6.,14.5)(6.,10.)(0.,){/Straight}{1}
\FALabel(4.93,12.25)[r]{\ssize$\Pd,\Ps$}
\FAProp(6.,5.5)(6.,10.)(0.,){/Straight}{-1}
\FALabel(4.93,7.75)[r]{\ssize$\Pu$}
\FAVert(14.,17.5){0}
\FAVert(14.,7.5){0}
\FAVert(6.,14.5){0}
\FAVert(6.,5.5){0}
\FAVert(6.,10.){0}
\end{diagram}
}
\vspace*{1.5ex}
\centering{
\begin{diagram}
\FADiagram{}
\FAProp(0.,15.)(4.,12.5)(0.,){/Straight}{1}
\FALabel(-1.,15.)[r]{\ssize$\Pu$}
\FAProp(0.,5.)(4.,5.5)(0.,){/Straight}{-1}
\FALabel(-1.,5.)[r]{\ssize$\bar\Pu$}
\FAProp(10.,12.5)(4.,12.5)(0.,){/Sine}{0}
\FALabel(7.5,11.5)[t]{\ssize$\PZ,\gamma$}
\FAProp(14.,17.5)(10.,12.5)(0.,){/Sine}{0}
\FALabel(12.,16.)[br]{\ssize$\PWplus$}
\FAProp(14.,17.5)(20.,20)(0.,){/Straight}{1}
\FALabel(21.,20)[l]{\ssize$\Pne$}
\FAProp(14.,17.5)(20.,15)(0.,){/Straight}{-1}
\FALabel(21.,15)[l]{\ssize$\Pe^+$}
\FAProp(14.,7.5)(10.,12.5)(0.,){/Sine}{0}
\FALabel(13.,10.)[bl]{\ssize$\PWminus$}
\FAProp(14.,7.5)(20.,5.)(0.,){/Straight}{-1}
\FALabel(21.,5.)[l]{\ssize$\Pnmubar$}
\FAProp(14.,7.5)(20.,10.)(0.,){/Straight}{1}
\FALabel(21.,10.)[l]{\ssize$\Pmu^-$}
\FAProp(20.,0.)(4.,5.5)(0.,){/Cycles}{0}
\FALabel(21,0.)[l]{\ssize$\Pg$}
\FAProp(4.,12.5)(4.,5.5)(0.,){/Straight}{1}
\FALabel(3.,9)[r]{\ssize$\Pu$}
\FAVert(14.,7.5){0}
\FAVert(14.,17.5){0}
\FAVert(10.,12.5){0}
\FAVert(4.,12.5){0}
\FAVert(4.,5.5){0}
\end{diagram}
\hspace*{2em}
\begin{diagram}
\FADiagram{}
\FAProp(0.,15.)(4.,14.5)(0.,){/Straight}{1}
\FALabel(-1.,15.)[r]{\ssize$\Pu$}
\FAProp(0.,5.)(4.,7.5)(0.,){/Straight}{-1}
\FALabel(-1.,5.)[r]{\ssize$\bar\Pu$}
\FAProp(10.,7.5)(4.,7.5)(0.,){/Sine}{0}
\FALabel(7.,6.5)[t]{\ssize$\PZ,\gamma$}
\FAProp(14.,12.5)(10.,7.5)(0.,){/Sine}{0}
\FALabel(12.,11.)[br]{\ssize$\PWplus$}
\FAProp(14.,12.5)(20.,15.)(0.,){/Straight}{1}
\FALabel(21.,15)[l]{\ssize$\Pne$}
\FAProp(14.,12.5)(20.,10.)(0.,){/Straight}{-1}
\FALabel(21.,10)[l]{\ssize$\Pe^+$}
\FAProp(14.,2.5)(10.,7.5)(0.,){/Sine}{0}
\FALabel(13.,5)[bl]{\ssize$\PWminus$}
\FAProp(14.,2.5)(20.,0.)(0.,){/Straight}{-1}
\FALabel(21.,0.)[l]{\ssize$\Pnmubar$}
\FAProp(14.,2.5)(20.,5.)(0.,){/Straight}{1}
\FALabel(21.,5.)[l]{\ssize$\Pmu^-$}
\FAProp(20.,20.)(4.,14.5)(0.,){/Cycles}{0}
\FALabel(21,20.)[l]{\ssize$\Pg$}
\FAProp(4.,14.5)(4.,7.5)(0.,){/Straight}{1}
\FALabel(3.,11)[r]{\ssize$\Pu$}
\FAVert(14.,2.5){0}
\FAVert(14.,12.5){0}
\FAVert(10.,7.5){0}
\FAVert(4.,14.5){0}
\FAVert(4.,7.5){0}
\end{diagram}
}
\caption{Diagrams with two resonant W propagators in the partonic subprocess \mbox{$\Pu\Pubar\to\Pne\Pep\Pmum\Pnmubar\Pg$}.}
\label{fi:dy:WWdoubleresonant}
\end{figure}

Aside from this doubly-resonant group, a number of diagrams that in
general contain only one possibly resonant gauge-boson propagator contributes. Their topologies are characterized by the situation that one neutral gauge boson decays to a lepton--antilepton pair with the second pair attached to this leptonic fermion chain via a $\PW$ boson. All diagrams for this configuration are given in \reffi{fi:dy:WWsingleresonant} for one specific partonic channel. The configuration with both the Z and the W propagator being resonant is kinematically allowed, but does not significantly contribute, since the corresponding phase-space region is too small.

\begin{figure}
\centering{
\begin{diagram}
\FADiagram{}
\FAProp(0.,15.)(4.,12.5)(0.,){/Straight}{1}
\FALabel(-1.,15.)[r]{\ssize$\Pu$}
\FAProp(0.,5.)(4.,5.5)(0.,){/Straight}{-1}
\FALabel(-1.,5.)[r]{\ssize$\bar\Pu$}
\FAProp(4.,12.5)(4.,5.5)(0.,){/Straight}{1}
\FALabel(3.,9)[r]{\ssize$\Pu$}
\FAProp(20.,0.)(4.,5.5)(0.,){/Cycles}{0}
\FALabel(21,0.)[l]{\ssize$\Pg$}
\FAProp(8.,12.5)(4.,12.5)(0.,){/Sine}{0}
\FALabel(6,13.5)[b]{\ssize$\PZ,\gamma$}
\FAProp(8.,12.5)(20.,20)(0.,){/Straight}{1}
\FALabel(21,20)[l]{\ssize$\Pmu^-$}
\FAProp(16.,12.5)(20.,15)(0.,){/Straight}{1}
\FALabel(21,15)[l]{\ssize$\Pne$}
\FAProp(12.,10.)(16.,12.5)(0.,){/Sine}{0}
\FALabel(14.25,12.5)[br]{\ssize$\PWplus$}
\FAProp(16.,12.5)(20.,10)(0.,){/Straight}{-1}
\FALabel(21,10)[l]{\ssize$\Pe^+$}
\FAProp(8.,12.5)(12.,10)(0.,){/Straight}{-1}
\FALabel(9.75,10.25)[tr]{\ssize$\Pmu$}
\FAProp(12.,10)(20.,5)(0.,){/Straight}{-1}
\FALabel(21,5)[l]{\ssize$\Pnmubar$}
\FAVert(16.,12.5){0}
\FAVert(12.,10.){0}
\FAVert(8.,12.5){0}
\FAVert(4.,12.5){0}
\FAVert(4.,5.5){0}
\end{diagram}
\hspace*{2em}
\begin{diagram}
\FADiagram{}
\FAProp(0.,15.)(4.,12.5)(0.,){/Straight}{1}
\FALabel(-1.,15.)[r]{\ssize$\Pu$}
\FAProp(0.,5.)(4.,5.5)(0.,){/Straight}{-1}
\FALabel(-1.,5.)[r]{\ssize$\bar\Pu$}
\FAProp(4.,12.5)(4.,5.5)(0.,){/Straight}{1}
\FALabel(3.,9)[r]{\ssize$\Pu$}
\FAProp(20.,0.)(4.,5.5)(0.,){/Cycles}{0}
\FALabel(21,0.)[l]{\ssize$\Pg$}
\FAProp(8.,12.5)(4.,12.5)(0.,){/Sine}{0}
\FALabel(6.,13.5)[b]{\ssize$\PZ$}
\FAProp(12.,15.)(20.,20.)(0.,){/Straight}{1}
\FALabel(21,20)[l]{\ssize$\Pmu^-$}
\FAProp(8.,12.5)(12.,15)(0.,){/Straight}{1}
\FALabel(11.,13.)[tl]{\ssize$\Pnmu$}
\FAProp(16.,12.5)(20.,15)(0.,){/Straight}{1}
\FALabel(21,15)[l]{\ssize$\Pne$}
\FAProp(12.,15.)(16.,12.5)(0.,){/Sine}{0}
\FALabel(14.5,14.)[bl]{\ssize$\PWplus$}
\FAProp(16.,12.5)(20.,10)(0.,){/Straight}{-1}
\FALabel(21,10)[l]{\ssize$\Pe^+$}
\FAProp(8.,12.5)(20.,5)(0.,){/Straight}{-1}
\FALabel(21,5)[l]{\ssize$\Pnmubar$}
\FAVert(16.,12.5){0}
\FAVert(12.,15.){0}
\FAVert(8.,12.5){0}
\FAVert(4.,12.5){0}
\FAVert(4.,5.5){0}
\end{diagram}
\hspace*{2em}
\begin{diagram}
\FADiagram{}
\FAProp(0.,15.)(4.,14.5)(0.,){/Straight}{1}
\FALabel(-1.,15.)[r]{\ssize$\Pu$}
\FAProp(0.,5.)(4.,7.5)(0.,){/Straight}{-1}
\FALabel(-1.,5.)[r]{\ssize$\bar\Pu$}
\FAProp(4.,14.5)(4.,7.5)(0.,){/Straight}{1}
\FALabel(3.,11.)[r]{\ssize$\Pu$}
\FAProp(20.,20.)(4.,14.5)(0.,){/Cycles}{0}
\FALabel(21,20.)[l]{\ssize$\Pg$}
\FAProp(8.,7.5)(4.,7.5)(0.,){/Sine}{0}
\FALabel(6.,6.5)[t]{\ssize$\PZ,\gamma$}
\FAProp(8.,7.5)(20.,15.)(0.,){/Straight}{1}
\FALabel(21.,15.)[l]{\ssize$\Pmu^-$}
\FAProp(16.,7.5)(20.,10.)(0.,){/Straight}{1}
\FALabel(21.,10.)[l]{\ssize$\Pne$}
\FAProp(12.,5.)(16.,7.5)(0.,){/Sine}{0}
\FALabel(14.25,7.5)[br]{\ssize$\PWplus$}
\FAProp(16.,7.5)(20.,5.)(0.,){/Straight}{-1}
\FALabel(21,5.)[l]{\ssize$\Pe^+$}
\FAProp(8.,7.5)(12.,5.)(0.,){/Straight}{-1}
\FALabel(10.25,4.75)[tr]{\ssize$\Pmu$}
\FAProp(12.,5.)(20.,0.)(0.,){/Straight}{-1}
\FALabel(21,0.)[l]{\ssize$\Pnmubar$}
\FAVert(16.,7.5){0}
\FAVert(12.,5.){0}
\FAVert(8.,7.5){0}
\FAVert(4.,14.5){0}
\FAVert(4.,7.5){0}
\end{diagram}
}
\vspace*{1.5em}
\centering{
\begin{diagram}
\FADiagram{}
\FAProp(0.,15.)(4.,14.5)(0.,){/Straight}{1}
\FALabel(-1.,15.)[r]{\ssize$\Pu$}
\FAProp(0.,5.)(4.,7.5)(0.,){/Straight}{-1}
\FALabel(-1.,5.)[r]{\ssize$\bar\Pu$}
\FAProp(4.,7.5)(4.,14.5)(0.,){/Straight}{-1}
\FALabel(3.,11)[r]{\ssize$\Pu$}
\FAProp(20.,20.)(4.,14.5)(0.,){/Cycles}{0}
\FALabel(21.,20.)[l]{\ssize$\Pg$}
\FAProp(8.,7.5)(4.,7.5)(0.,){/Sine}{0}
\FALabel(6.,6.5)[t]{\ssize$\PZ$}
\FAProp(12.,10.)(20.,15.)(0.,){/Straight}{1}
\FALabel(21.,15)[l]{\ssize$\Pmu^-$}
\FAProp(8.,7.5)(12.,10.)(0.,){/Straight}{1}
\FALabel(11.,8.)[tl]{\ssize$\Pnmu$}
\FAProp(16.,7.5)(20.,10.)(0.,){/Straight}{1}
\FALabel(21.,10.)[l]{\ssize$\Pne$}
\FAProp(12.,10.)(16.,7.5)(0.,){/Sine}{0}
\FALabel(14.5,9.)[bl]{\ssize$\PWplus$}
\FAProp(16.,7.5)(20.,5.)(0.,){/Straight}{-1}
\FALabel(21.,5.)[l]{\ssize$\Pe^+$}
\FAProp(8.,7.5)(20.,0.)(0.,){/Straight}{-1}
\FALabel(21.,0.)[l]{\ssize$\Pnmubar$}
\FAVert(16.,7.5){0}
\FAVert(12.,10.){0}
\FAVert(8.,7.5){0}
\FAVert(4.,14.5){0}
\FAVert(4.,7.5){0}
\end{diagram}
\hspace*{2em}
\begin{diagram}
\FADiagram{}
\FAProp(0.,15.)(4.,12.5)(0.,){/Straight}{1}
\FALabel(-1.,15.)[r]{\ssize$\Pu$}
\FAProp(0.,5.)(4.,5.5)(0.,){/Straight}{-1}
\FALabel(-1.,5.)[r]{\ssize$\bar\Pu$}
\FAProp(4.,12.5)(4.,5.5)(0.,){/Straight}{1}
\FALabel(3.,9)[r]{\ssize$\Pu$}
\FAProp(20.,0.)(4.,5.5)(0.,){/Cycles}{0}
\FALabel(21,0.)[l]{\ssize$\Pg$}
\FAProp(8.,12.5)(4.,12.5)(0.,){/Sine}{0}
\FALabel(6,13.5)[b]{\ssize$\PZ$}
\FAProp(8.,12.5)(20.,20)(0.,){/Straight}{1}
\FALabel(21,20)[l]{\ssize$\Pne$}
\FAProp(16.,12.5)(20.,15)(0.,){/Straight}{1}
\FALabel(21,15)[l]{\ssize$\Pmu^-$}
\FAProp(12.,10.)(16.,12.5)(0.,){/Sine}{0}
\FALabel(14.25,12.5)[br]{\ssize$\PWminus$}
\FAProp(16.,12.5)(20.,10)(0.,){/Straight}{-1}
\FALabel(21,10)[l]{\ssize$\Pnmubar$}
\FAProp(8.,12.5)(12.,10)(0.,){/Straight}{-1}
\FALabel(9.75,10.25)[tr]{\ssize$\Pne$}
\FAProp(12.,10)(20.,5)(0.,){/Straight}{-1}
\FALabel(21,5)[l]{\ssize$\Pe^+$}
\FAVert(16.,12.5){0}
\FAVert(12.,10.){0}
\FAVert(8.,12.5){0}
\FAVert(4.,12.5){0}
\FAVert(4.,5.5){0}
\end{diagram}
\hspace*{2em}
\begin{diagram}
\FADiagram{}
\FAProp(0.,15.)(4.,12.5)(0.,){/Straight}{1}
\FALabel(-1.,15.)[r]{\ssize$\Pu$}
\FAProp(0.,5.)(4.,5.5)(0.,){/Straight}{-1}
\FALabel(-1.,5.)[r]{\ssize$\bar\Pu$}
\FAProp(4.,12.5)(4.,5.5)(0.,){/Straight}{1}
\FALabel(3.,9)[r]{\ssize$\Pu$}
\FAProp(20.,0.)(4.,5.5)(0.,){/Cycles}{0}
\FALabel(21,0.)[l]{\ssize$\Pg$}
\FAProp(8.,12.5)(4.,12.5)(0.,){/Sine}{0}
\FALabel(6.,13.5)[b]{\ssize$\PZ,\gamma$}
\FAProp(12.,15.)(20.,20.)(0.,){/Straight}{1}
\FALabel(21,20)[l]{\ssize$\Pne$}
\FAProp(8.,12.5)(12.,15)(0.,){/Straight}{1}
\FALabel(11.,13.)[tl]{\ssize$\Pe$}
\FAProp(16.,12.5)(20.,15)(0.,){/Straight}{1}
\FALabel(21,15)[l]{\ssize$\Pmu^-$}
\FAProp(12.,15.)(16.,12.5)(0.,){/Sine}{0}
\FALabel(14.5,14.)[bl]{\ssize$\PWminus$}
\FAProp(16.,12.5)(20.,10)(0.,){/Straight}{-1}
\FALabel(21,10)[l]{\ssize$\Pnmu$}
\FAProp(8.,12.5)(20.,5)(0.,){/Straight}{-1}
\FALabel(21,5)[l]{\ssize$\Pe^+$}
\FAVert(16.,12.5){0}
\FAVert(12.,15.){0}
\FAVert(8.,12.5){0}
\FAVert(4.,12.5){0}
\FAVert(4.,5.5){0}
\end{diagram}
}
\vspace*{1.5em}
\centering{
\begin{diagram}
\FADiagram{}
\FAProp(0.,15.)(4.,14.5)(0.,){/Straight}{1}
\FALabel(-1.,15.)[r]{\ssize$\Pu$}
\FAProp(0.,5.)(4.,7.5)(0.,){/Straight}{-1}
\FALabel(-1.,5.)[r]{\ssize$\bar\Pu$}
\FAProp(4.,14.5)(4.,7.5)(0.,){/Straight}{1}
\FALabel(3.,11.)[r]{\ssize$\Pu$}
\FAProp(20.,20.)(4.,14.5)(0.,){/Cycles}{0}
\FALabel(21,20.)[l]{\ssize$\Pg$}
\FAProp(8.,7.5)(4.,7.5)(0.,){/Sine}{0}
\FALabel(6.,6.5)[t]{\ssize$\PZ$}
\FAProp(8.,7.5)(20.,15.)(0.,){/Straight}{1}
\FALabel(21.,15.)[l]{\ssize$\Pne$}
\FAProp(16.,7.5)(20.,10.)(0.,){/Straight}{1}
\FALabel(21.,10.)[l]{\ssize$\Pmu^-$}
\FAProp(12.,5.)(16.,7.5)(0.,){/Sine}{0}
\FALabel(14.25,7.5)[br]{\ssize$\PWminus$}
\FAProp(16.,7.5)(20.,5.)(0.,){/Straight}{-1}
\FALabel(21,5.)[l]{\ssize$\Pnmubar$}
\FAProp(8.,7.5)(12.,5.)(0.,){/Straight}{-1}
\FALabel(10.25,4.75)[tr]{\ssize$\Pne$}
\FAProp(12.,5.)(20.,0.)(0.,){/Straight}{-1}
\FALabel(21,0.)[l]{\ssize$\Pe^+$}
\FAVert(16.,7.5){0}
\FAVert(12.,5.){0}
\FAVert(8.,7.5){0}
\FAVert(4.,14.5){0}
\FAVert(4.,7.5){0}
\end{diagram}
\hspace*{2em}
\begin{diagram}
\FADiagram{}
\FAProp(0.,15.)(4.,14.5)(0.,){/Straight}{1}
\FALabel(-1.,15.)[r]{\ssize$\Pu$}
\FAProp(0.,5.)(4.,7.5)(0.,){/Straight}{-1}
\FALabel(-1.,5.)[r]{\ssize$\bar\Pu$}
\FAProp(4.,7.5)(4.,14.5)(0.,){/Straight}{-1}
\FALabel(3.,11)[r]{\ssize$\Pu$}
\FAProp(20.,20.)(4.,14.5)(0.,){/Cycles}{0}
\FALabel(21.,20.)[l]{\ssize$\Pg$}
\FAProp(8.,7.5)(4.,7.5)(0.,){/Sine}{0}
\FALabel(6.,6.5)[t]{\ssize$\PZ,\gamma$}
\FAProp(12.,10.)(20.,15.)(0.,){/Straight}{1}
\FALabel(21.,15)[l]{\ssize$\Pne$}
\FAProp(8.,7.5)(12.,10.)(0.,){/Straight}{1}
\FALabel(11.,8.)[tl]{\ssize$\Pe$}
\FAProp(16.,7.5)(20.,10.)(0.,){/Straight}{1}
\FALabel(21.,10.)[l]{\ssize$\Pmu^-$}
\FAProp(12.,10.)(16.,7.5)(0.,){/Sine}{0}
\FALabel(14.5,9.)[bl]{\ssize$\PWminus$}
\FAProp(16.,7.5)(20.,5.)(0.,){/Straight}{-1}
\FALabel(21.,5.)[l]{\ssize$\Pnmubar$}
\FAProp(8.,7.5)(20.,0.)(0.,){/Straight}{-1}
\FALabel(21.,0.)[l]{\ssize$\Pe^+$}
\FAVert(16.,7.5){0}
\FAVert(12.,10.){0}
\FAVert(8.,7.5){0}
\FAVert(4.,14.5){0}
\FAVert(4.,7.5){0}
\end{diagram}
}
\caption{Diagrams with only one resonant W propagator in the partonic subprocess \mbox{$\Pu\Pubar\to\Pne\Pep\Pmum\Pnmubar\Pg$}.}
\label{fi:dy:WWsingleresonant}
\end{figure}

The singularities appearing at the poles of the LO gauge-boson propagators 
are treated by 
introducing decay widths in the propagators.  More precisely,
the gauge-boson widths are introduced in terms of complex gauge-boson 
masses according to the complex-mass scheme at LO, which is described 
in \citeres{Denner:1999gp,Denner:2005fg}. 
Thus, in the LO amplitudes we perform the substitutions
\begin{eqnarray}
\label{eq:dy:complexmasses}
\MW^2\to\MW^2-\ri\GW\MW\;,\hspace*{.5cm}\MZ^2\to\MZ^2-\ri\GZ\MZ\;.
\end{eqnarray}
To preserve gauge invariance, the gauge-boson masses have to be treated as complex quantities everywhere, in particular in the definition of the weak mixing angle,
\begin{eqnarray}
\label{eq:dy:complexmixingangle}
\cos^2\theta_\mathrm{W}=\frac{\MW^2}{\MZ^2}\to\frac{\MW^2-\ri\GW\MW}{\MZ^2-\ri\GZ\MZ}\;,
\end{eqnarray}
which renders all couplings complex that are
derived from this quantity. The values of the gauge-boson widths are calculated at NLO QCD level with vanishing fermion masses,
\begin{eqnarray}
\label{eq:dy:GammaW}
\GW&=&\frac{\alpha}6\MW\biggl[\sum_\Pl(g^-_{\PW\Pnl\Pl})^2+N_{\rm{c}}\sum_{\Pu,\Pd}(g^-_{\PW\Pu\Pd})^2\biggl(1+\frac{\als(\MZ)}{\pi}\biggr)\biggr]\;,\\
\label{eq:dy:GammaZ}
\GZ&=&\frac{\alpha}6\MZ\biggl[\sum_\Pl\left((g^+_{\PZ\Pl\Pl})^2+(g^-_{\PZ\Pl\Pl})^2
+(g^-_{\PZ\nu_\Pl\nu_\Pl})^2\right)
\nonumber\\
&&\hspace*{.8cm}+N_{\rm{c}}\sum_{\Pq\neq\Pt}\left((g^+_{\PZ\Pq\Pq})^2+(g^-_{\PZ\Pq\Pq})^2\right)\biggl(1+\frac{\als(\MZ)}{\pi}\biggr)\biggr],
\end{eqnarray}
where \mbox{$N_{\rm{c}}=3$} is the number of quark colours 
and $g^\pm_{\PV ff'}$ denote the chiral couplings of $\PV=\gamma,\PZ,\PW$
to the fermions $f,f'$,
\begin{equation}
g^{\sigma}_{\gamma ff} = -Q_f, \qquad
g^{\sigma}_{\PZ ff} = 
\frac{\delta_{\sigma-}I^3_{\PW,f}-\sin^2\theta_\mathrm{W}\,Q_f}{\sin\theta_\mathrm{W}\cos\theta_\mathrm{W}}, \qquad
g^{\sigma}_{\PW ff'} = \frac{\delta_{\sigma-}}{\sqrt{2}\,\sin\theta_\mathrm{W}}.
\label{eq:chiralcouplings}
\end{equation}
Here $Q_f$ and $I^3_{\PW,f}=\pm\frac{1}{2}$ are the relative
electric charge and the third component of the weak isospin of $f$, 
respectively.
In \refeq{eq:dy:GammaW}, the sums run over all three lepton generations and the two light-quark families, in \refeq{eq:dy:GammaZ} over the three charged leptons and neutrinos and the five light quarks.
The W and Z~decay widths are calculated using real masses $\MW$ and $\MZ$,
and the strong coupling is set to the measured value $\als(\MZ)$ quoted
below. Further improvements would go beyond NLO QCD accuracy.

\subsection{Narrow-width approximation}
In the naive narrow-width approximation (NWA), the produced $\PW$
bosons are treated as on-shell particles in the production
process. The leptonic decays are assumed to be isotropic in the rest frames of the respective $\PW$ bosons, because their spin information from the production process is dropped. Therefore, the 
squared matrix elements of the WW+jet-production subprocesses can be used without modifications by only 
multiplying them with 
the respective branching ratios for the $\PW$-boson decays, which are given by
\begin{eqnarray}
\mathrm{BR}_{\PW^+\to\Pnl\Pl^+}=\frac{\Gamma_{\PW^+\to\Pnl\Pl^+}}{\GW}\;,\hspace*{0.5cm}\mathrm{BR}_{\PW^-\to \Pl^-\Pnlbar}=\frac{\Gamma_{\PW^-\to \Pl^-\Pnlbar}}{\GW}\;.
\end{eqnarray}
The partial widths for leptonic $\PW$ decays are
\begin{eqnarray}
\Gamma_{\PW^+\to\Pnl\Pl^+}&=&\Gamma_{\PW^-\to \Pl^-\Pnlbar}
=\frac{\alpha}6\MW(g^-_{\PW\Pnl\Pl})^2\;,
\end{eqnarray}
since the leptonic decays do not receive NLO QCD corrections. 

\begin{figure}
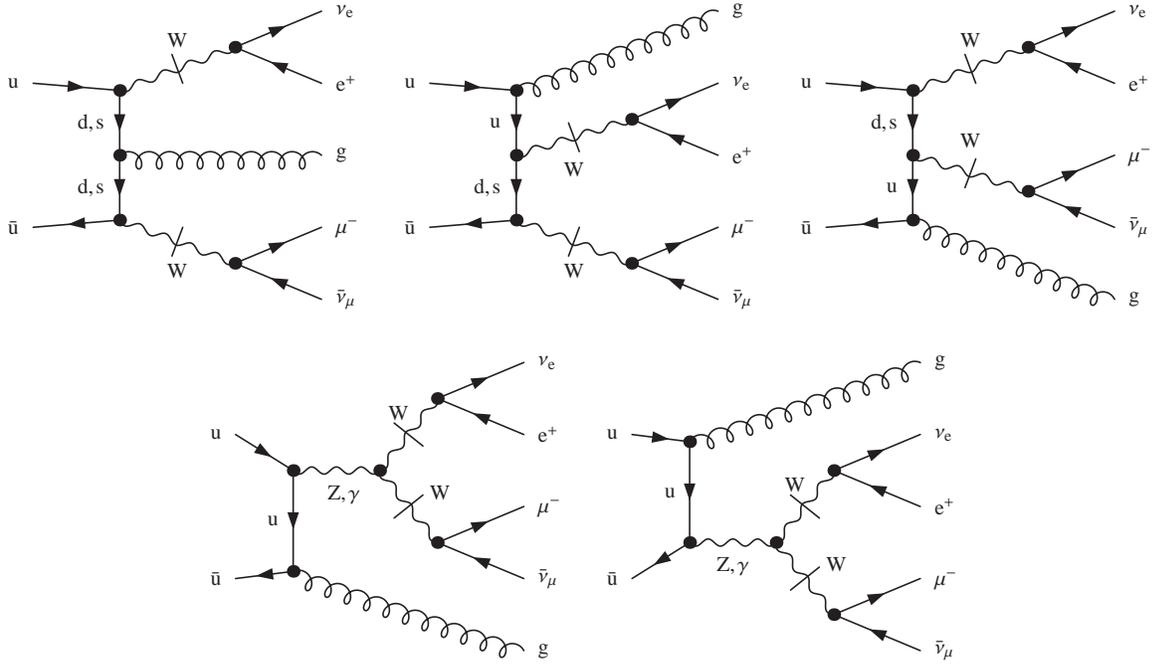

\centering{
\begin{diagram}
\FADiagram{}
\FAProp(0.,15.)(6.,14.5)(0.,){/Straight}{1}
\FALabel(-1.,15.)[r]{\ssize$\Pu$}
\FAProp(0.,5.)(6.,5.5)(0.,){/Straight}{-1}
\FALabel(-1.,5.)[r]{\ssize$\bar\Pu$}
\FAProp(14.,17.5)(6.,14.5)(0.,){/Sine}{0}
\FAProp(9.625,17.)(10.375,15.)(0.,){/Straight}{0}
\FALabel(10,17.5)[b]{\ssize$\PWplus$}
\FAProp(14.,17.5)(20.,20)(0.,){/Straight}{1}
\FALabel(21,20)[l]{\ssize$\Pne$}
\FAProp(14.,17.5)(20.,15)(0.,){/Straight}{-1}
\FALabel(21,15)[l]{\ssize$\Pe^+$}
\FAProp(14.,2.5)(6.,5.5)(0.,){/Sine}{0}
\FAProp(9.625,3.)(10.375,5.)(0.,){/Straight}{0}
\FALabel(10,2.5)[t]{\ssize$\PWminus$}
\FAProp(14.,2.5)(20.,0)(0.,){/Straight}{-1}
\FALabel(21,0)[l]{\ssize$\Pnmubar$}
\FAProp(14.,2.5)(20.,5)(0.,){/Straight}{1}
\FALabel(21,5)[l]{\ssize$\Pmu^-$}
\FAProp(20.,10.)(6.,10.)(0.,){/Cycles}{0}
\FALabel(21,10.)[l]{\ssize$\Pg$}
\FAProp(6.,14.5)(6.,10.)(0.,){/Straight}{1}
\FALabel(4.93,12.25)[r]{\ssize$\Pd,\Ps$}
\FAProp(6.,5.5)(6.,10.)(0.,){/Straight}{-1}
\FALabel(4.93,7.75)[r]{\ssize$\Pd,\Ps$}
\FAVert(14.,17.5){0}
\FAVert(14.,2.5){0}
\FAVert(6.,14.5){0}
\FAVert(6.,5.5){0}
\FAVert(6.,10.){0}
\end{diagram}
\hspace*{2em}
\begin{diagram}
\FADiagram{}
\FAProp(0.,15.)(6.,14.5)(0.,){/Straight}{1}
\FALabel(-1.,15.)[r]{\ssize$\Pu$}
\FAProp(0.,5.)(6.,5.5)(0.,){/Straight}{-1}
\FALabel(-1.,5.)[r]{\ssize$\bar\Pu$}
\FAProp(14.,12.5)(6.,10.)(0.,){/Sine}{0}
\FAProp(9.625,12.25)(10.375,10.25)(0.,){/Straight}{0}
\FALabel(10,9.5)[t]{\ssize$\PWplus$}
\FAProp(14.,12.5)(20.,15)(0.,){/Straight}{1}
\FALabel(21,15)[l]{\ssize$\Pne$}
\FAProp(14.,12.5)(20.,10)(0.,){/Straight}{-1}
\FALabel(21,10)[l]{\ssize$\Pe^+$}
\FAProp(14.,2.5)(6.,5.5)(0.,){/Sine}{0}
\FAProp(9.625,3.)(10.375,5.)(0.,){/Straight}{0}
\FALabel(10,2.5)[t]{\ssize$\PWminus$}
\FAProp(14.,2.5)(20.,0)(0.,){/Straight}{-1}
\FALabel(21,0)[l]{\ssize$\Pnmubar$}
\FAProp(14.,2.5)(20.,5)(0.,){/Straight}{1}
\FALabel(21,5)[l]{\ssize$\Pmu^-$}
\FAProp(20.,20.)(6.,14.5)(0.,){/Cycles}{0}
\FALabel(21,20.)[l]{\ssize$\Pg$}
\FAProp(6.,14.5)(6.,10.)(0.,){/Straight}{1}
\FALabel(4.93,12.25)[r]{\ssize$\Pu$}
\FAProp(6.,5.5)(6.,10.)(0.,){/Straight}{-1}
\FALabel(4.93,7.75)[r]{\ssize$\Pd,\Ps$}
\FAVert(14.,12.5){0}
\FAVert(14.,2.5){0}
\FAVert(6.,10.){0}
\FAVert(6.,5.5){0}
\FAVert(6.,14.5){0}
\end{diagram}
\hspace*{2em}
\begin{diagram}
\FADiagram{}
\FAProp(0.,15.)(6.,14.5)(0.,){/Straight}{1}
\FALabel(-1.,15.)[r]{\ssize$\Pu$}
\FAProp(0.,5.)(6.,5.5)(0.,){/Straight}{-1}
\FALabel(-1.,5.)[r]{\ssize$\bar\Pu$}
\FAProp(14.,17.5)(6.,14.5)(0.,){/Sine}{0}
\FAProp(9.625,17.)(10.375,15.)(0.,){/Straight}{0}
\FALabel(10.,17.5)[b]{\ssize$\PWplus$}
\FAProp(14.,17.5)(20.,20)(0.,){/Straight}{1}
\FALabel(21.,20)[l]{\ssize$\Pne$}
\FAProp(14.,17.5)(20.,15)(0.,){/Straight}{-1}
\FALabel(21.,15)[l]{\ssize$\Pe^+$}
\FAProp(14.,7.5)(6.,10.)(0.,){/Sine}{0}
\FAProp(9.625,7.75)(10.375,9.75)(0.,){/Straight}{0}
\FALabel(10.,10.5)[b]{\ssize$\PWminus$}
\FAProp(14.,7.5)(20.,5.)(0.,){/Straight}{-1}
\FALabel(21.,5.)[l]{\ssize$\Pnmubar$}
\FAProp(14.,7.5)(20.,10.)(0.,){/Straight}{1}
\FALabel(21.,10.)[l]{\ssize$\Pmu^-$}
\FAProp(20.,0.)(6.,5.5)(0.,){/Cycles}{0}
\FALabel(21.,0.)[l]{\ssize$\Pg$}
\FAProp(6.,14.5)(6.,10.)(0.,){/Straight}{1}
\FALabel(4.93,12.25)[r]{\ssize$\Pd,\Ps$}
\FAProp(6.,5.5)(6.,10.)(0.,){/Straight}{-1}
\FALabel(4.93,7.75)[r]{\ssize$\Pu$}
\FAVert(14.,17.5){0}
\FAVert(14.,7.5){0}
\FAVert(6.,14.5){0}
\FAVert(6.,5.5){0}
\FAVert(6.,10.){0}
\end{diagram}
}
\vspace*{1em}
\centering{
\begin{diagram}
\FADiagram{}
\FAProp(0.,15.)(4.,12.5)(0.,){/Straight}{1}
\FALabel(-1.,15.)[r]{\ssize$\Pu$}
\FAProp(0.,5.)(4.,5.5)(0.,){/Straight}{-1}
\FALabel(-1.,5.)[r]{\ssize$\bar\Pu$}
\FAProp(10.,12.5)(4.,12.5)(0.,){/Sine}{0}
\FALabel(7.5,11.5)[t]{\ssize$\PZ,\gamma$}
\FAProp(14.,17.5)(10.,12.5)(0.,){/Sine}{0}
\FAProp(11.,15.8)(13.,14.2)(0.,){/Straight}{0}
\FALabel(12.,16.)[br]{\ssize$\PWplus$}
\FAProp(14.,17.5)(20.,20)(0.,){/Straight}{1}
\FALabel(21.,20)[l]{\ssize$\Pne$}
\FAProp(14.,17.5)(20.,15)(0.,){/Straight}{-1}
\FALabel(21.,15)[l]{\ssize$\Pe^+$}
\FAProp(14.,7.5)(10.,12.5)(0.,){/Sine}{0}
\FAProp(11.,9.2)(13.,10.8)(0.,){/Straight}{0}
\FALabel(13.5,10.5)[bl]{\ssize$\PWminus$}
\FAProp(14.,7.5)(20.,5.)(0.,){/Straight}{-1}
\FALabel(21.,5.)[l]{\ssize$\Pnmubar$}
\FAProp(14.,7.5)(20.,10.)(0.,){/Straight}{1}
\FALabel(21.,10.)[l]{\ssize$\Pmu^-$}
\FAProp(20.,0.)(4.,5.5)(0.,){/Cycles}{0}
\FALabel(21,0.)[l]{\ssize$\Pg$}
\FAProp(4.,12.5)(4.,5.5)(0.,){/Straight}{1}
\FALabel(3.,9)[r]{\ssize$\Pu$}
\FAVert(14.,7.5){0}
\FAVert(14.,17.5){0}
\FAVert(10.,12.5){0}
\FAVert(4.,12.5){0}
\FAVert(4.,5.5){0}
\end{diagram}
\hspace*{2em}
\begin{diagram}
\FADiagram{}
\FAProp(0.,15.)(4.,14.5)(0.,){/Straight}{1}
\FALabel(-1.,15.)[r]{\ssize$\Pu$}
\FAProp(0.,5.)(4.,7.5)(0.,){/Straight}{-1}
\FALabel(-1.,5.)[r]{\ssize$\bar\Pu$}
\FAProp(10.,7.5)(4.,7.5)(0.,){/Sine}{0}
\FALabel(7.,6.5)[t]{\ssize$\PZ,\gamma$}
\FAProp(14.,12.5)(10.,7.5)(0.,){/Sine}{0}
\FAProp(11.,10.8)(13.,9.2)(0.,){/Straight}{0}
\FALabel(12.,11.)[br]{\ssize$\PWplus$}
\FAProp(14.,12.5)(20.,15.)(0.,){/Straight}{1}
\FALabel(21.,15)[l]{\ssize$\Pne$}
\FAProp(14.,12.5)(20.,10.)(0.,){/Straight}{-1}
\FALabel(21.,10)[l]{\ssize$\Pe^+$}
\FAProp(14.,2.5)(10.,7.5)(0.,){/Sine}{0}
\FAProp(11.,4.2)(13.,5.8)(0.,){/Straight}{0}
\FALabel(13.5,5.5)[bl]{\ssize$\PWminus$}
\FAProp(14.,2.5)(20.,0.)(0.,){/Straight}{-1}
\FALabel(21.,0.)[l]{\ssize$\Pnmubar$}
\FAProp(14.,2.5)(20.,5.)(0.,){/Straight}{1}
\FALabel(21.,5.)[l]{\ssize$\Pmu^-$}
\FAProp(20.,20.)(4.,14.5)(0.,){/Cycles}{0}
\FALabel(21,20.)[l]{\ssize$\Pg$}
\FAProp(4.,14.5)(4.,7.5)(0.,){/Straight}{1}
\FALabel(3.,11)[r]{\ssize$\Pu$}
\FAVert(14.,2.5){0}
\FAVert(14.,12.5){0}
\FAVert(10.,7.5){0}
\FAVert(4.,14.5){0}
\FAVert(4.,7.5){0}
\end{diagram}
}
\caption{NWA diagrams with two on-shell gauge-boson propagators for the partonic process \mbox{$\Pu\Pubar\to\Pne\Pep\Pmum\Pnmubar\Pg$} at LO. The on-shell propagators are denoted by slashed propagator lines.}
\label{fi:dy:WWnarrowwidth}
\end{figure}

Consequently, the approximation of on-shell gauge bosons restricts the diagrams contributing to the subprocesses to those containing two resonant $\PW$ propagators, which stand in one-to-one correspondence to the pure gauge-boson pair-production diagrams in \reffi{fi:dy:WWdoubleresonant}. The analogous NWA diagrams are collected in \reffi{fi:dy:WWnarrowwidth} for one LO subprocess.

\subsection{Improved narrow-width approximation}

In the comparison that is presented in \refse{se:nu:approximations}
for LO calculations, the naive  NWA turns out to 
be not satisfactory.
While the integrated cross sections show modest deviations between the
full amplitude calculation and the naive NWA, large discrepancies arise in some phase-space regions if differential cross sections
are considered. 

An appropriate compromise between the full calculation and the 
naive NWA is obtained by an improved NWA (iNWA). 
Here, the gauge bosons are still treated as on-shell particles, but their spin information is kept. 
It is used to improve the description of the leptonic $\PW$-boson
 decays, which are not isotropic in the respective $\PW$-boson rest
 frames owing to the $V-A$ structure of the decay. Details on
 how to keep the spin information of the decaying gauge bosons are
 provided in 
\refapp{se:onshelldecays}.

The diagrams that are relevant for the iNWA are the same as for the NWA,
which are shown in \reffi{fi:dy:WWnarrowwidth}
for one specific channel in LO.
The modification of amplitudes is essentially the same for LO subprocesses and all contributions to the NLO QCD cross section: 
the polarization vector of each of the outgoing gauge bosons is replaced by the leptonic currents of its decay products. 
For the rest of the production
amplitude, the momenta of the gauge bosons are set on shell. The Breit--Wigner propagators arising in the absolute square of the amplitudes are separated. They are replaced by delta functions with an appropriate normalization obtained by integrating over the Breit--Wigner propagator in the limit $\GV\to0$. 

Since the iNWA turns out to reproduce the full calculation to sufficient accuracy in the LO comparison of \refse{se:nu:approximations}, this strategy for describing the leptonic decays is applied for the NLO QCD calculations.

\section{Numerical results}
\label{se:numres}

\subsection{Setup and input parameters}
\label{se:nr:setup}
For the numerical evaluations, 
the following SM parameters \cite{Eidelman:2004wy} are used,
\begin{eqnarray}
\MW&=&80.425\GeV\;,\hspace*{.5cm}\MZ=91.1876\GeV\;,\hspace*{.5cm}
\GF=1.16637\times 10^{-5}\GeV^{-2}\;, 
\nn\\
\MH&=&150\GeV\;,\hspace*{.5cm}\GH=0.017\GeV\;,\hspace*{.5cm}m_\Pt=174.3\GeV\;.
\label{eq:nr:numsetup}
\end{eqnarray}
The total Higgs-boson width $\GH$ has been calculated with 
{\sl Hdecay}~\cite{Djouadi:1997yw}. 
The only contribution from a Higgs boson at NLO QCD is given by the loop 
diagams in \reffi{fi:fermWWvertexdirect}. Since these diagrams turn out 
to contribute less than $0.3\%$ to the cross sections for the chosen
Higgs mass, their impact is not investigated further.%
\footnote{For $\MH>2\MW$ the included Higgs-boson effects account for 
the interference
between a $\PH(\to\PW\PW)+\mathrm{jet}$ signal and the coherent 
irreducible $\PW\PW+\mathrm{jet}$ background. A proper description of the
signal requires, of course, also the squared resonance diagrams and
contributions from $\Pg\Pg$ fusion with the corresponding radiative
corrections.}
The electromagnetic coupling $\alpha$ is evaluated from \refeq{eq:nr:numsetup} via
\begin{eqnarray}
\alpha=\alpha_{\mathrm{G}_\mu}=\frac{\sqrt2\GF\MW^2\left(1-\frac{\MW^2}{\MZ^2}\right)}{\pi}\;.
\end{eqnarray}
This choice of $\alpha$ absorbs some universal corrections to the 
electroweak coupling (running of $\alpha$
to the electroweak scale, leading corrections
to the $\rho$-parameter), 
as is explained, e.g., in \citere{Dittmaier:2001ay}. 
The widths of the weak gauge bosons are
calculated according to
\refeqsa{eq:dy:GammaW}{eq:dy:GammaZ}---in 
case of the production of stable weak gauge bosons they
are, of course, set to zero---from all decay channels at
NLO QCD with the fermion masses neglected.
With the value of the strong coupling at
the scale $\MZ$ taken from \citere{Amsler:2008zz}, 
\begin{eqnarray}
\als(\MZ)=0.1176\;,
\end{eqnarray}
the calculated widths are
\begin{eqnarray}
\GW=2.0996\GeV\;,\hspace*{.5cm}\GZ=2.5097\GeV\;.
\end{eqnarray}

The SM parameters already used in the publications on WW+jet production
\cite{Bern:2008ef,Dittmaier:2007th,Dittmaier:2007iq,Dittmaier:2008pc}
are not replaced by the more recent values \cite{Amsler:2008zz} in order to facilitate comparisons; the numerical impact of the slightly changed 
measured value of $\MW$ is negligible anyway at the required accuracy.

The values for the strong coupling in the amplitude calculation are evaluated according to a 1-loop-running at LO and a 2-loop-running at NLO as described in \citere{Ellis_Stirling_Webber},
\begin{eqnarray}
\alpha_{\mathrm{s,1-loop}}(\muren)&=&\frac{1}{\frac{33-2\Nf}{12\pi}\,\ln\frac{\muren^2}{\Lambda_{\mathrm{QCD}}^2}}\;,\\
\alpha_{\mathrm{s,2-loop}}(\muren)&=&\alpha_{\mathrm{s,1-loop}}(\muren)\left(1-\frac{6(153-19\Nf)}{(33-2\Nf)^2}\,\frac{\ln\left(\ln\frac{\muren^2}{\Lambda_{\mathrm{QCD}}^2}\right)}{\ln\frac{\muren^2}{\Lambda_{\mathrm{QCD}}^2}}\right),
\end{eqnarray}
where $\muren$ is the renormalization scale, $\Nf$ the number of active 
light-quark flavours, and $\Lambda_{\mathrm{QCD}}$ the QCD scale parameter. The
values of $\Lambda_{\mathrm{QCD}}$ are chosen as prescribed by the applied PDF
sets: In the five-flavour scheme, the PDFs of CTEQ6
\cite{Pumplin:2002vw,Stump:2003yu} are used with \mbox{$\Nf=5$}, namely CTEQ6L1 with
$\Lambda_{\mathrm{QCD}}=165\MeV$ at LO and CTEQ6M with
$\Lambda_{\mathrm{QCD}}=226\MeV$ at NLO. In the four-flavour scheme, the PDFs
of MRST2004 \cite{Martin:2006qz} with \mbox{$\Nf=4$} are taken, namely
MRST2004F4LO with \mbox{$\Lambda_{\mathrm{QCD}}=220\MeV$} at LO and MRST2004F4NLO
with \mbox{$\Lambda_{\mathrm{QCD}}=347\MeV$} at NLO. The renormalization of $\als$
is performed as described in \refse{se:virtual}, i.e.\ with the heavy-quark loops in the gluon self-energy decoupled: In the five-flavour scheme only the top-quark loop is
decoupled, in the four-flavour scheme both the top- and the
bottom-quark loops are decoupled.

The Cabbibo angle in the approximation \refeq{eq:pr:CKMmatrix} for the CKM matrix is set to
\begin{eqnarray}
\thetac=0.227\;,
\end{eqnarray}
the explicit entries are calculated from this.

To give an IR-safe definition of the cross section
we apply the successive-combination jet
algorithm of \citere{Ellis:1993tq} with \mbox{$R=1$} to decide whether two
final-state partons can be resolved as two separated jets or whether they have to
be combined to only one jet. A dependence on the
specific jet algorithm only arises through the real-emission subprocesses,
because the LO and all other NLO contributions contain only one parton
in the final state which is identified with the jet.
To resolve the additional hadronic jet, a cut on the transverse
momentum is applied. This renders the cross section also IR safe which 
would otherwise diverge already at LO. Different
values are used for $p_{\mathrm{T,jet,cut}}$ which are explicitly given for
the respective results.

If the weak-gauge-boson decays are included, a set of additional cuts is applied which is in general not necessary for the finiteness of cross sections, but provides results that are closer to the experimental situation. These additional cuts are
\begin{eqnarray}
\label{eq:nr:cutetapT}
|\eta_{\mathrm{jet}}|<4.5\;,\hspace*{.5cm}p_{\mathrm{T,lepton}}>25\GeV\;,\hspace*{.5cm}|\eta_{\mathrm{lepton}}|<2.5\;,
\end{eqnarray}
where $\eta$ is the pseudo-rapidity and $p_{\mathrm{T}}$ the transverse momentum of the respective particle. All leptonic cuts are applied only to charged leptons, of course. As an additional condition the missing transverse momentum,
which is deduced from the sum of the neutrino momenta, is required to obey
\begin{eqnarray}
\label{eq:nr:cutpTmiss}
p_{\mathrm{T,miss}}>25\GeV\;.
\end{eqnarray}
The missing transverse momentum is particularly of interest if WW+jet is considered as a background process for SUSY searches. Here, the
missing transverse momentum mimics the 
lightest supersymmetric particle that also leaves the detector undetected.

\def\RR{R}
Furthermore, isolation cuts are applied in order to separate all visible leptons from each other and from hadronic jets,
\begin{eqnarray}
\label{eq:nr:cutR}
\RR_{\mathrm{lepton,jet}}>0.4\;,\hspace*{.5cm}\RR_{\mathrm{lepton,lepton}}>0.2\;,
\end{eqnarray}
where $\RR_{ab}=\sqrt{(\Delta\varphi_{ab})^2+(\Delta\eta_{ab})^2}$. The
angular distance between the two particles $a,b$
in the transverse plane is denoted by $\Delta\varphi_{ab}$, and
$\Delta\eta_{ab}$ is the difference of their pseudo-rapidities.

\subsection{Results on integrated cross sections}

We start our discussion of numerical results with integrated cross sections.
In order to render these cross sections insensitive to the decay of the 
W~bosons, we do not apply the cuts on the decay leptons specified in 
Eqs.~\refeq{eq:nr:cutetapT}--\refeq{eq:nr:cutR} in this section.

\subsubsection{LO cross sections}
\label{se:nr:LOcrosssections}
The LO cross sections get contributions from $\Pq\Pqbar$, $\Pq\Pg$, and $\Pg\Pqbar$ initial states. In this section, the relevance of the respective partonic channels is discussed both for results on proton--proton collisions at $\sqrt{s}=14\TeV$ (LHC setup) and on proton--antiproton collisions at $\sqrt{s}=1.96\TeV$ (Tevatron setup). Scale variations of a factor 10 around the central scale, which is chosen to be $\MW$, are considered. Here and in the following sections, the common scale $\mu=\muren=\mufact$ is used, i.e.\ renormalization and factorization scales are set equal and are varied simultaneously.

\begin{figure}
\centering
\includegraphics[bb = 170 410 450 660, scale = .8]{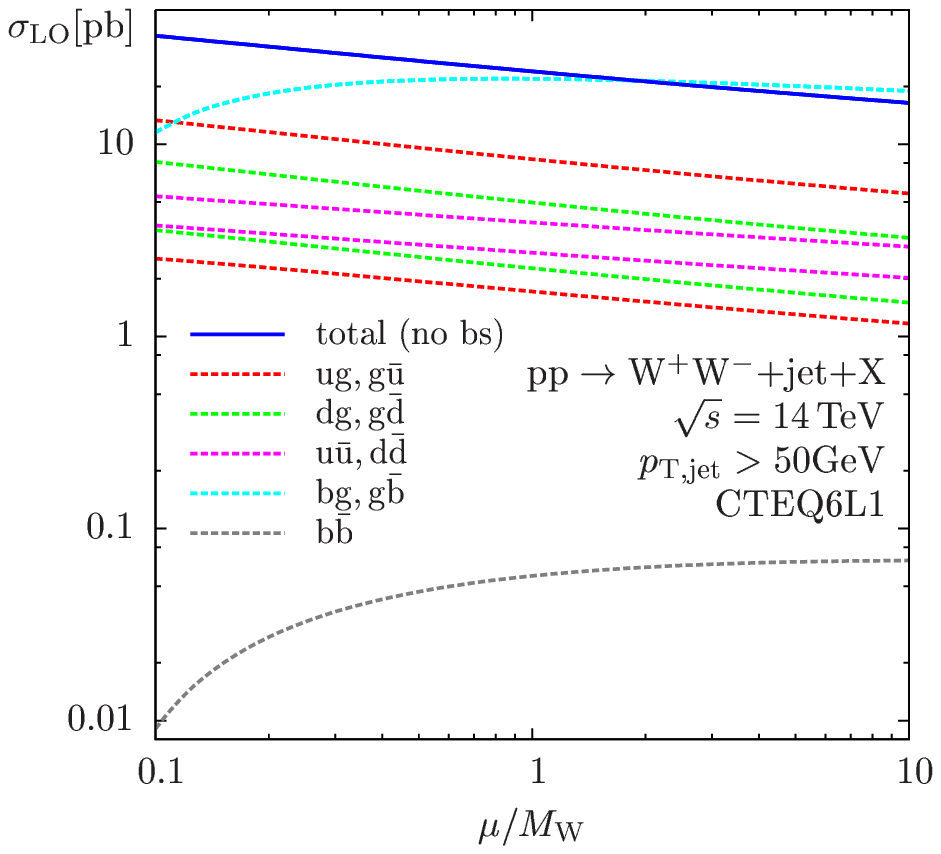}
\includegraphics[bb = 160 410 410 660, scale = .8]{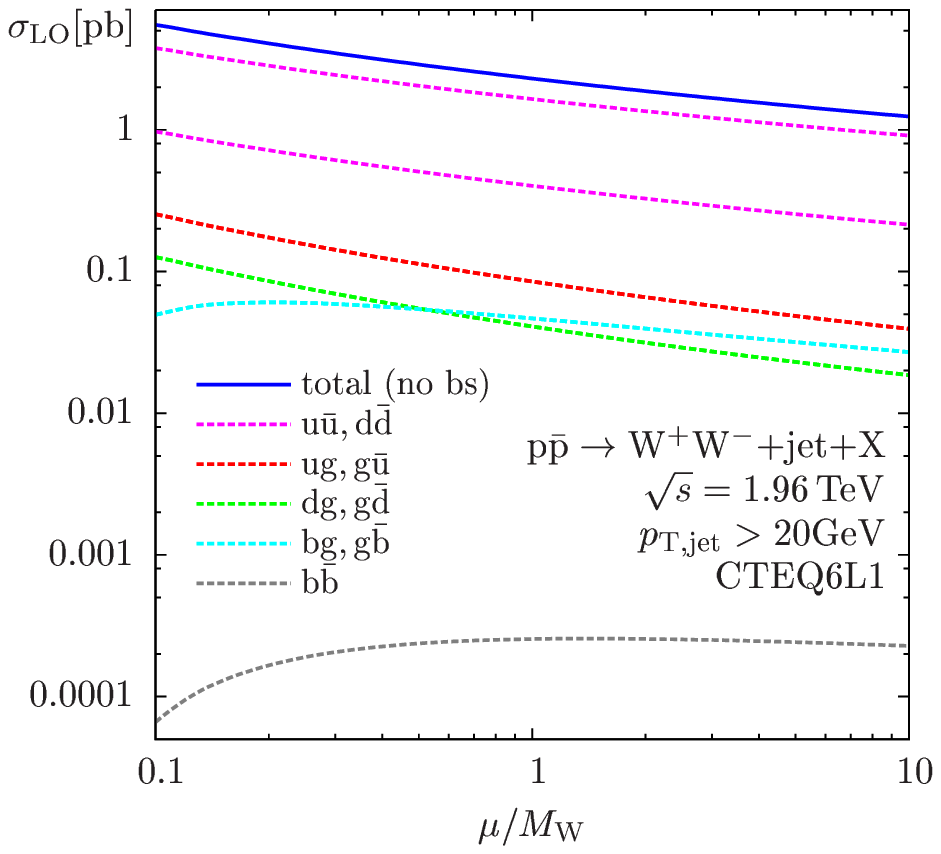}
\caption[Partonic contributions to the LO cross sections of WW+jet at the LHC and the Tevatron]{Partonic contributions to the LO cross sections of WW+jet at the LHC (l.h.s.) and the Tevatron (r.h.s.): 
The variation of the common scale
$\mu=\mufact=\muren$ is shown for all channels. If
more than one curve with identical colours is given, the first one
mentioned in the key corresponds to the upper curve. The contribution
named ``total'' does not contain contributions from external bottom
quarks.}  
\label{fi:nr:LOcrosssectionsWW}
\end{figure}

\reffi{fi:nr:LOcrosssectionsWW} shows the results for the various 
partonic contributions to WW+jet production, evaluated in the five-flavour
scheme. 
At the LHC, each partonic process involves at most one valence quark. 
Therefore, the gluon flux, which is essentially larger at the LHC compared to Tevatron, in general 
leads to larger contributions from $\Pq\Pg$ channels compared to $\Pq\Pqbar$ annihilation, whereas the $\Pg\Pqbar$ initial states contribute significantly less since no valence partons are involved here. The situation at Tevatron is 
different, because the proton--antiproton collisions provide $\Pq\Pqbar$ contributions with two valence partons that dominate the cross sections. 
The scale dependence of the cross sections at Tevatron turns out to be stronger 
than at LHC, which is due to the factorization-scale variation. While the renormalization-scale variation only reflects the running of the strong coupling $\als$, which is the same in both cases, the factorization scale dependence is quite flat for nearly all channels at LHC in the considered range. At Tevatron, however, an increase of roughly the same order as the one arising from the renormalization-scale dependence is found in the direction of lower scales. A detailed analysis of the scale dependence of LO cross sections to WW+jet production is provided in \citere{Kallweit:2006dt}.

In case of the LHC, the partonic contributions with up-type quarks dominate 
over those with down-type quarks, which can be understood by comparing the valence-quark PDFs in the protons. The antiquark PDFs of the lightest generation result in an inverted order of the $\Pg\Pqbar$ channels at the LHC where the $\Pg\Pdbar$ channel prevails the $\Pg\Pubar$ contribution. For Tevatron, the described valence-PDF effect occurs twice in the $\Pq\Pqbar$ channel, leading roughly to a factor 4 between $\Pu\Pubar$ and $\Pd\Pdbar$. The channels $\Pq\Pg$ and $\Pg\Pqbar$ deliver exactly the same contribution to the integrated cross sections due to the charge-conjugation-invariant hadronic initial state. Naturally, the channels involving valence-\mbox{\antiquarks} dominate, and therefore the described PDF effect leads again to larger contributions from $\Pu\Pg$/$\Pg\Pubar$ compared to $\Pd\Pg$/$\Pg\Pdbar$. 

In addition, for both colliders the channels involving bottom flavours are shown in \reffi{fi:nr:LOcrosssectionsWW}. As expected, $\Pb\bPb$ annihilation
is numerically negligible in both cases, whereas the $\Pb\Pg$ channel and the $\Pg\bPb$ channel---whose contributions to the integrated cross section are the same---each account roughly for the same amount as all the subprocesses involving \mbox{\antiquarks} of the two light generations at the
LHC. As already explained in \refse{se:pr:bquarks}, this is due to the
fact that these subprocesses actually describe resonant and
non-resonant $\PWm\Pt$ and $\PWp\bPt$ production, respectively, with
the top decays included. In addition the bottom PDF is due to the
  gluon splitting $\Pg\to\Pb\bPb$. The bg-flux is thus rather large at
  the LHC due
  to the large gluon lumionsity. As mentioned earlier the
  corresponding contributions should be attributed to different
  processes, therefore  these channels and the respective NLO
corrections are not taken into account. 
For Tevatron, their numerical impact is not that large, since the centre-of-mass (CM)
energy to produce top resonances in addition to a $\PW$ boson is
rarely available and the gluon flux is small. The respective channels are, however, treated in the same way as for the LHC. The different shapes of the scale-variation curves with bottom flavours result from a strong decrease of the bottom PDFs for smaller values of the factorization scale.

\subsubsection{NLO QCD cross sections}
\label{se:nr:NLOproduction}
After analyzing the individual contributions to the LO cross sections in the previous section, the effects of NLO QCD corrections on the WW+jet cross sections are discussed here. For the numerical results of this and the following sections, two different definitions of the NLO observables are used. The one observable is defined more inclusively by only requiring at least one hard jet with a minimum transverse momentum after application of 
the jet algorithm.  The more exclusively defined observable applies a veto on a second separable hard jet and describes therefore genuine WW+jet production. To this end, real-correction events with two jets 
that fulfill the $p_{\mathrm{T,jet,cut}}$ condition and
are not combined by the jet algorithm are not counted in the more exclusive observable. The phase-space regions in the real-correction subprocesses that are relevant for curing infrared singularities from the virtual corrections are not influenced, since the applied restriction only refers to genuine WW+2jets events. Therefore, the difference between the results for the two NLO observables is precisely given by the respective LO observable of WW+2jets production---evaluated, however, 
in the NLO setup (NLO PDFs, 2-loop running of $\alpha_{\mathrm{s}}$).

\begin{figure}
\centering
\includegraphics[bb = 200 380 450 700, scale = .8]{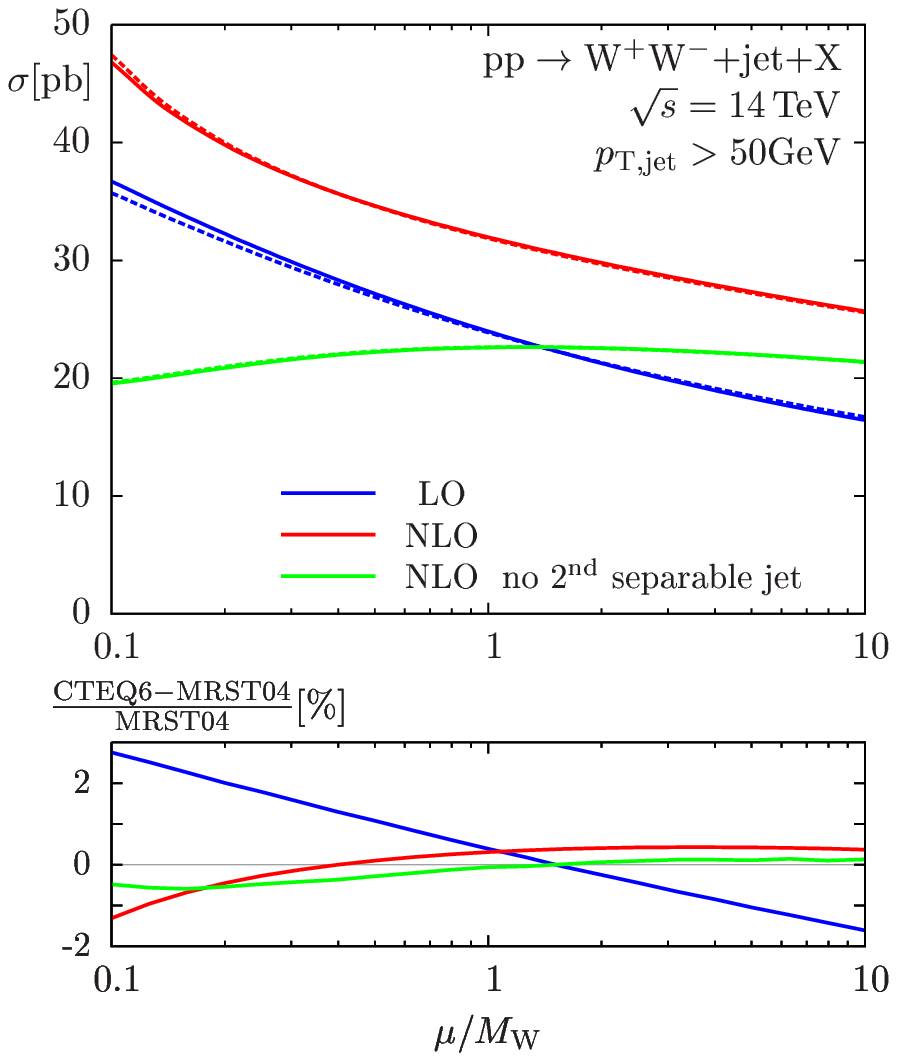}
\includegraphics[bb = 160 380 410 700, scale = .8]{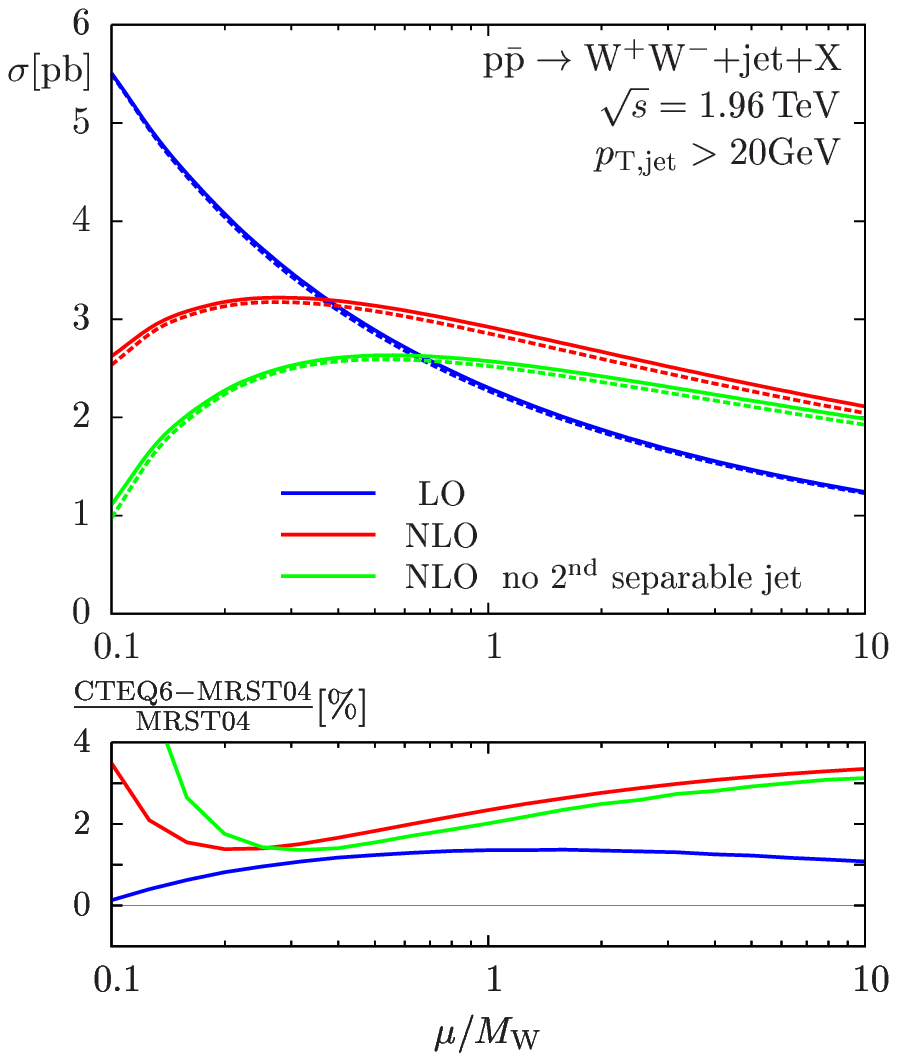}\\[-1em]
\caption{Comparison of WW+jet production cross sections in the LHC setup with \mbox{$p_{\mathrm{T,jet}}>50\GeV$} and for Tevatron with \mbox{$p_{\mathrm{T,jet}}>20\GeV$}: The straight lines show the results calculated with the five-flavour PDFs of CTEQ6, the dashed lines those calculated with the four-flavour PDFs of MRST2004F4. Contributions from external bottom \mbox{\antiquarks} are omitted, as described in \refse{se:pr:bquarks}.} 
\label{fi:nr:pdfcomare}
\end{figure}

\paragraph{Four-flavour versus five-flavour scheme and scale dependence of NLO QCD cross sections}
\label{se:nr:pdfcomparison}
We start the discussion of the scale dependence of the NLO cross sections with a comparison between the results obtained in the four-flavour and the five-flavour schemes, 
respectively, as described in \refse{se:pr:bquarks}. 
The NLO QCD cross sections for WW+jet production are presented in \reffi{fi:nr:pdfcomare} for the LHC setup with $p_{\mathrm{T,jet,cut}}=50\GeV$ and for the Tevatron setup with $p_{\mathrm{T,jet,cut}}=20\GeV$, both for the four-flavour calculations with MRST2004F4 PDFs and the five-flavour calculations with CTEQ6 PDFs.
In the four-flavour calculation, loop diagrams involving (massive) 3rd-generation quarks are included. However, their contributions account for less than 
$0.2\%$ to the integrated cross sections.

In case of the LHC, the relative deviations between the NLO cross sections in the two schemes behave as follows: 
while the deviation significantly increases in direction of lower scales at LO, which leads, however, to relative deviations of less than 3\% 
in the whole range, it is nearly flat for both NLO observables where the results deviate by less than 2\%. For the Tevatron, the comparison between the two approaches reveals
deviations of less than 2\% at LO over the full range of scale variations. The NLO cross sections turn out to be slightly larger in the CTEQ6 approach, but on a level of less than 4\% for the whole range of interest. The smallest scales depicted may be ignored, because they obviously describe a region where the NLO calculation is not a good approximation anymore, as one can see from the steep decline of the absolute values for the NLO cross sections in that range of scales.

Since the results evaluated with different PDF sets can in general not be
expected to agree exactly---not only in the present case where different
descriptions of the bottom \mbox{\antiquark} are applied---this analysis can be
seen as a confirmation that the approximation that is applied in the
five-flavour calculation delivers fully acceptable results. Moreover, the
deviations between the two approaches are far below the expected experimental
errors. The good agreement has been further confirmed by performing the same
comparison for $p_{\mathrm{T,jet,cut}}=100\GeV$ at LHC and 
for $p_{\mathrm{T,jet,cut}}=50\GeV$ at Tevatron, yielding the same level of agreement. In addition, also differential cross sections have been compared with no significant deviations showing up in any phase-space regions. Numerical results of these further comparisons are omitted here, because they are not supposed to provide new insight. In the following, only the five-flavour scheme with CTEQ6 PDFs is used.

\begin{figure}
\centering
\includegraphics[bb = 200 410 450 660, scale = .8]{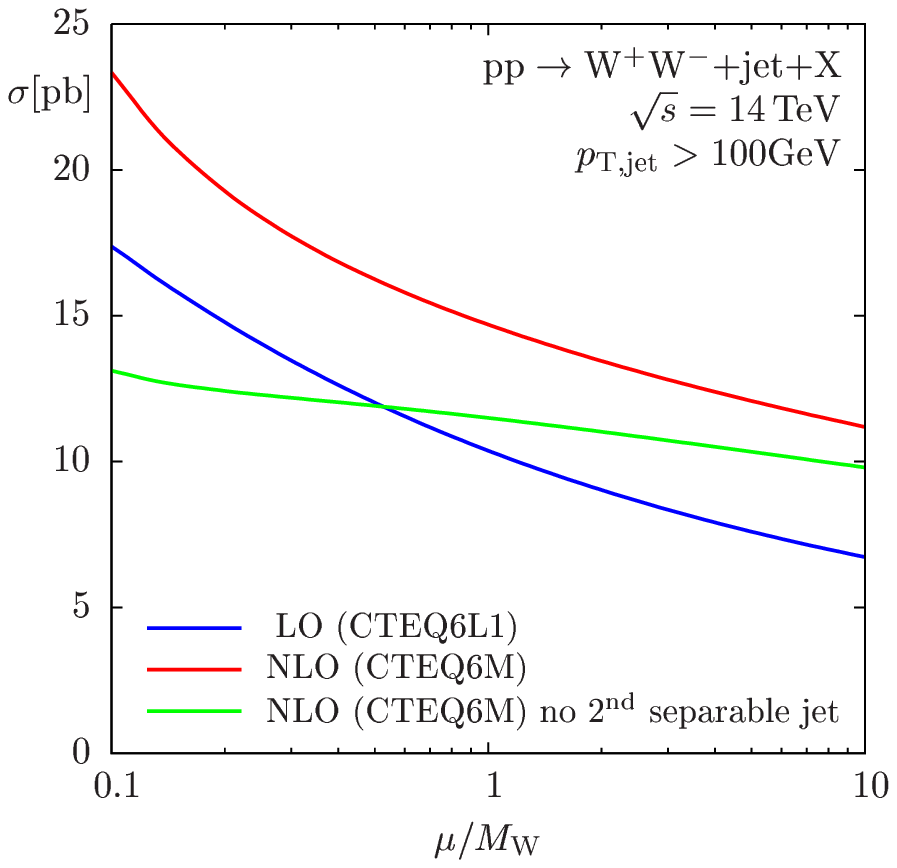}
\includegraphics[bb = 160 410 410 660, scale = .8]{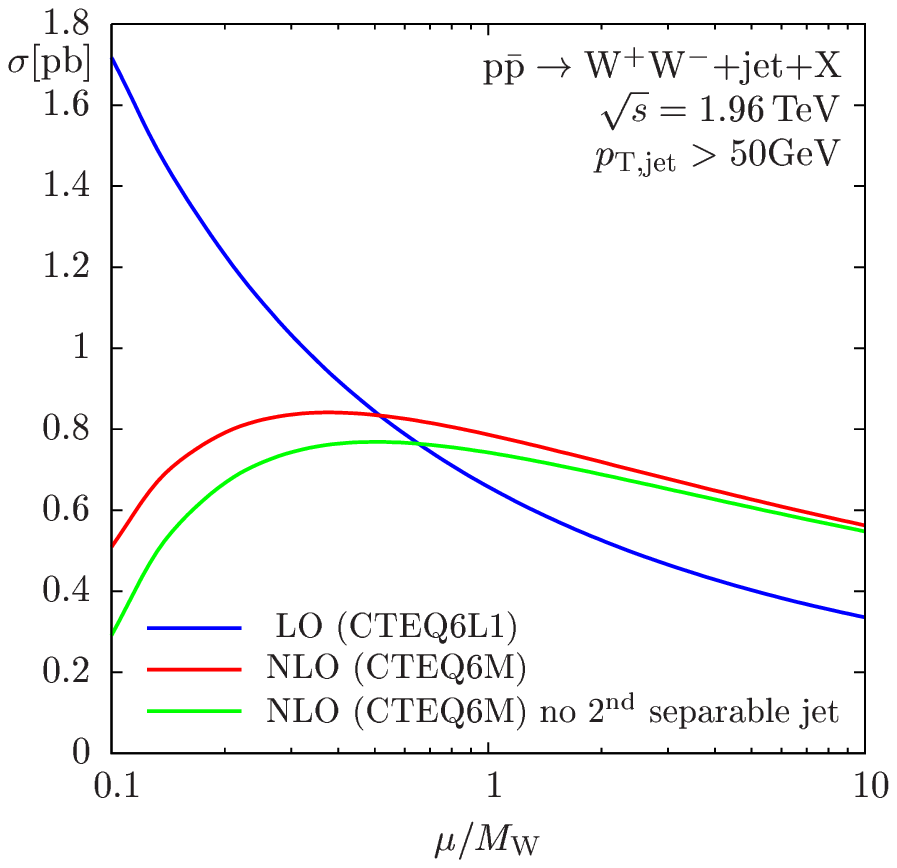}
\caption{Scale dependence of the WW+jet cross sections with \mbox{$\mu=\mu_{\mathrm{fact}}=\mu_{\mathrm{ren}}$}, where \mbox{$p_{\mathrm{T,jet}}>100\GeV$} is applied for the LHC, and \mbox{$p_{\mathrm{T,jet}}>50\GeV$} for the Tevatron.}
\label{fi:nr:WWproductionNLO}
\end{figure}

\begin{sloppypar}
Considering the scale dependence in the transition from LO to NLO at the LHC
(see \reffi{fi:nr:pdfcomare}), only a modest reduction is observed if
$\PW$-boson pairs in association with two hard jets are taken into
account. This large residual scale dependence is mainly due to the $\Pq\Pg$
channels, followed by those contributions with two valence quarks in the initial
state that are present in the real corrections, but not at LO. The scale
dependence can be significantly suppressed upon applying the veto of
having ``no $2^{\mbox{\footnotesize nd}}$ separable jet''. The relevance of a jet veto in order to suppress the scale dependence at NLO was also realized for genuine $\PW$-pair production at hadron colliders~\cite{Dixon:1999di}. A reduction of the difference between the two curves, which represents---as mentioned in the beginning of this section---the contribution of genuine WW+2jets events, is also achieved by increasing the value of the cut on $p_{\mathrm{T,jet}}$, which is illustrated in \reffi{fi:nr:WWproductionNLO} for WW+jet production with \mbox{$p_{\mathrm{T,jet,cut}}>100\GeV$} in the LHC setup and \mbox{$p_{\mathrm{T,jet,cut}}>50\GeV$} for the Tevatron. Explicit numbers for the cross sections---both with and without leptonic W decays--- are collected in \reftasx{ta:nr:crosssectionsLHC50}{ta:nr:crosssectionsTev50}.
\begin{table}
\begin{center}
\begin{tabular}{|c|c|c|c|}
\hline
\multicolumn{4}{|c|}{$ \Pp\Pp \to \PWp\PWm+ \mathrm{jet}+X$ @ $14\TeV$}\\
\hline
$\mu=\muren=\mufact$&$\sigma_{\mathrm{LO}}[\pba]$ & $\sigma_{\mathrm{NLO,excl}}[\pba]$ & $\sigma_{\mathrm{NLO,incl}}[\pba]$\\
\hline
$0.5\,\MW$ & $27.17638 \,(89)$ & $22.216 \,(15)$ & $34.641 \,(14)$\\
$1\,\MW$ & $23.97398 \,(79)$ & $22.606 \,(12)$ & $31.970 \,(11)$\\
$2\,\MW$ & $21.26027 \,(70)$ & $22.5646 \,(97)$ & $29.7959 \,(89)$\\

\hline
\multicolumn{4}{|c|}{$ \Pp\Pp \to \PWp(\to\Pne\Pe^+)\PWm(\to\Pmu^-\Pnmubar) + \mathrm{jet}+X$ @ $14\TeV$}\\
\hline
$\mu=\muren=\mufact$&$\sigma_{\mathrm{LO}}[\fb]$ & $\sigma_{\mathrm{NLO,excl}}[\fb]$ & $\sigma_{\mathrm{NLO,incl}}[\fb]$\\
\hline
$0.5\,\MW$ & $118.849 \,(11)$ & $89.227 \,(99)$ & $148.826 \,(68)$\\
$1\,\MW$ & $104.9482 \,(94)$ & $93.789 \,(76)$ & $138.808 \,(56)$\\
$2\,\MW$ & $93.1789 \,(83)$ & $95.493 \,(60)$ & $130.401 \,(57)$\\

\hline
\end{tabular}
\caption{WW+jet cross sections at the LHC, calculated in the
  five-flavour scheme with CTEQ6 PDFs. The cut on the transverse
  momentum of the jet is set to $50\GeV$. In the lower part of the table
  where the decays of the W bosons are included
  further cuts according to \refeqsx{eq:nr:cutetapT}{eq:nr:cutR} are applied.}
\label{ta:nr:crosssectionsLHC50}
\end{center}
\end{table}
\begin{table}
\begin{center}
\begin{tabular}{|c|c|c|c|}
\hline
\multicolumn{4}{|c|}{$ \Pp\Pp \to \PWp\PWm+ \mathrm{jet}+X$ @ $14\TeV$}\\
\hline
$\mu=\muren=\mufact$&$\sigma_{\mathrm{LO}}[\pba]$ & $\sigma_{\mathrm{NLO,excl}}[\pba]$ & $\sigma_{\mathrm{NLO,incl}}[\pba]$\\
\hline
$0.5\,\MW$ & $12.01935 \,(44)$ & $11.8943 \,(76)$ & $16.2377 \,(73)$\\
$1\,\MW$ & $10.37216 \,(39)$ & $11.4827 \,(57)$ & $14.6889 \,(57)$\\
$2\,\MW$ & $9.02371 \,(34)$ & $11.0114 \,(46)$ & $13.4405 \,(44)$\\

\hline
\multicolumn{4}{|c|}{$ \Pp\Pp \to \PWp(\to\Pne\Pe^+)\PWm(\to\Pmu^-\Pnmubar) + \mathrm{jet}+X$ @ $14\TeV$}\\
\hline
$\mu=\muren=\mufact$&$\sigma_{\mathrm{LO}}[\fb]$ & $\sigma_{\mathrm{NLO,excl}}[\fb]$ & $\sigma_{\mathrm{NLO,incl}}[\fb]$\\
\hline
$0.5\,\MW$ & $61.6020 \,(66)$ & $57.471 \,(40)$ & $79.722 \,(45)$\\
$1\,\MW$ & $53.2170 \,(57)$ & $56.734 \,(47)$ & $73.199 \,(39)$\\
$2\,\MW$ & $46.3585 \,(50)$ & $55.192 \,(38)$ & $67.680 \,(28)$\\

\hline
\end{tabular}
\caption{As in \refta{ta:nr:crosssectionsLHC50}, but with $p_{\mathrm{T,jet,cut}}=100\GeV$.}
\label{ta:nr:crosssectionsLHC100}
\end{center}
\end{table}
\begin{table}
\begin{center}
\begin{tabular}{|c|c|c|c|}
\hline
\multicolumn{4}{|c|}{$ \Pp\bPp \to \PWp\PWm + \mathrm{jet}+X$ @ $1.96\TeV$}\\
\hline
$\mu=\muren=\mufact$&$\sigma_{\mathrm{LO}}[\pba]$ & $\sigma_{\mathrm{NLO,excl}}[\pba]$ & $\sigma_{\mathrm{NLO,incl}}[\pba]$\\
\hline
$0.5\,\MW$ & $2.88967 \,(29)$ & $2.6343 \,(32)$ & $3.1417 \,(30)$\\
$1\,\MW$ & $2.30259 \,(24)$ & $2.5761 \,(24)$ & $2.9257 \,(21)$\\
$2\,\MW$ & $1.87503 \,(20)$ & $2.4208 \,(18)$ & $2.6688 \,(16)$\\

\hline
\multicolumn{4}{|c|}{$ \Pp\bPp \to \PWp(\to\Pne\Pe^+)\PWm(\to\Pmu^-\Pnmubar) + \mathrm{jet}+X$ @ $1.96\TeV$}\\
\hline
$\mu=\muren=\mufact$&$\sigma_{\mathrm{LO}}[\fb]$ & $\sigma_{\mathrm{NLO,excl}}[\fb]$ & $\sigma_{\mathrm{NLO,incl}}[\fb]$\\
\hline
$0.5\,\MW$ & $13.7407 \,(21)$ & $12.2366 \,(88)$ & $14.8002 \,(91)$\\
$1\,\MW$ & $10.9245 \,(17)$ & $12.0582 \,(72)$ & $13.8052 \,(73)$\\
$2\,\MW$ & $8.8789 \,(14)$ & $11.3695 \,(68)$ & $12.6152 \,(68)$\\

\hline
\end{tabular}
\caption{As in \refta{ta:nr:crosssectionsLHC50}, but at the Tevatron with $p_{\mathrm{T,jet,cut}}=20\GeV$.}
\label{ta:nr:crosssectionsTev20}
\end{center}
\end{table}
\begin{table}
\begin{center}
\begin{tabular}{|c|c|c|c|}
\hline
\multicolumn{4}{|c|}{$ \Pp\bPp \to \PWp\PWm + \mathrm{jet}+X$ @ $1.96\TeV$}\\
\hline
$\mu=\muren=\mufact$&$\sigma_{\mathrm{LO}}[\fb]$ & $\sigma_{\mathrm{NLO,excl}}[\fb]$ & $\sigma_{\mathrm{NLO,incl}}[\fb]$\\
\hline
$0.5\,\MW$ & $0.843556 \,(80)$ & $0.76867 \,(81)$ & $0.83508 \,(80)$\\
$1\,\MW$ & $0.657787 \,(64)$ & $0.74241 \,(58)$ & $0.78664 \,(55)$\\
$2\,\MW$ & $0.525619 \,(52)$ & $0.68820 \,(44)$ & $0.71884 \,(44)$\\

\hline
\multicolumn{4}{|c|}{$ \Pp\bPp \to \PWp(\to\Pne\Pe^+)\PWm(\to\Pmu^-\Pnmubar) + \mathrm{jet}+X$ @ $1.96\TeV$}\\
\hline
$\mu=\muren=\mufact$&$\sigma_{\mathrm{LO}}[\fb]$ & $\sigma_{\mathrm{NLO,excl}}[\fb]$ & $\sigma_{\mathrm{NLO,incl}}[\fb]$\\
\hline
$0.5\,\MW$ & $4.55105 \,(67)$ & $4.0799 \,(36)$ & $4.4484 \,(36)$\\
$1\,\MW$ & $3.54156 \,(53)$ & $3.9594 \,(27)$ & $4.2037 \,(27)$\\
$2\,\MW$ & $2.82493 \,(43)$ & $3.6774 \,(21)$ & $3.8466 \,(21)$\\

\hline
\end{tabular}
\caption{As in \refta{ta:nr:crosssectionsLHC50}, but at the Tevatron with $p_{\mathrm{T,jet,cut}}=50\GeV$.}
\label{ta:nr:crosssectionsTev50}
\end{center}
\end{table}

\end{sloppypar}

In general, the influence of the restriction on genuine WW+jet production via the described jet veto is not that large in the Tevatron setup (see \reffi{fi:nr:pdfcomare}). This can be understood from the lower 
CM energy at the Tevatron: The energy for producing a second hard jet is available less frequently here, so that a stronger suppression of WW+2jets events is obtained. The fact that the difference between the two NLO observables strongly decreases when going to higher $p_{\mathrm{T,jet,cut}}$ values---as shown on the right-hand side of \reffi{fi:nr:WWproductionNLO}---confirms this interpretation.

\begin{figure}
\centering
\includegraphics[bb = 200 380 450 700, scale = .8]{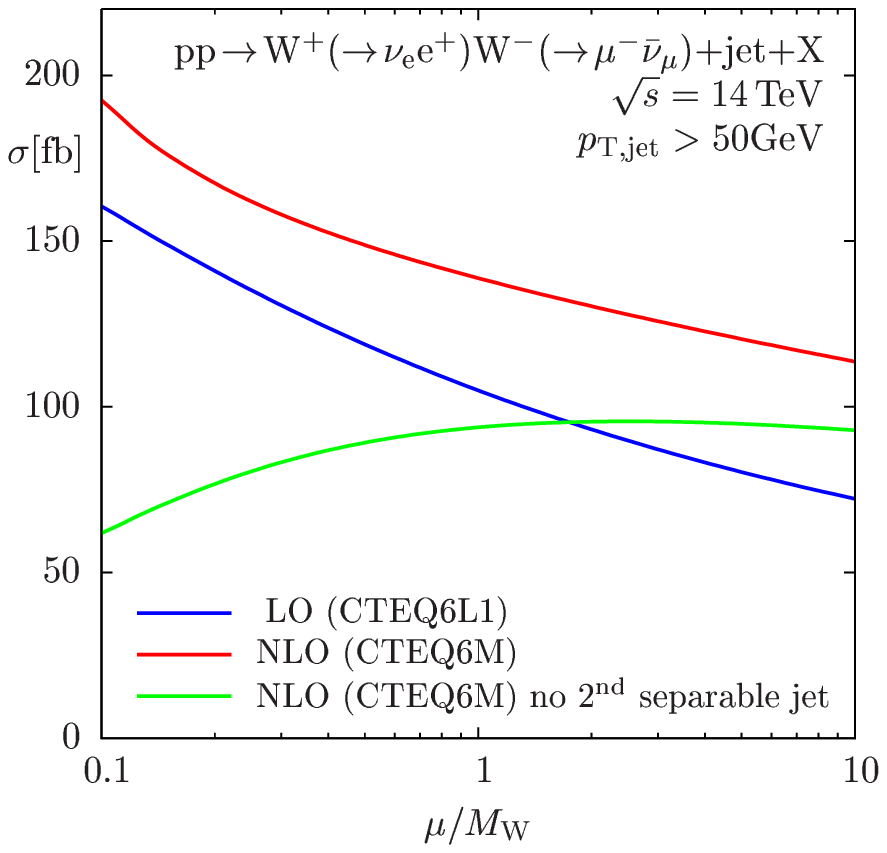}
\includegraphics[bb = 160 380 410 700, scale = .8]{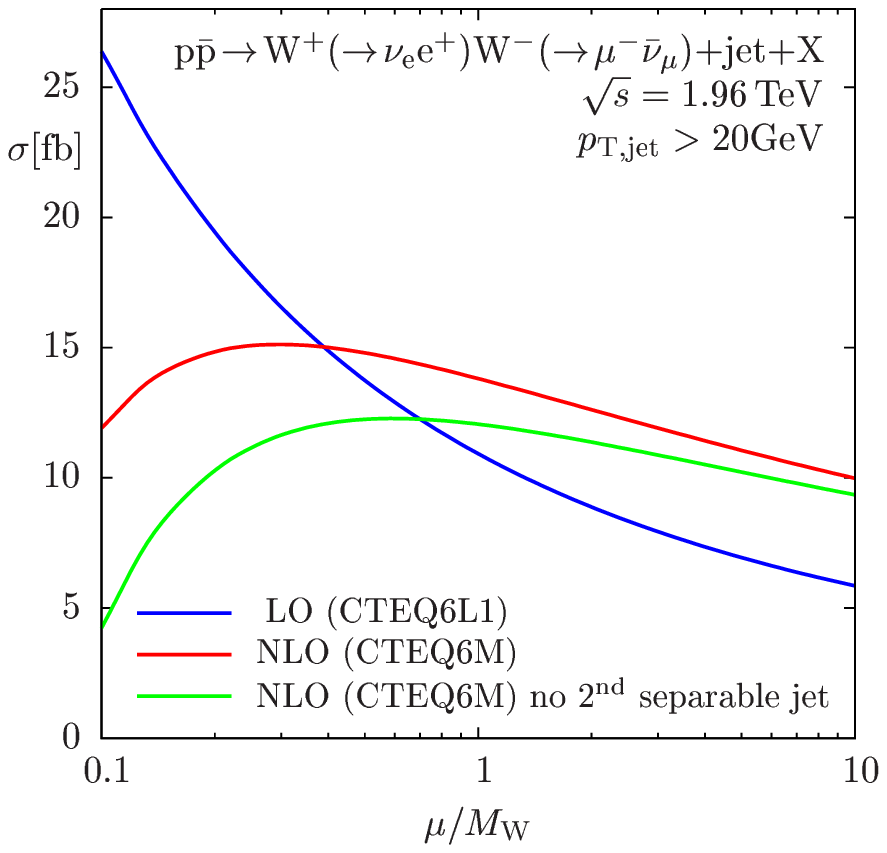}\\[-2em]
\includegraphics[bb = 200 380 450 700, scale = .8]{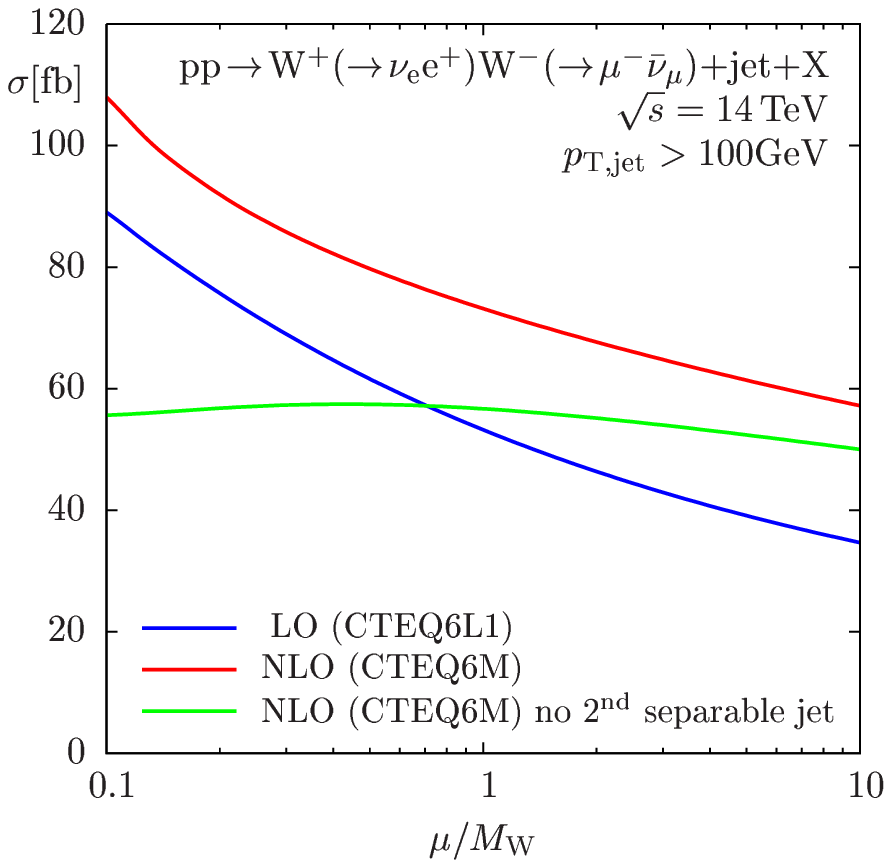}
\includegraphics[bb = 160 380 410 700, scale = .8]{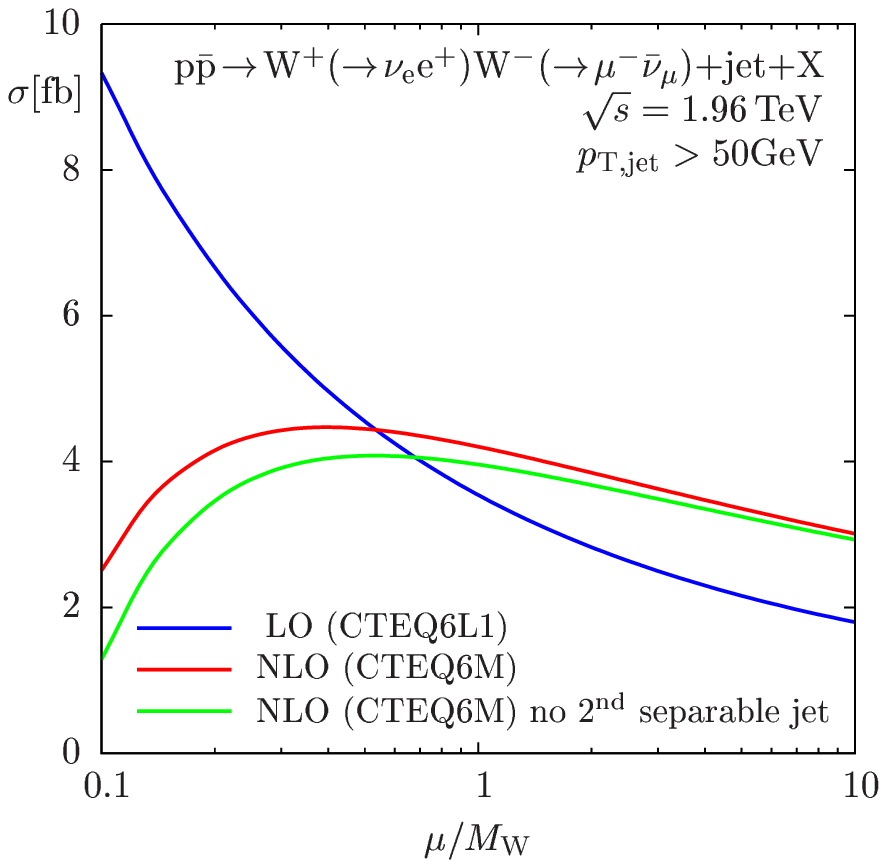}\\[-1em]
\caption{Scale dependence of the WW+jet cross sections with W decays included
  and further cuts applied according to \refeqsx{eq:nr:cutetapT}{eq:nr:cutR}. The five-flavour scheme with CTEQ6 PDFs is
  used. In the LHC setup, the results are given for
  $p_{\mathrm{T,jet,cut}}=50\GeV$ (upper left plot) and for
  $p_{\mathrm{T,jet,cut}}=100\GeV$ (lower left plot). For the Tevatron we show
  results for $p_{\mathrm{T,jet,cut}}=20\GeV$ (upper right plot) and for
  $p_{\mathrm{T,jet,cut}}=50\GeV$ (lower right plot).}
\label{fi:nr:scalevariationdecay}
\end{figure}

In \reffi{fi:nr:scalevariationdecay}, the respective plots are shown with the
W decays included via the iNWA and with further cuts applied according to
\refeqsx{eq:nr:cutetapT}{eq:nr:cutR}. Qualitatively, the features
discussed above can be observed as well.

\paragraph{Dependence on the transverse-momentum cut on the jet}
\begin{figure}
\centering
\includegraphics[bb = 200 380 450 650, scale = .8]{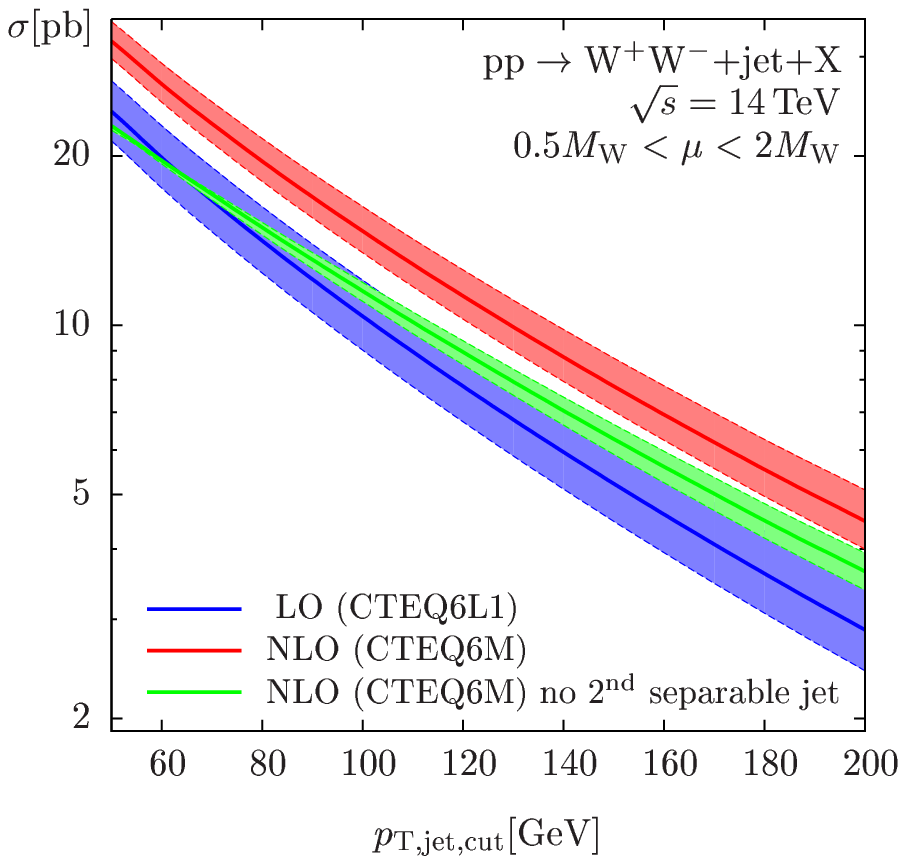}
\includegraphics[bb = 160 380 410 650, scale = .8]{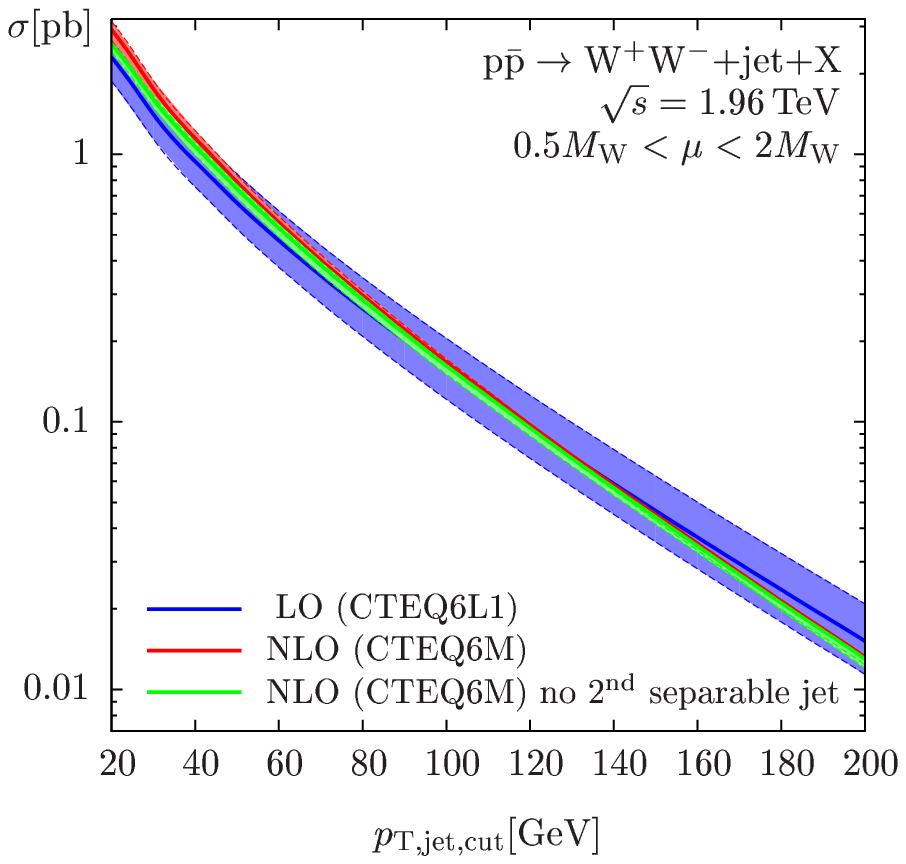}\\
\includegraphics[bb = 200 380 450 650, scale = .8]{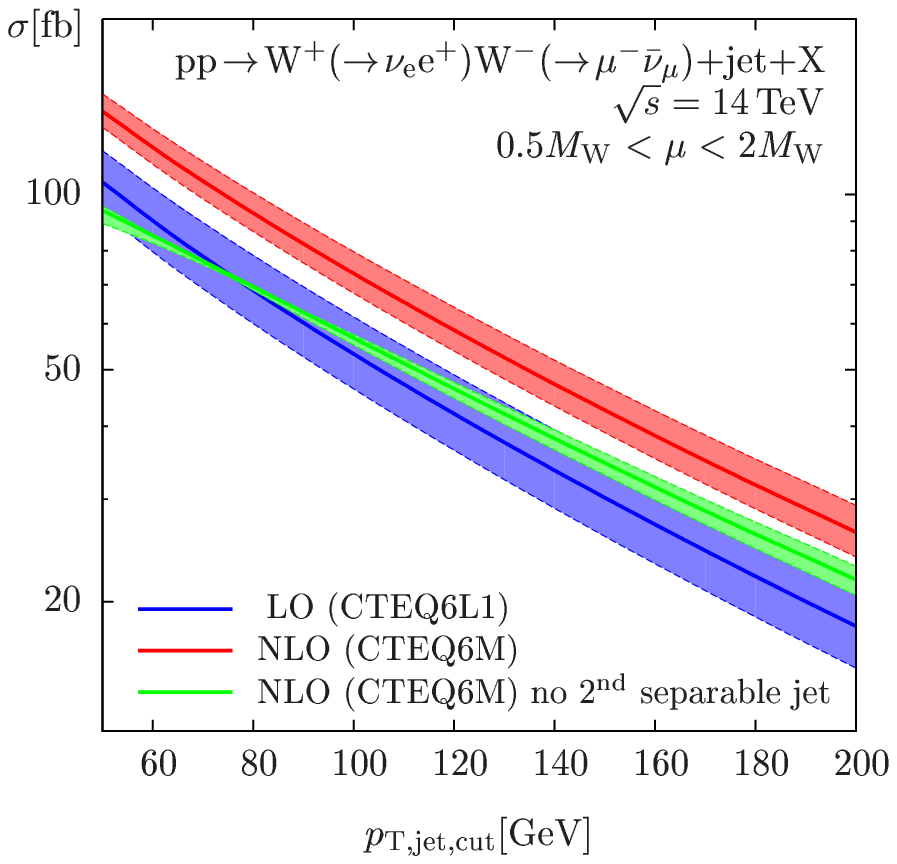}
\includegraphics[bb = 160 380 410 650, scale = .8]{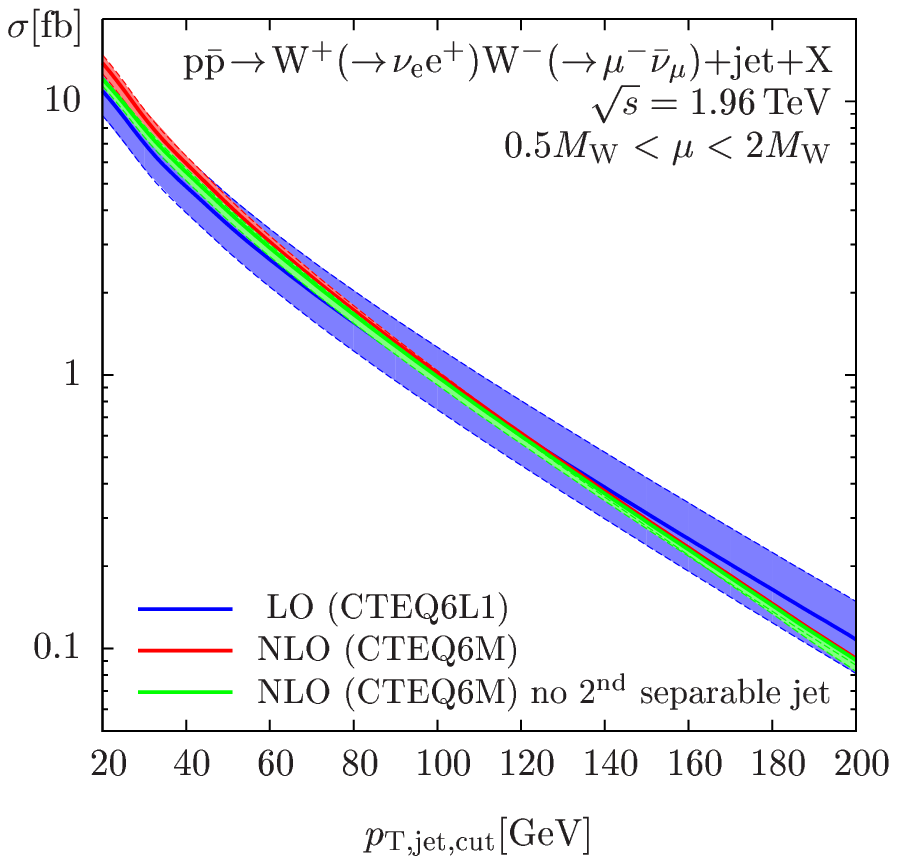}\\[-1em]
\caption[Variation of the WW+jet cross sections with $p_{\mathrm{T,jet,cut}}$
  at the LHC.]{Variation of the WW+jet cross sections with
  $p_{\mathrm{T,jet,cut}}$ at the LHC (l.h.s.) and at the Tevatron (r.h.s.):
The bands correspond to a scale variation, again with
\mbox{$\mu=\mu_{\mathrm{fact}}=\mu_{\mathrm{ren}}$}, by a factor 2 around the
central scale. The red and the green curves correspond to the different
definitions of the NLO observable, where the red one depicts the more
inclusive quantity. The upper plots correspond to stable W's, the lower
  plots include leptonic W decays with the additional cuts of
  \refeqsx{eq:nr:cutetapT}{eq:nr:cutR} applied.}
\label{fi:nr:WWproductionpTjetcut}
\end{figure}

To close the discussion of the NLO cross sections for WW+jet 
that are not sensitive to the decays of the $\PW$ bosons, the dependence on the cut applied to the transverse momentum of the jet is considered here. In \reffi{fi:nr:WWproductionpTjetcut}, the $p_\mathrm{T,jet,cut}$ dependence is shown for the LHC and the Tevatron setup, respectively. In order to introduce a measure for the scale uncertainties of the cross sections, bands are depicted that correspond to a variation of \mbox{$\mu=\muren=\mufact$} by a factor 2 around the central scale. A crossing of the curves for differing scale values leads to vanishing band widths at some points, which is an artifact of how the results are depicted and should not be misinterpreted as a vanishing scale uncertainty. 
In the plots of \reffi{fi:nr:WWproductionpTjetcut}, the band corresponding to the more inclusive cross section is partially covered by the more exclusive one. Both curves are, however, shown here in order to confirm the statement that the effect of genuine WW+2jets events decreases with an increasing value of $p_\mathrm{T,jet,cut}$. 
This becomes manifest in the overlap of the two bands especially for large cut values at the Tevatron, but is also evident in the plots for the LHC setup from the convergence of the two NLO bands when going to larger values for $p_\mathrm{T,jet,cut}$. Eventually, the $p_\mathrm{T,jet,cut}$ variation reflects the behaviour discussed in the previous paragraph: For Tevatron, a considerable reduction of the scale uncertainty is achieved when going from LO to NLO, whereas this reduction is only mild for LHC unless WW+2jets events are vetoed.

\subsection{Results on differential cross sections}

Now we mainly focus on differential distributions in observables
defined from the W~decay products, i.e.\
the cuts on the decay leptons specified in 
Eqs.~\refeq{eq:nr:cutetapT}--\refeq{eq:nr:cutR} are applied.

\subsubsection{LO analysis of the different decay descriptions}
\label{se:nu:approximations}
The integrated WW+jet cross section as defined above is not sensitive
to the decays of the $\PW$ bosons. An inclusion of leptonic decays
into the calculation of WW+jet production can be performed in
different ways. In this section, a comparison of LO results for the
three strategies discussed in \refse{se:decays} is performed, which
are a full amplitude calculation within the complex-mass
scheme, the naive NWA, and 
its improved version iNWA that treats the $\PW$ bosons as on-shell particles, but keeps spin correlations. The aim of this discussion is to find an adequate approximation in order to avoid performing the full amplitude calculation but still to obtain an appropriate description of the decays. 

Considering only
integrated cross sections, both approximations reproduce the full
results in the LHC setup at the expected accuracy of $\mathcal{O}(\GV/\MV)$. More
precisely, we find a deviation of about $3\%$
for the naive NWA and of less than $1\%$ for the improved version with
$p_{\mathrm{T,jet,cut}}=50,100\GeV$. In the Tevatron setup, the naive
NWA predicts cross sections deviating by roughly $15\%$ ($9\%$) for
$p_{\mathrm{T,jet,cut}}=20\GeV$ ($50\GeV$), while the iNWA is good within 
$1\%$.
The differences between the two approximations are due to the ignored spin correlations in the naive NWA, which lead to different distributions of the decay leptons over the phase space. These changes become manifest in the integrated cross sections if cuts on the decay products are applied. 
In general, both the naive and the improved version of the NWA give the same integrated cross sections 
if no leptonic cuts are applied. This is due to the fact that taking the spin correlations into account only causes a (W-polarization-dependent)
redistribution of the momenta of the decay leptons, 
which does not influence the result as long as the full decay phase spaces are
integrated over.

\begin{figure}
\centering
\includegraphics[bb = 200 380 450 700, scale = .8]{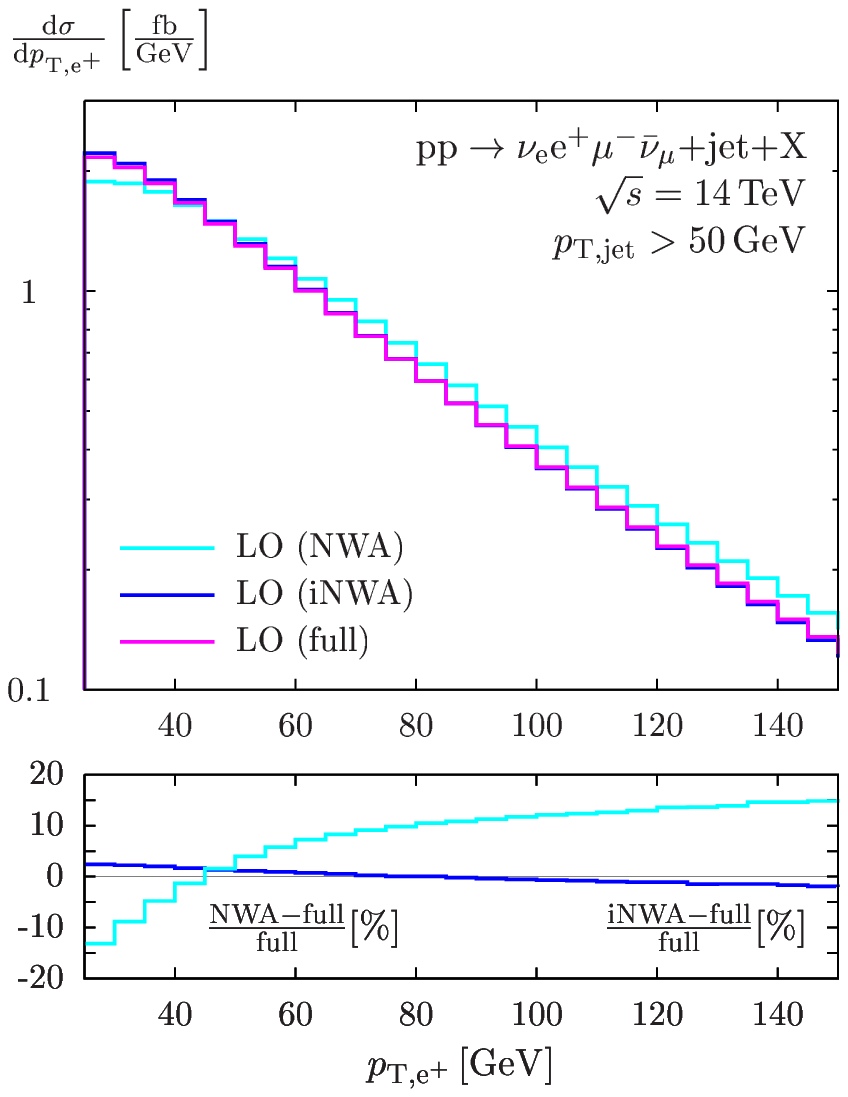}
\includegraphics[bb = 160 380 410 700, scale = .8]{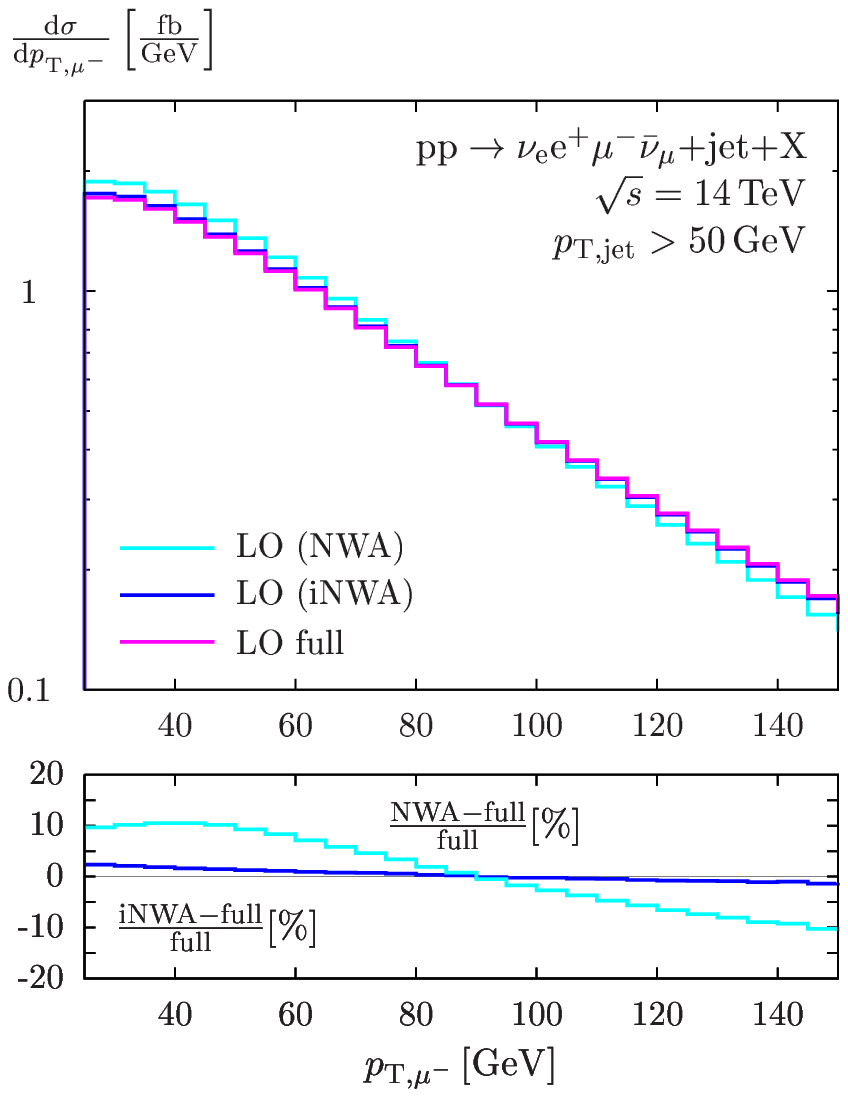}\\[-2em]
\caption{Comparison of different $\PW$-decay descriptions in the $p_{\mathrm{T}}$ distributions of each of the two decay leptons. The LO cross sections are evaluated at \mbox{$\mu=\mu_{\mathrm{fact}}=\mu_{\mathrm{ren}}=\MW$} for the full amplitude calculation, the naive NWA, and the improved NWA.}
\label{fi:nr:decayLOdistWWa}
\end{figure}

\begin{figure}
\centering
\includegraphics[bb = 200 380 450 700, scale = .8]{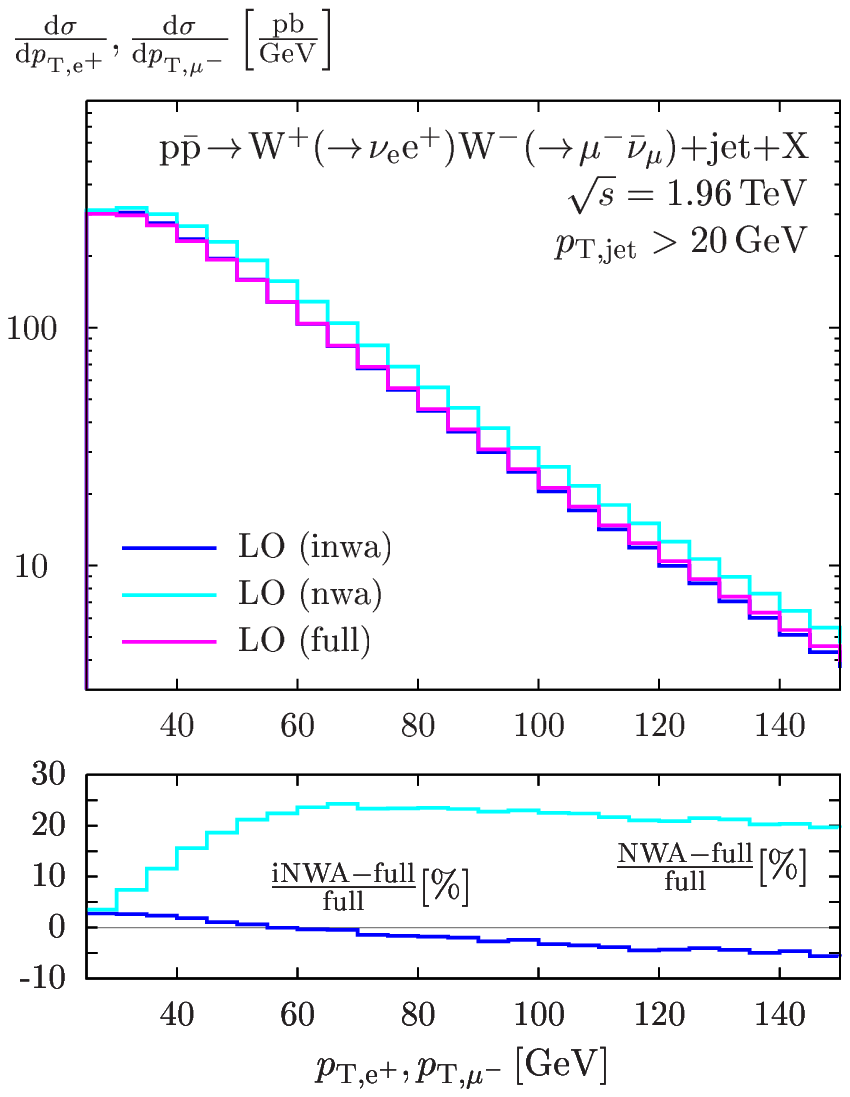}
\includegraphics[bb = 160 380 410 700, scale = .8]{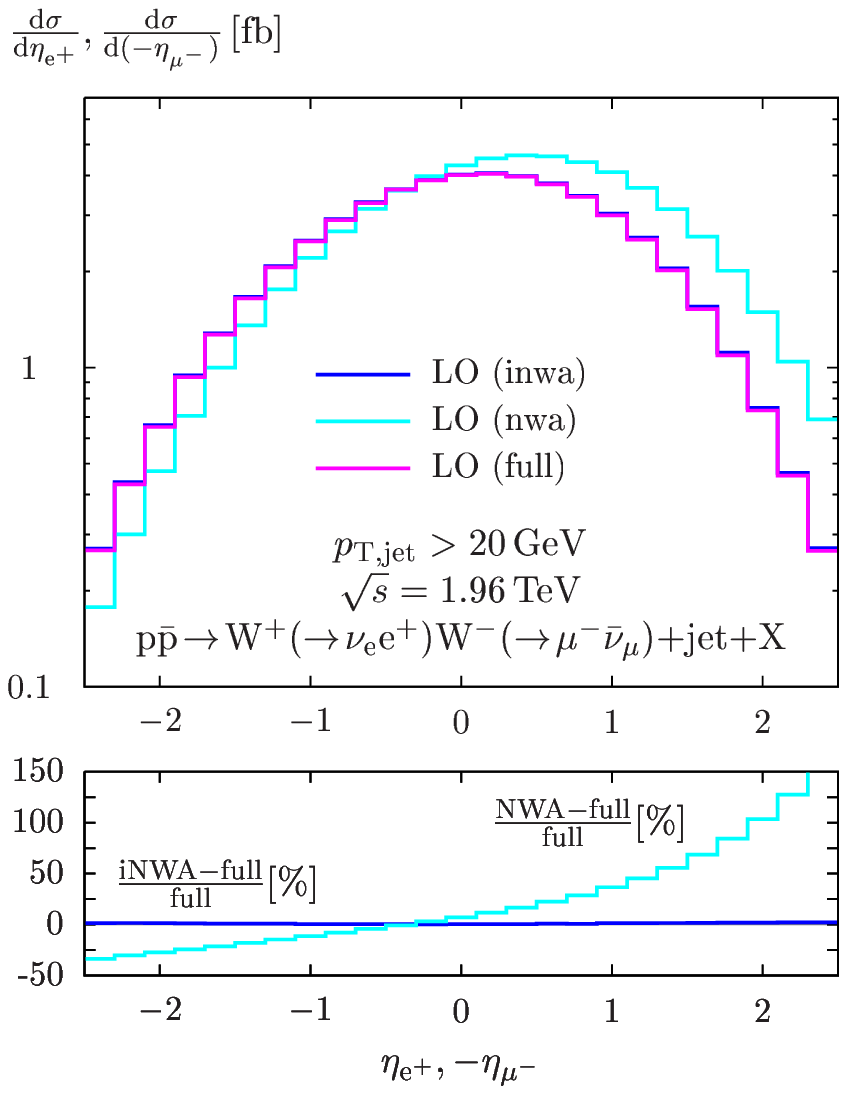}\\[-2em]
\caption{Comparison of different $\PW$-decay descriptions in the
  $p_{\mathrm{T}}$ and $\eta$ distributions of each of the two decay leptons. The LO
  cross sections are evaluated at
  \mbox{$\mu=\mu_{\mathrm{fact}}=\mu_{\mathrm{ren}}=\MW$} for the full
  amplitude calculation, the naive NWA, and the improved NWA at the Tevatron.}
\label{fi:nr:decayLOdistWWd}
\end{figure}

Distributions in specific variables are, however, strongly affected, which can
be seen from the differential cross sections for the LHC given in \reffisb{fi:nr:decayLOdistWWa}{fi:nr:decayLOdistWWb}{fi:nr:decayLOdistWWc}. 
\begin{figure}
\centering
\includegraphics[bb = 200 500 450 700, scale = .8]{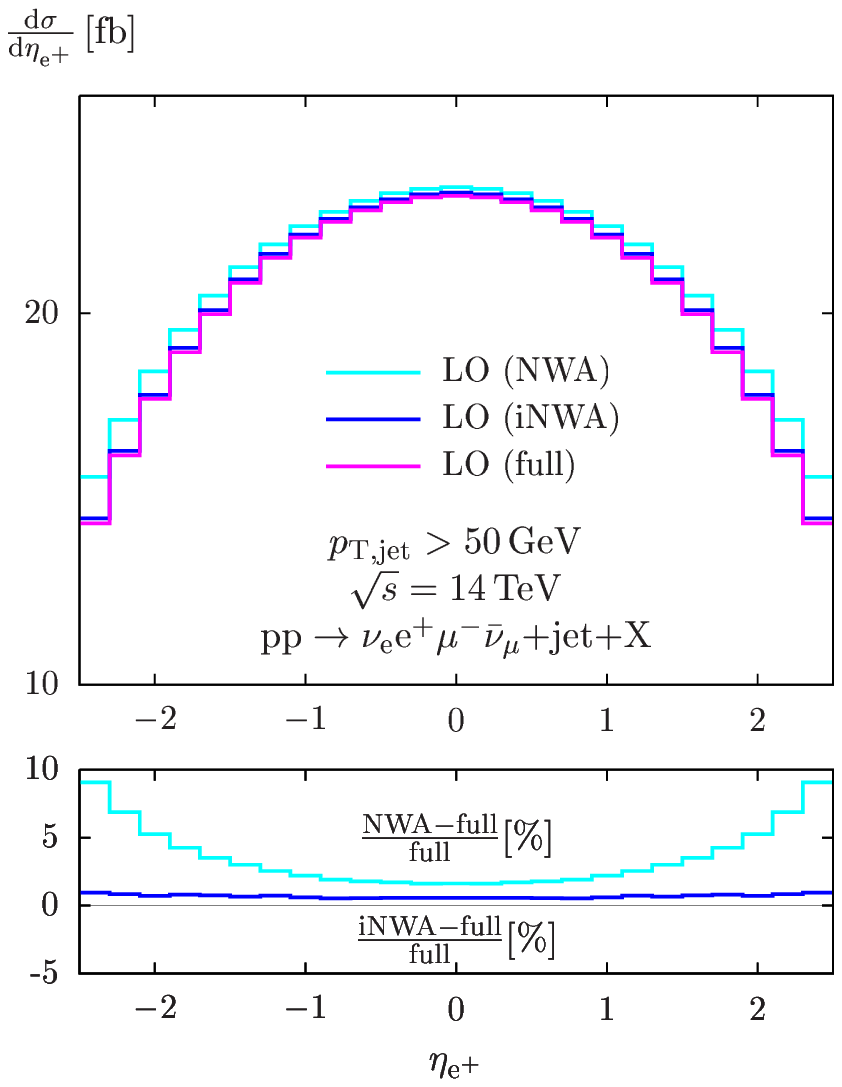}
\includegraphics[bb = 160 500 410 700, scale = .8]{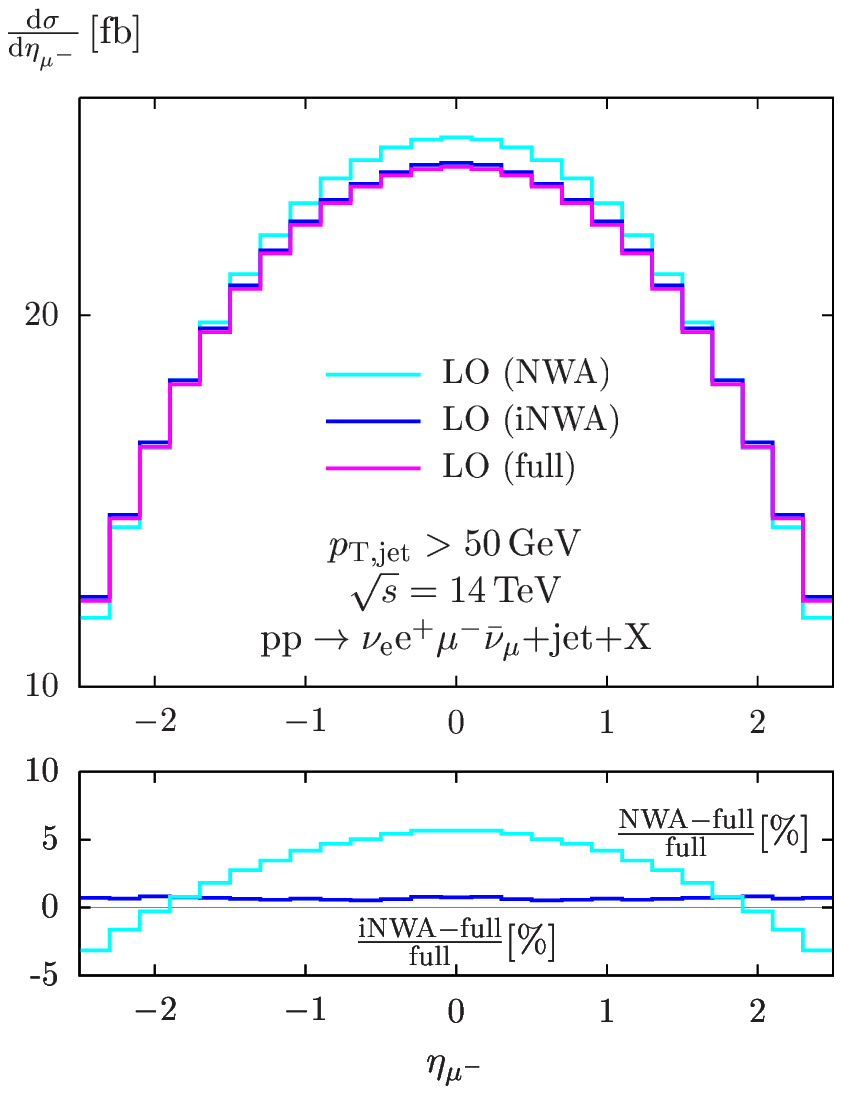}\\[2.8cm]
\caption{As in \reffi{fi:nr:decayLOdistWWa}, but for the $\eta$ distributions of the two decay leptons.}
\label{fi:nr:decayLOdistWWb}
\vspace*{5ex}
\centering
\includegraphics[bb = 200 380 450 700, scale = .8]{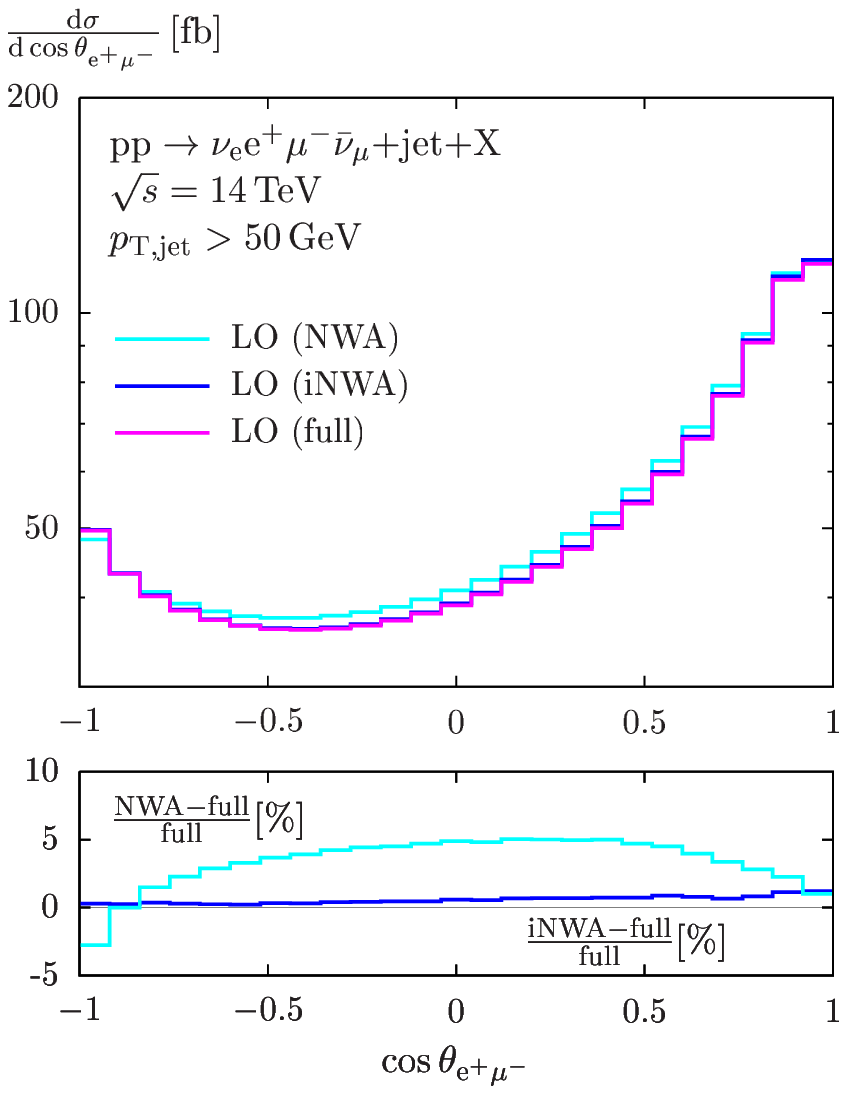}
\includegraphics[bb = 160 380 410 700, scale = .8]{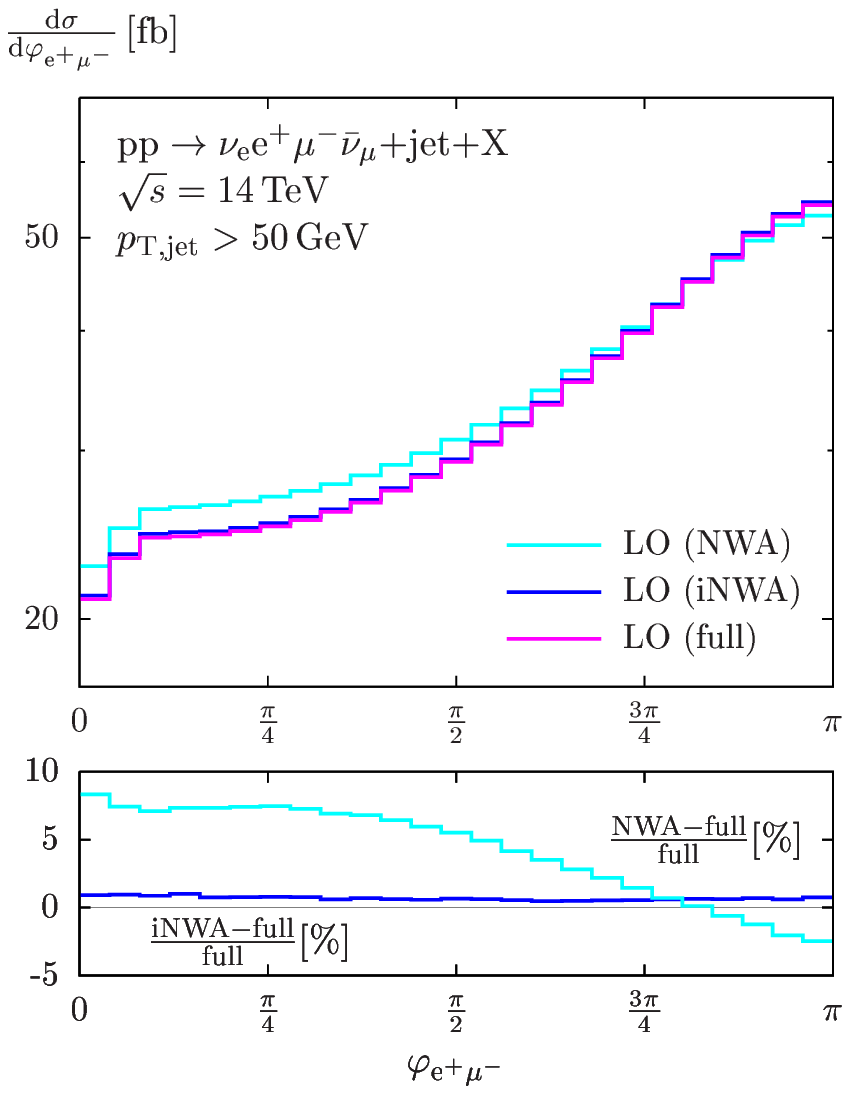}\\[-1.5em]
\caption{As in \reffi{fi:nr:decayLOdistWWa}, but for the cosine of the opening angle, $\cos\theta$, and the angle $\varphi$ in the transverse plane between the two leptons.}
\label{fi:nr:decayLOdistWWc}
\end{figure}
Only the distributions for decay leptons are shown there, because these are, naturally, mainly affected by the different decay descriptions. Considering the distributions of transverse momentum $p_\mathrm{T}$ and pseudo-rapidity $\eta$ of each of the two decay leptons, the improved NWA delivers a very accurate reproduction of the full calculation, whereas the naive version deviates by up to 15\% in some phase-space regions. The distributions of the angles between the two leptons---$\varphi$ denotes the angle in the transverse plane and $\cos\theta$ the cosine of the 
angle between the two leptons---resulting from the full amplitude calculation
are also in general reproduced more precisely by the improved NWA. At the
  Tevatron the situation is quite similar, which can be read off \reffisa{fi:nr:decayLOdistWWd}{fi:nr:decayLOdistWWe}: While the improved NWA
reproduces the full result quite well, the predictions of the naive NWA
deviate quite strongly---in particular in case of the pseudo-rapidity
distributions of the leptons.
\begin{figure}
\centering
\includegraphics[bb = 200 380 450 700, scale = .8]{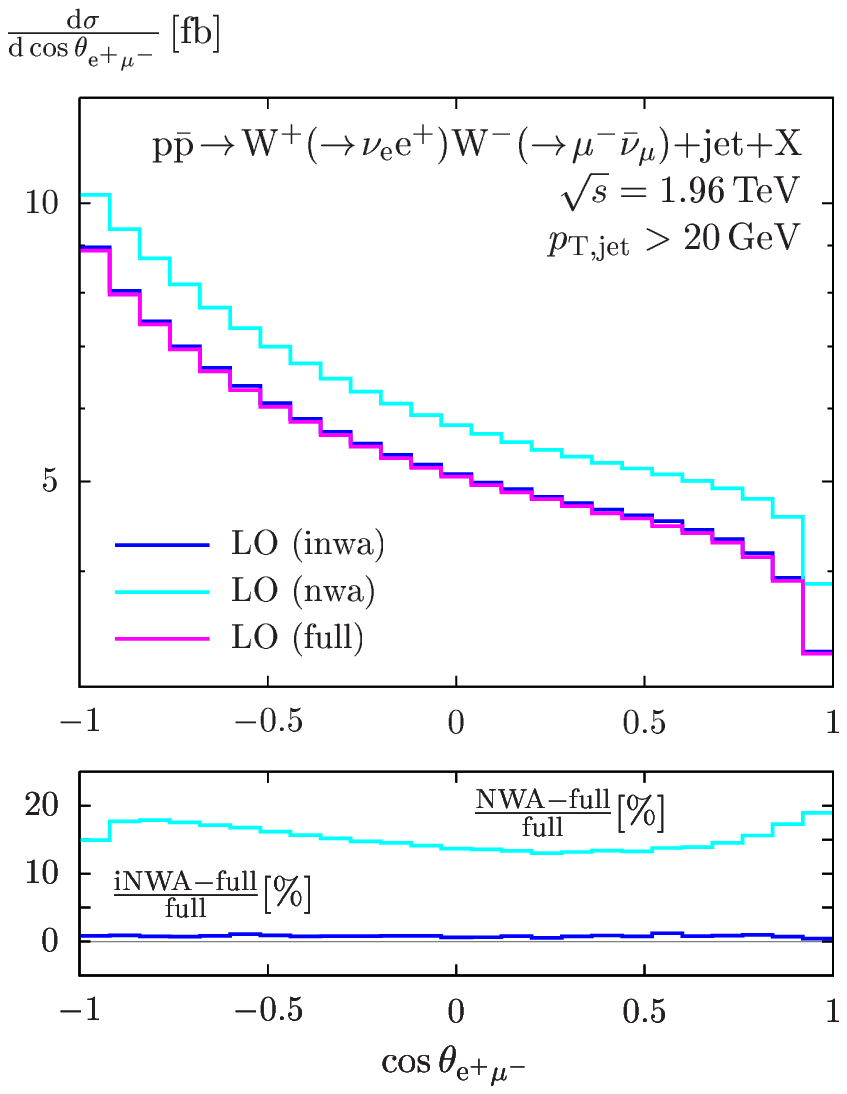}
\includegraphics[bb = 160 380 410 700, scale = .8]{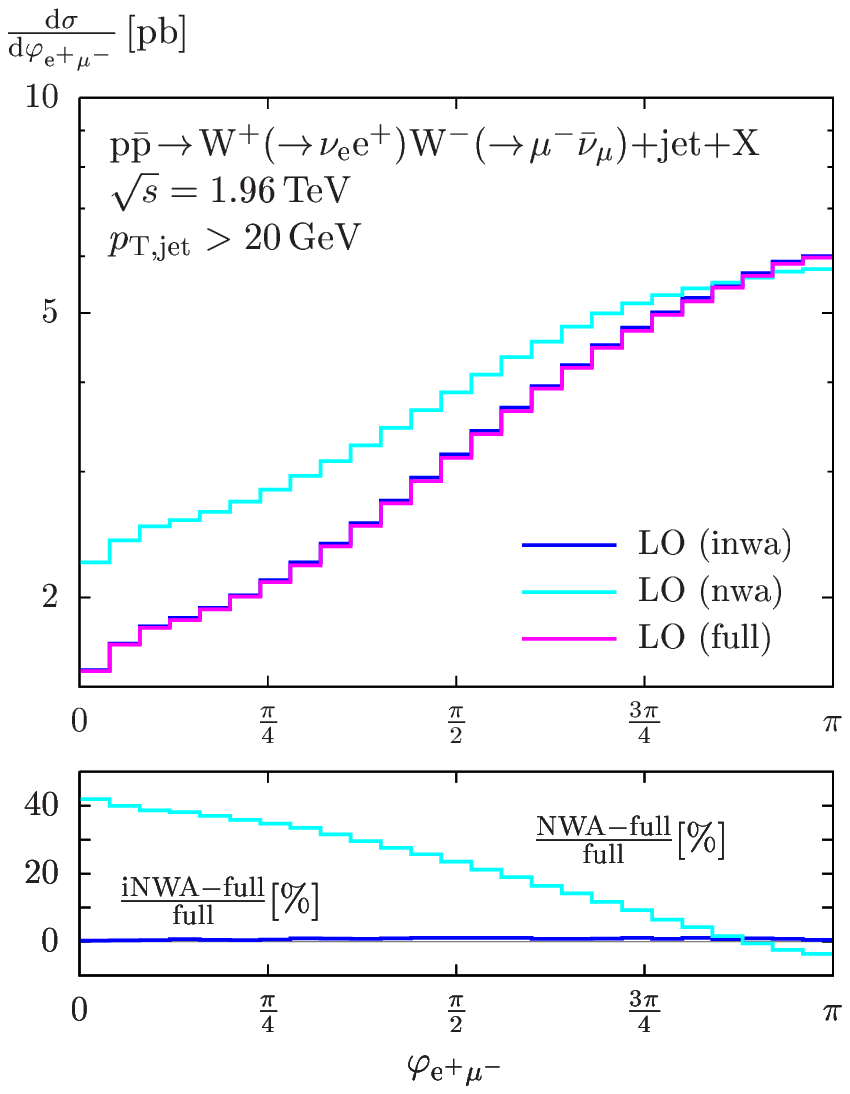}\\[-1.5em]
\caption{As in \reffi{fi:nr:decayLOdistWWd}, but for the cosine of the opening angle, $\cos\theta$, and the angle $\varphi$ in the transverse plane between the two leptons.}
\label{fi:nr:decayLOdistWWe}
\end{figure}

The analysis of this section justifies the application of the improved NWA for the NLO QCD calculations to \mbox{\tppWWj} with leptonic decays.

\subsubsection{Differential NLO cross sections at the LHC}
\label{se:nr:decayNLOdist}
\begin{figure}
\centering
\includegraphics[bb = 200 520 450 700, scale = .8]{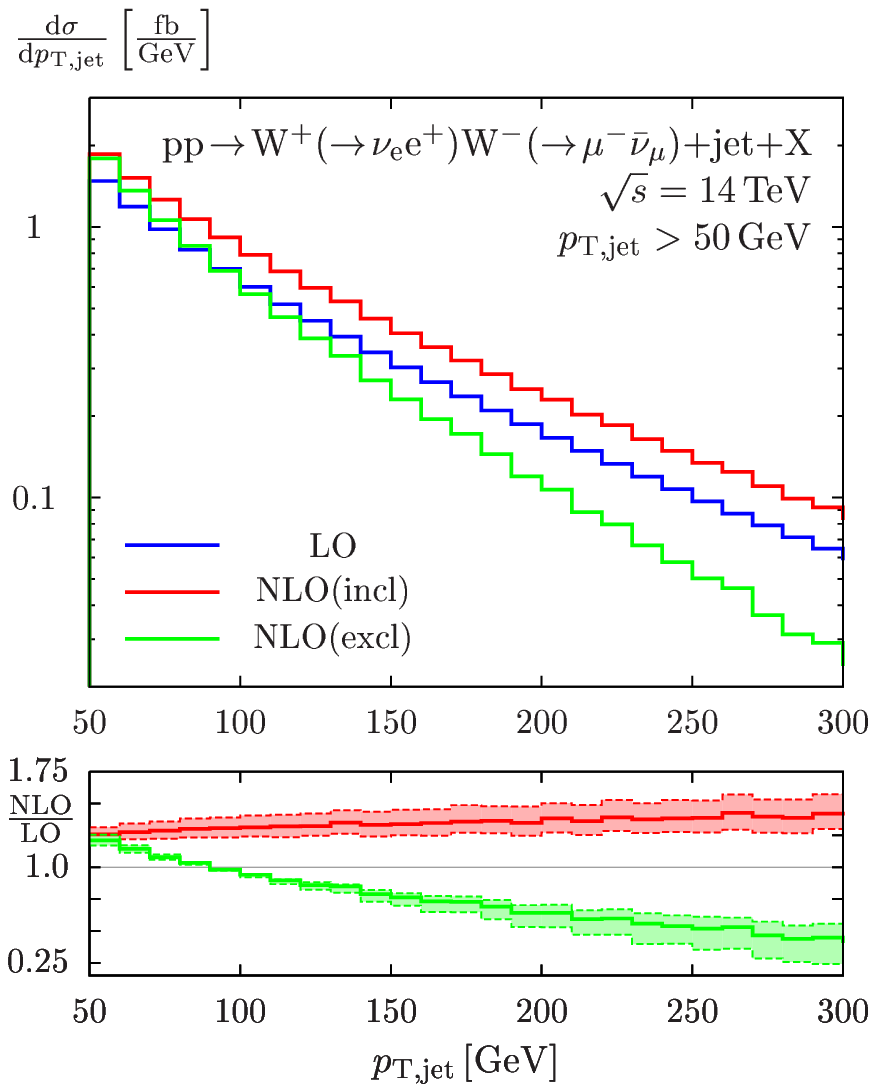}
\includegraphics[bb = 160 520 410 700, scale = .8]{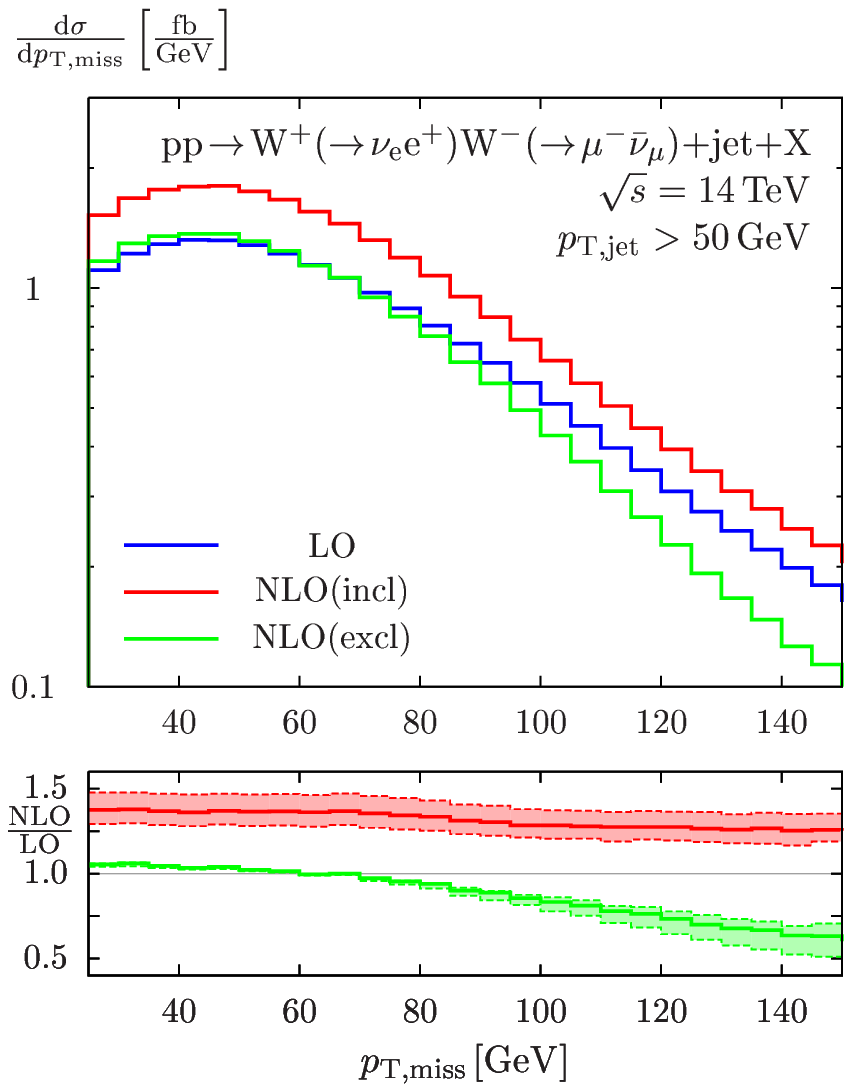}\\[4cm]
\includegraphics[bb = 200 400 450 700, scale = .8]{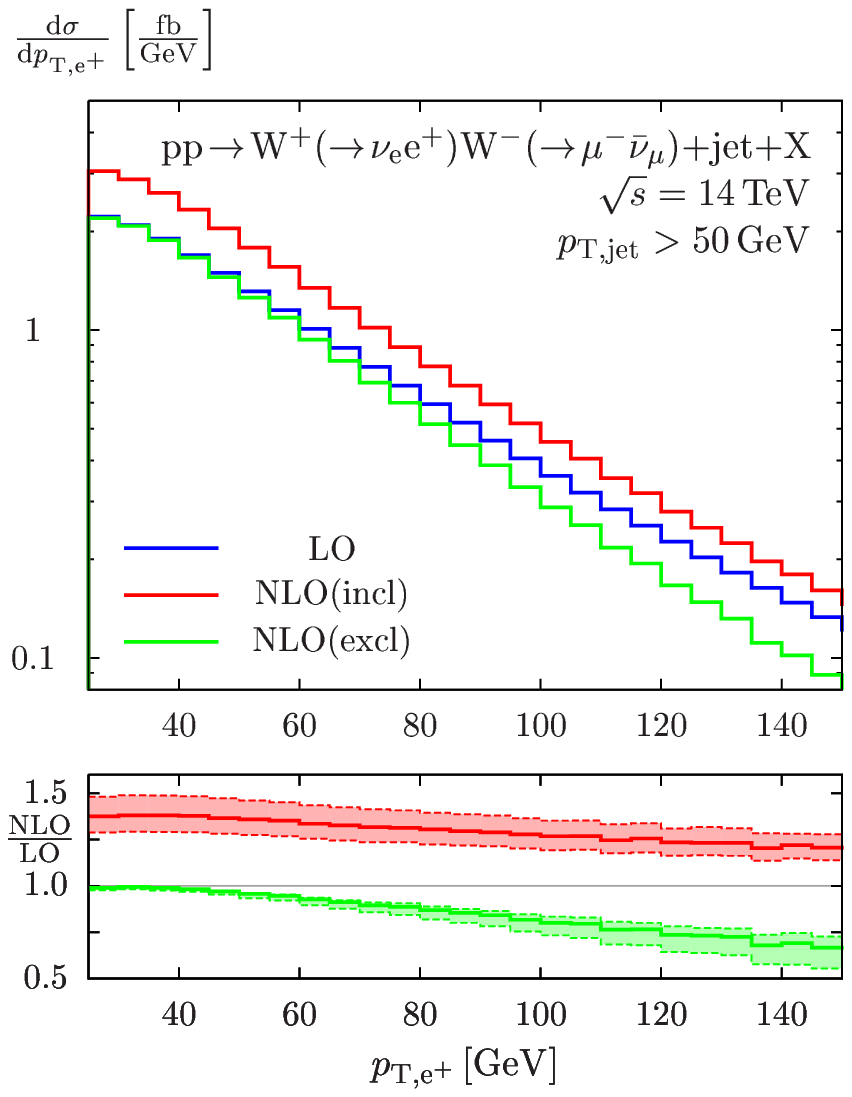}
\includegraphics[bb = 160 400 410 700, scale = .8]{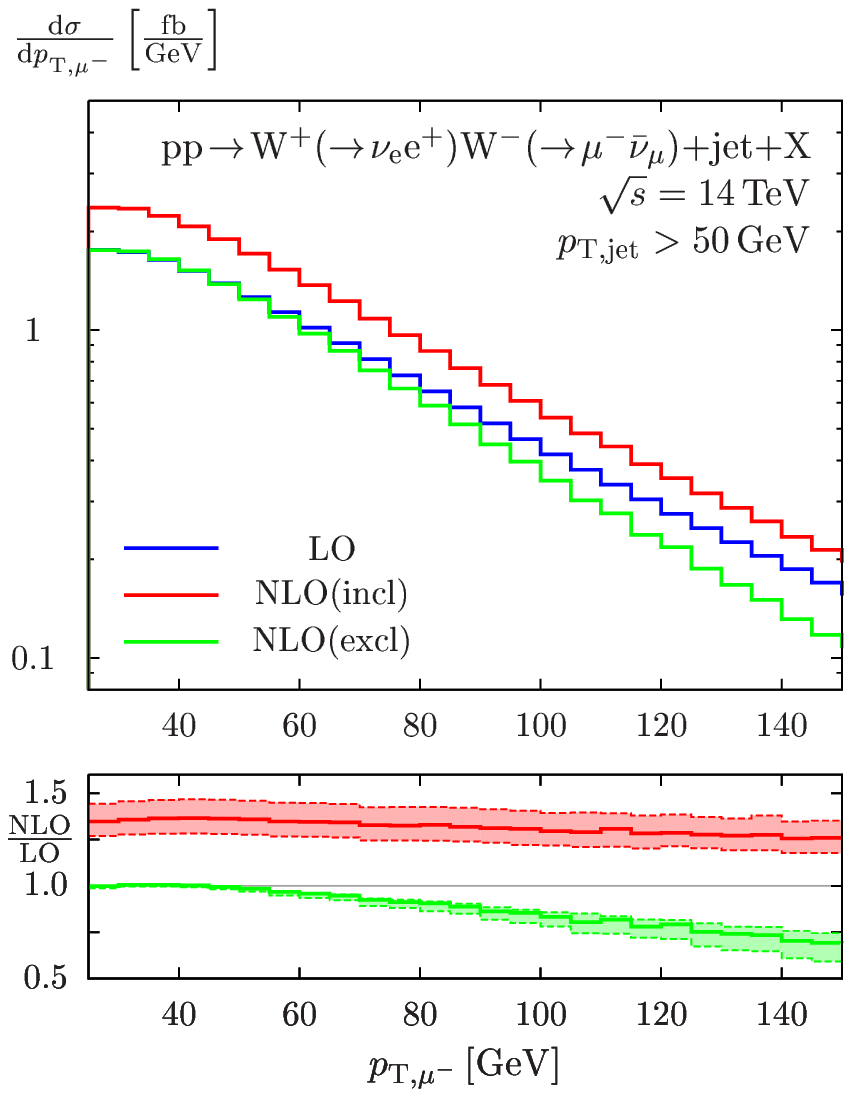}\\
\caption{Differential cross sections for WW+jet with decays included in the
  improved NWA at the LHC: The LO and NLO distributions are shown for
  \mbox{$\mu=\mu_{\rm{fact}}=\mu_{\rm{ren}}=\MW$}. The distributions for the
  transverse momenta $p_{\mathrm{T}}$  of the jet and of the decay leptons,
  and for the missing transverse momentum $p_{\mathrm{T,miss}}$ are
  depicted. The bands in the $K$-factors are precisely defined in the text.}
\label{fi:nr:decayNLOdistWWa}
\end{figure}
A survey of distributions for differential WW+jet cross sections with leptonic $\PW$ decays is given in \reffisx{fi:nr:decayNLOdistWWa}{fi:nr:decayNLOdistWWc} for the LHC setup with \mbox{$p_{\mathrm{T,jet,cut}}=50\GeV$} and $\mu=\muren=\mufact=\MW$. The relative size of the NLO corrections is represented in terms of a $K$-factor. The depicted bands correspond to a variation of the scale $\mu$ by a factor of 2 in the NLO quantities only, i.e.\ we show $\sigma_{\mathrm{NLO}}(\mu)/\sigma_{\mathrm{LO}}(\MW)$ with $0.5\MW<\mu<2\MW$.

The jet distributions are understood as distributions of the hardest jet. 
\reffi{fi:nr:decayNLOdistWWa} provides transverse-momentum
distributions of the hadronic jet, of the missing transverse momentum
due to \mbox{\antineutrinos} leaving the detector undetected 
(upper plots), and of the two charged decay leptons (lower plots).
For all $p_{\mathrm{T}}$ distributions, a tendency of the more exclusive NLO cross section to decrease faster than the LO cross section when going to higher values is evident.
To a large extent this is due to the fact that a fixed value is 
used for the renormalization scale. 
Instead, the transverse momentum of the jet seems a more appropriate
scale in the high-$p_{\mathrm{T}}$ tail,
since the only arising strong coupling concerns this jet. 
Using $\als(\MW)$ in the LO calculation
overestimates the contributions for large $p_{\mathrm{T,jet}}$ values
due to the ignored decrease of the QCD coupling. 
In other words, the large
  $p_{\mathrm{T,jet}}$ introduces an additional scale which is
  responsible for a large logarithmic enhancement. These corrections
  may be absorbed into the running of the strong coupling constant
  (and thus resummed to all orders) by
  using a more appropriate renormalization scale, i.e.\ a scale that is
  set through $p_{\mathrm{T,jet}}$.
Similar arguments can explain part of
the behaviour at small $p_{\mathrm{T}}$ 
observed in the respective plots of \reffi{fi:nr:decayNLOdistWWa}, but of course
the theoretical description in regions of small transverse momenta should
eventually be improved by dedicated resummations of higher-order QCD
corrections.

Since the leptonic transverse momenta are connected to $p_{\mathrm{T,jet}}$ by momentum conservation, the same argumentation also holds for these distributions.
The effects are, however, much stronger for the jet compared to the leptons due to the tendency of coloured particles to collinearly 
radiate further QCD partons. 
For the more inclusive NLO observable, this effect is overcompensated by genuine WW+2jets events which are---being actually LO contributions---also influenced by this effect. In the $p_{\mathrm{T}}$ distributions for the two decay leptons, a slight tendency to 
higher $p_{\mathrm{T}}$ can be found for the lepton. That a difference 
between lepton and antilepton
arises at all can be understood from the defined order of the $\PW$ bosons coupling to the fermion chain, because $t$-channel-like emissions of the $\PW$ bosons cause large contributions. 
Whereas a $t$-channel-like emission of a $\PWp$ can only arise in case of an incoming up quark or down antiquark, an analogous emission of a $\PWm$ always stems from an incoming down quark or up antiquark. The antilepton always results from the $\PWp$ decay and the lepton from the $\PWm$ decay. Due to the difference of the involved PDFs, the antilepton prefers lower transverse momenta compared to the lepton.

\begin{figure}
\centering
\includegraphics[bb = 200 520 450 700, scale = .8]{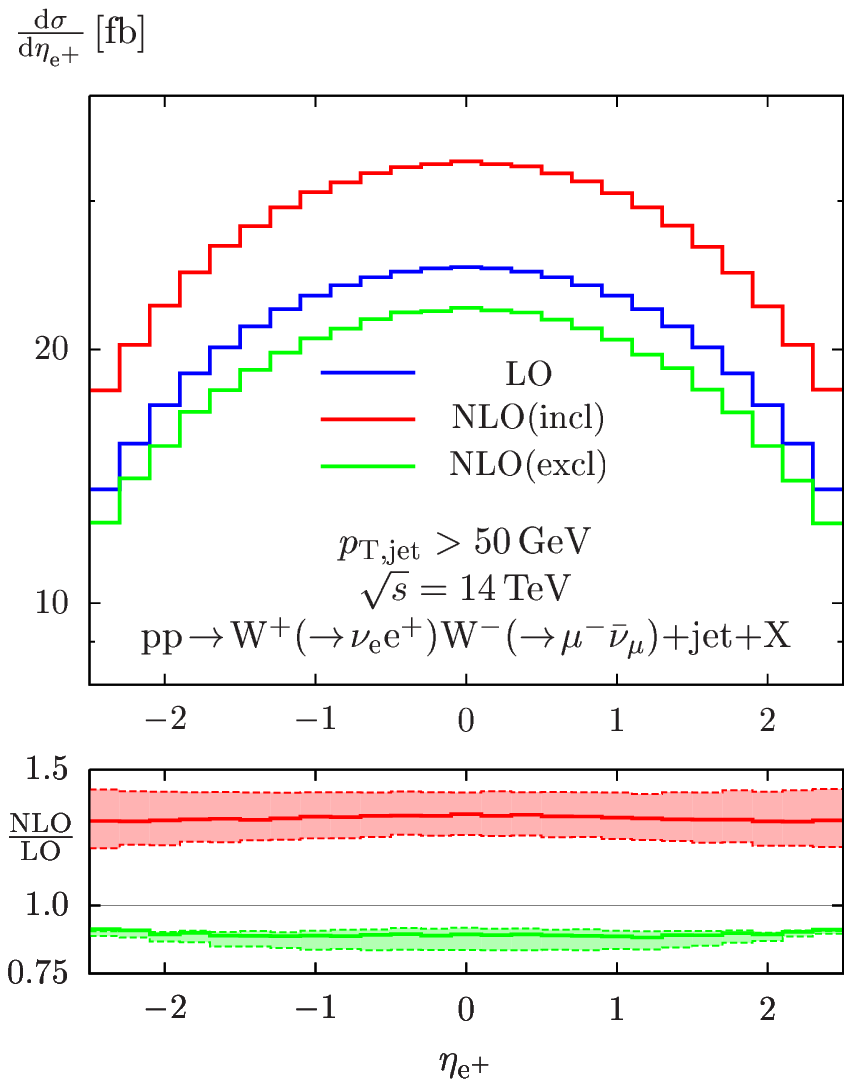}
\includegraphics[bb = 160 520 410 700, scale = .8]{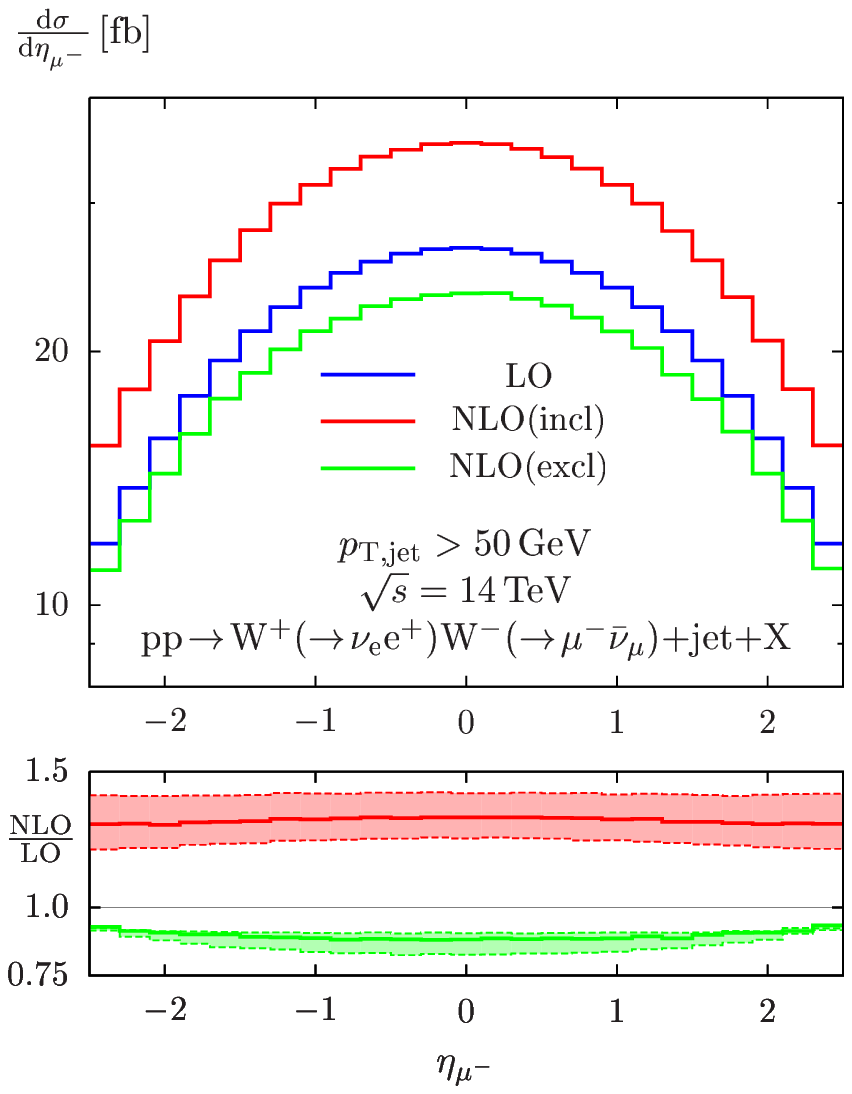}\\[4cm]
\includegraphics[bb = 200 400 450 700, scale = .8]{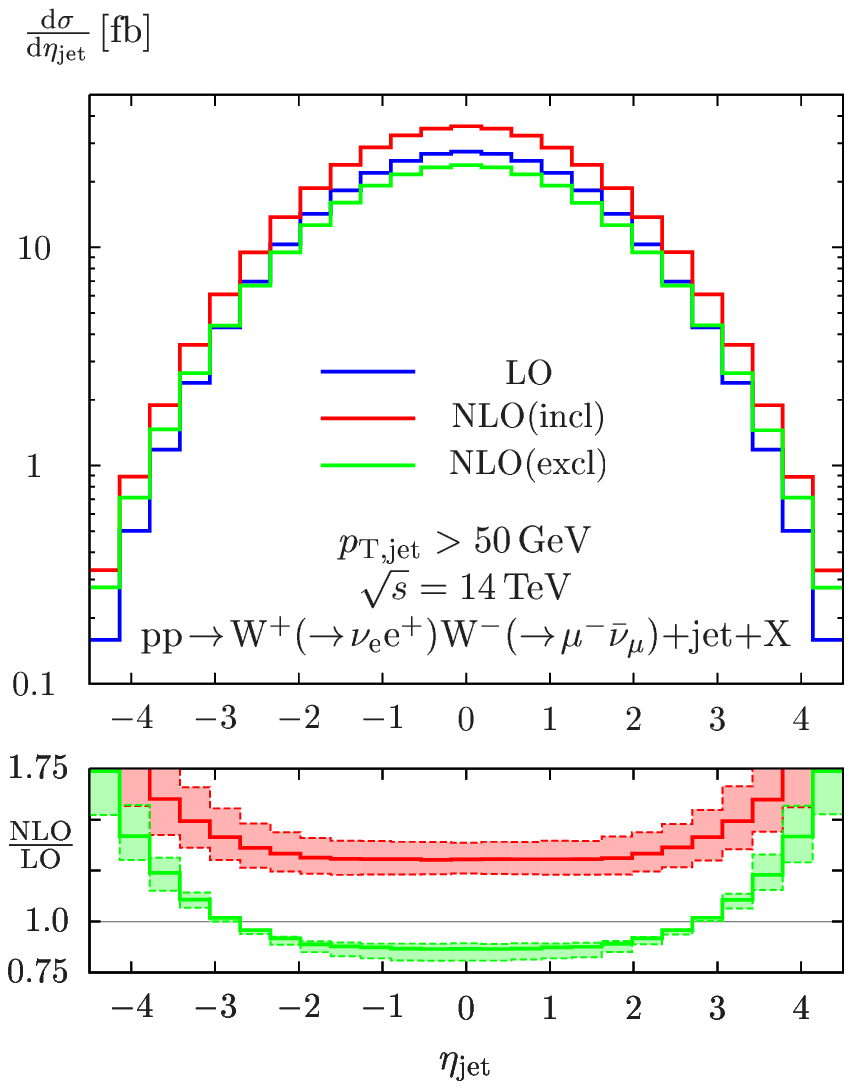}
\includegraphics[bb = 160 400 410 700, scale = .8]{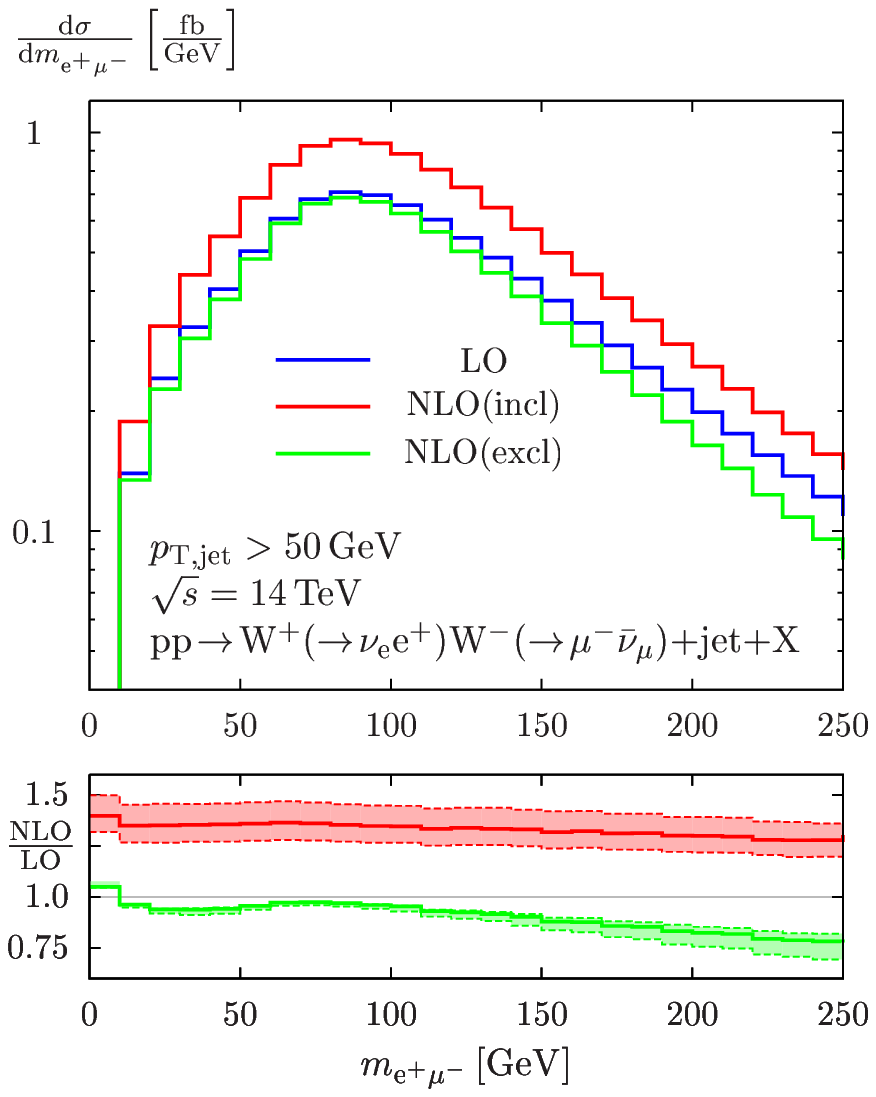}\\
\caption[As in \reffi{fi:nr:decayNLOdistWWa}, but for the pseudo-rapidity $\eta$ of
  the charged decay leptons and of the jet.]{As in
  \reffi{fi:nr:decayNLOdistWWa}, but for the pseudo-rapidity $\eta$ of the charged
  decay leptons (upper plots), the pseudo-rapidity of the jet
    (lower left plot), and the invariant mass of the charged leptons (lower
    right plot).}
\label{fi:nr:decayNLOdistWWb}
\end{figure}

An important impact of
this  effect can also be seen in the upper plots of
\reffi{fi:nr:decayNLOdistWWb} where distributions of the pseudo-rapidities of
the leptons are depicted. Here, the antilepton shows a slightly larger
tendency to small angles against the beam axes than the
lepton. Naturally, due to the symmetric hadronic initial state in
proton--proton collisions the pseudo-rapidity distributions are
symmetric. The size of the relative NLO corrections turns out to be
nearly independent of these quantities---well described by a
  constant $K$-factor. The symmetry property still holds for the
  pseudo-rapidity of the hadronic jet, which is shown in the lower
  plot of \reffi{fi:nr:decayNLOdistWWb}. The NLO corrections to this
  quantity, however, increase for larger 
pseudo-rapidities. These large positive corrections in the region of
large pseudo-rapidities can be understood by a simple statistical effect: NLO corrections in general redistribute events
via radiation.
Since most events show jets in the region of lower pseudo-rapidities, it is just more likely to redistribute events from the low-pseudo-rapidity region to the high-pseudo-rapidity region than vice versa. 
Apart from this small phase-space region, which is negligible anyway if the total amount of events is considered, the NLO corrections turn out to be nearly independent of the jet pseudo-rapidity as well.

The invariant-mass distribution of the charged leptons, which is shown in
  the lower right plot of \reffi{fi:nr:decayNLOdistWWb}, depicts the typical
  behaviour of two particles that are not resonantly
  produced: while the steep increase at low invarint masses is mainly due
  to the lepton-separation cut and the transverse-momentum cuts on the
  individual particles, the decrease to higher values reflects the
  behaviour of all variables involving a new energy scale, according to the 
  dependence of hadronic cross sections on the partonic centre-of-mass
  energy.

\begin{figure}
\centering
\includegraphics[bb = 200 380 450 700, scale = .8]{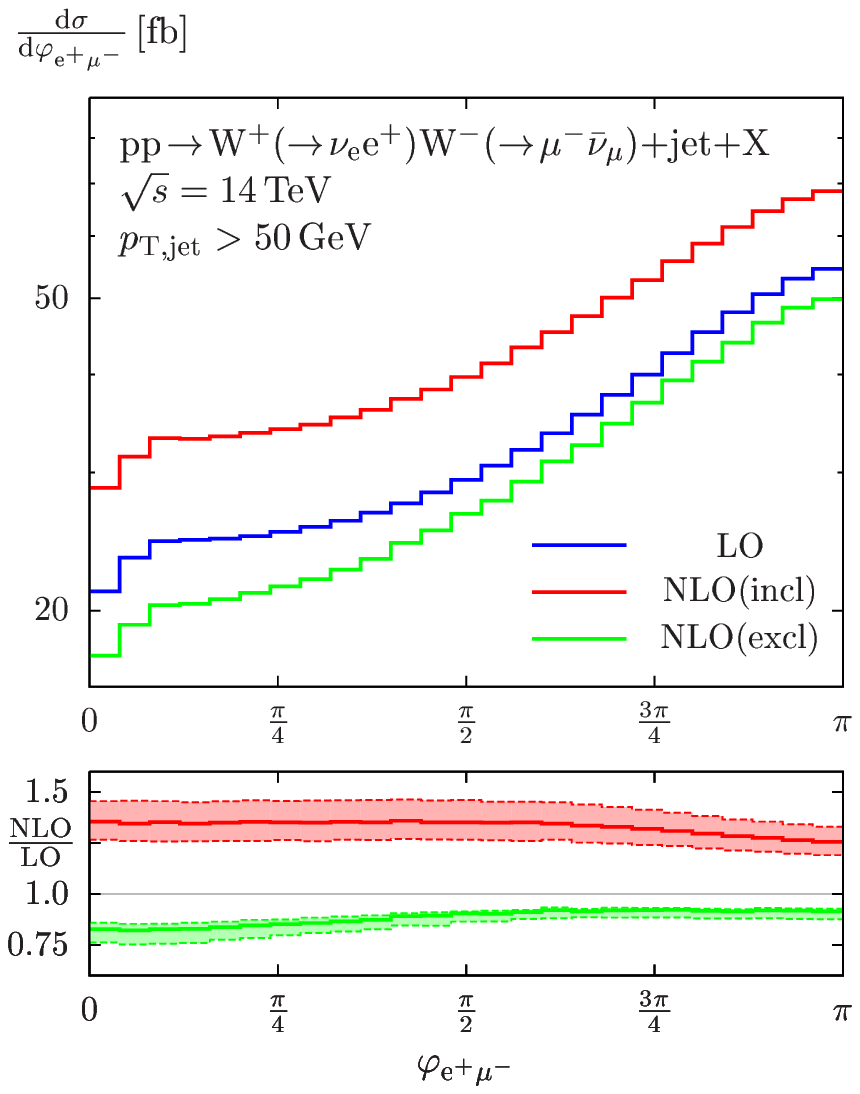}
\includegraphics[bb = 160 380 410 700, scale = .8]{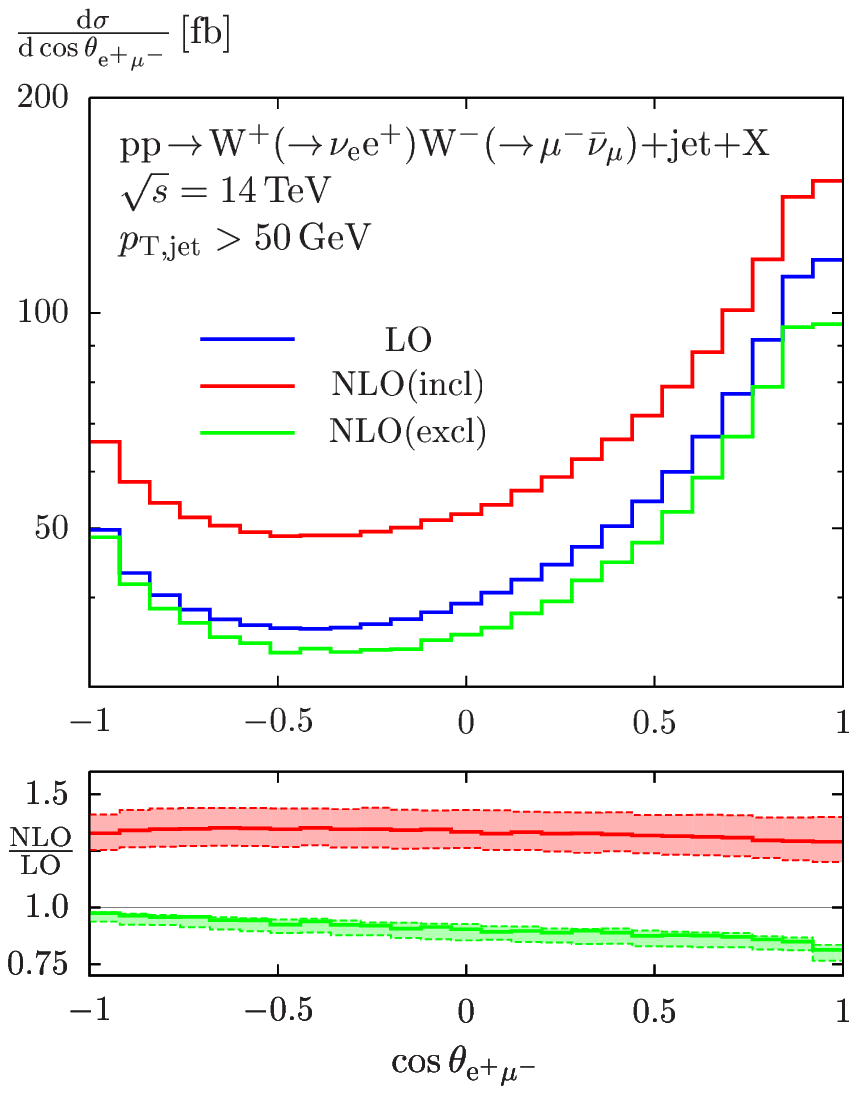}\\[-2em]
\caption{As in \reffi{fi:nr:decayNLOdistWWa}, but for the angle $\varphi$ in the
  transverse plane and the cosine of the opening angle $\theta$ between the two leptons.}
\label{fi:nr:decayNLOdistWWc}
\end{figure}
The two distributions in \reffi{fi:nr:decayNLOdistWWc}
depict the angular
correlations between the two decay leptons. 
The angle between the two leptons in the transverse plane is represented by $\varphi$, which is
a quantity invariant under boosts along the beam direction. By $\theta$ the
opening angle between these two leptons in the laboratory frame is denoted. Considering
the $\varphi$ distribution, i.e.\ ignoring the boost effect along the beam axes, the two charged leptons turn out to fly preferentially into
opposite directions. This is not surprising, since momentum
conservation forces the two $\PW$ bosons at least to show a tendency to
opposite directions which is mediated to their decay products by boost
effects. 
The angle between the leptons is, however, important for the distinction of the
background process WW+jet from the signal process $\PH(\to\PW\PW^\ast)$+jet in 
Higgs searches. 
This is due to the fact that the decay leptons of a $\PW$-boson pair arising from the decay of a
scalar Higgs particle show the---on the first view non-intuitive---tendency to fly
into the same direction. This property results from the spin correlation of the
$\PWp\PWm$ system, which is discussed in detail in \citere{Dittmar:1996sp}:
Since, in the rest frame of the Higgs boson, the 
helicities of the $\PW$ bosons are
correlated, 
and only a left-handed charged lepton and a right-handed charged antilepton
can arise from the 
W~decays, their emission in the same direction is
favoured. This correlation effect is, of course, smeared in
$\PH(\to\PW\PW^\ast)$+jet,  since the Higgs boson is in general boosted, but a remainder of the
effect should still be measurable. 

For WW+jet production, however, the 
helicities
of the two $\PW$ bosons are not correlated in this way,
as depicted in
\reffi{fi:nr:decayNLOdistWWc}, but tend to opposite directions. In the $\cos\theta$ distribution, this
preference is overcompensated by the boost effect along the beam axes, leading to a tendency in direction of small
opening angles. The dependence of the size of NLO corrections on both angles
turns out to be of the order of 10--20\%.

\subsubsection{Differential NLO cross sections at the Tevatron}
\begin{figure}
\centering
\includegraphics[bb = 200 520 450 700, scale = .8]{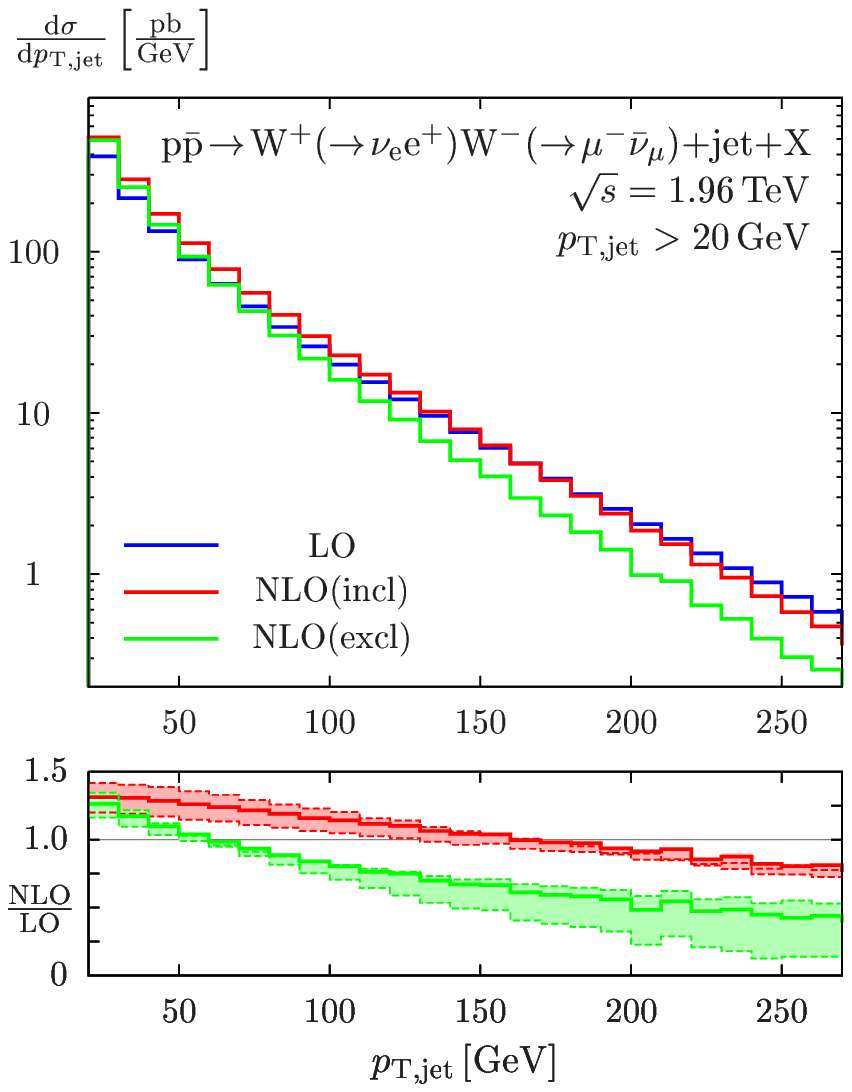}
\includegraphics[bb = 160 520 410 700, scale = .8]{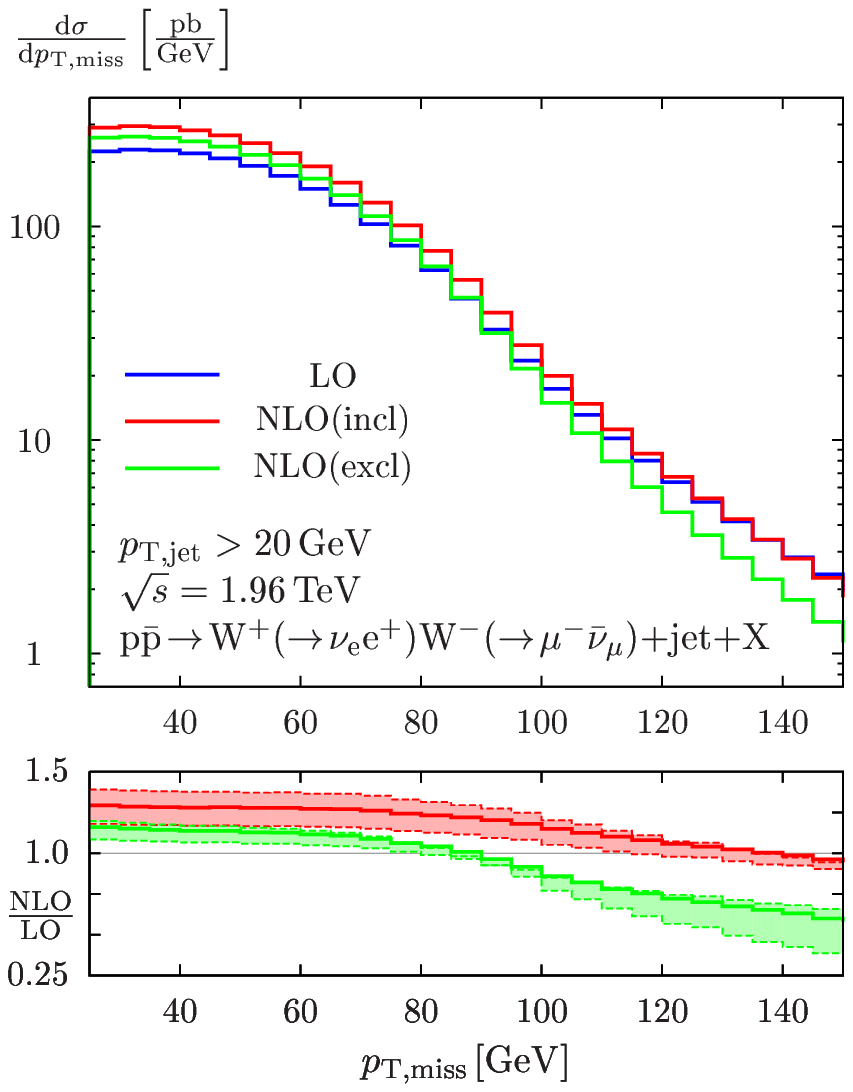}\\[4cm]
\includegraphics[bb = 200 400 450 700, scale = .8]{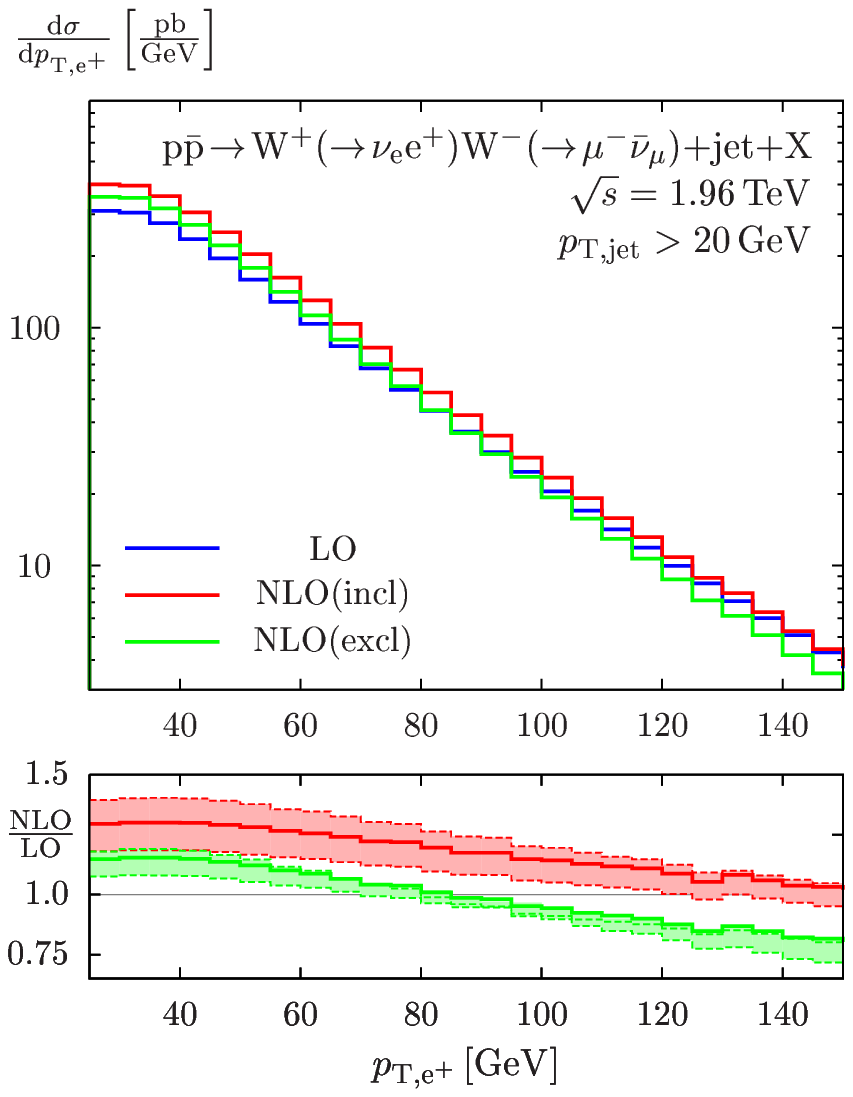}
\includegraphics[bb = 160 400 410 700, scale = .8]{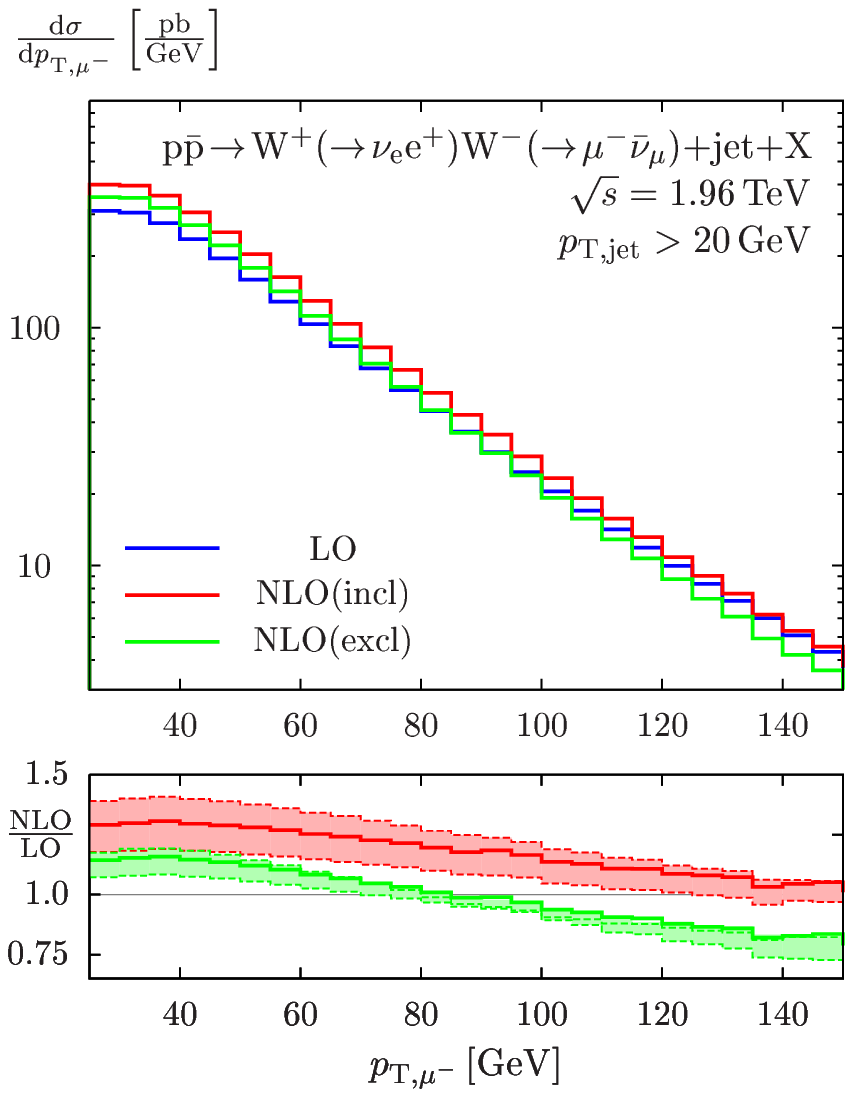}\\
\caption{Differential cross sections for WW+jet with decays included in the
  improved NWA at the Tevatron: The LO and NLO distributions are shown for
  \mbox{$\mu=\mu_{\rm{fact}}=\mu_{\rm{ren}}=\MW$}. The distributions for the
  transverse momenta $p_{\mathrm{T}}$  of the jet and of the decay leptons,
  and for the missing transverse momentum $p_{\mathrm{T,miss}}$ are
  depicted. The bands in the $K$-factors are defined as in the LHC case.}
\label{fi:nr:decayNLOdistWWaTev}
\end{figure}

Considering the same quantities for WW+jet production in the Tevatron setup,
most of the effects can be explained by the fact that proton--antiproton
collisions take place here instead of proton--proton collisions. Besides, the
lower CM energy compared to the LHC plays an important role. We start again
with the discussion of transverse-momentum distributions of the hard jet
and the decay leptons, and 
of the missing transverse momentum, which are shown in
\reffi{fi:nr:decayNLOdistWWaTev}. 
Here, the behaviour of the more exclusive
NLO observables can in principle be explained in the same way as for LHC: the
LO results overestimate the cross section at high scales and underestimate it
at low scales due to the fixed renormalization scale used in the
calculation. As already observed when considering integrated cross sections,
the difference between the two NLO observables is quite small at the Tevatron,
which can be understood from the smaller CM energy: In most cases, not enough
energy is available for the production of a second hard jet. Therefore, the LO
contributions of \mbox{\tppbWWjj} contained in the more inclusive NLO
observable only weaken the effect of negative corrections at large
$p_{\mathrm{T}}$, but do not overcompensate it as for the LHC
setup. The $p_{\mathrm{T}}$ distributions of the two decay leptons are
identical up to numerical fluctuations. This results from the fact
that the hadronic process \tppbWWj\ and the underlying model
(SM without $\mathcal{CP}$-violating phases) are 
invariant under $\mathcal{CP}$ transformation.
Having not introduced any such  $\mathcal{CP}$ violating contributions, we thus expect that the final 
state is also even under $\mathcal{CP}$ transformation.
Thus, in contrast to proton--proton collisions, PDF effects do not cause differences between the $p_{\mathrm{T}}$ distributions of the leptons, since the quark PDFs of the proton equal the antiquark PDFs of the 
antiproton. 

\begin{figure}
\centering
\includegraphics[bb = 200 520 450 700, scale = .8]{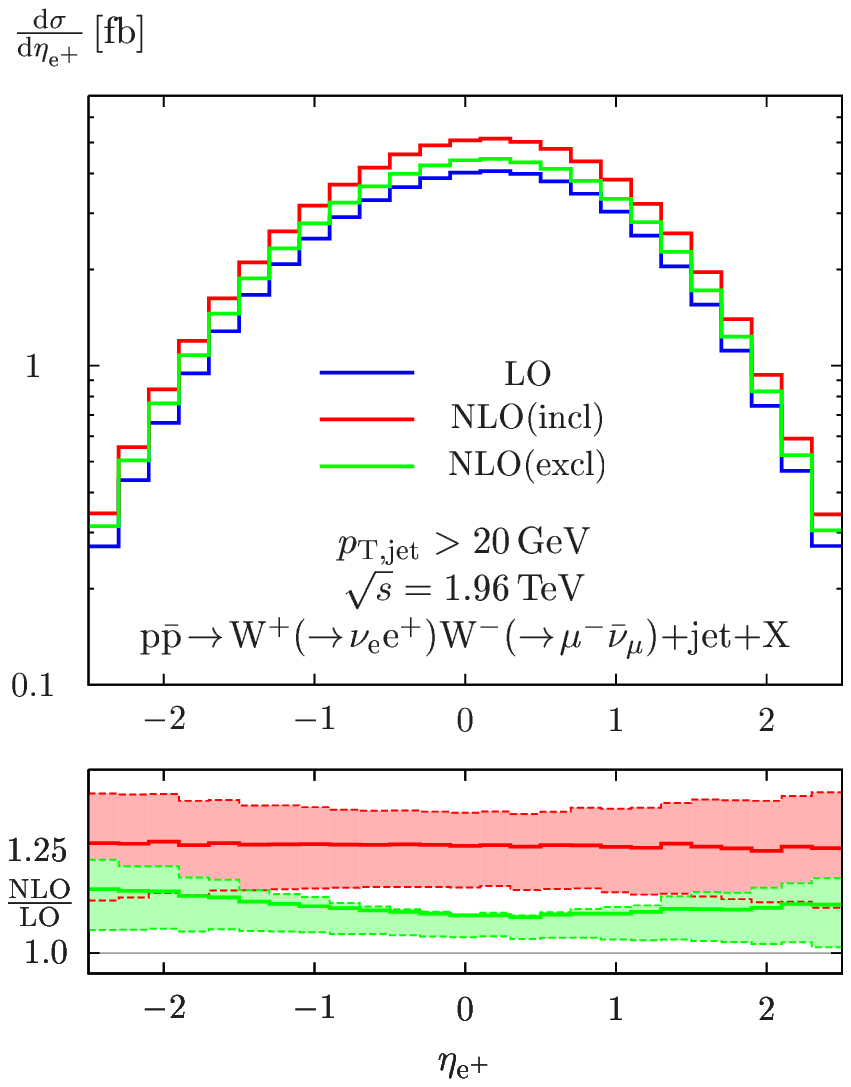}
\includegraphics[bb = 160 520 410 700, scale = .8]{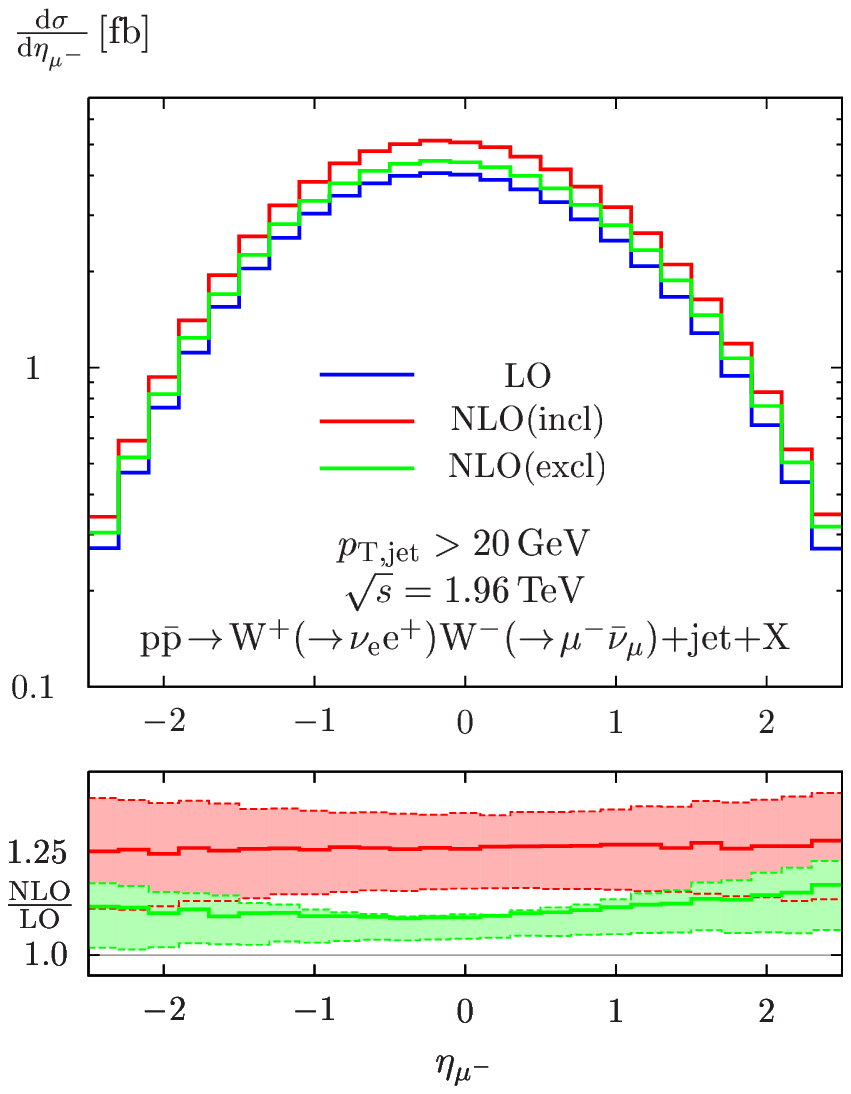}\\[4cm]
\includegraphics[bb = 200 400 450 700, scale = .8]{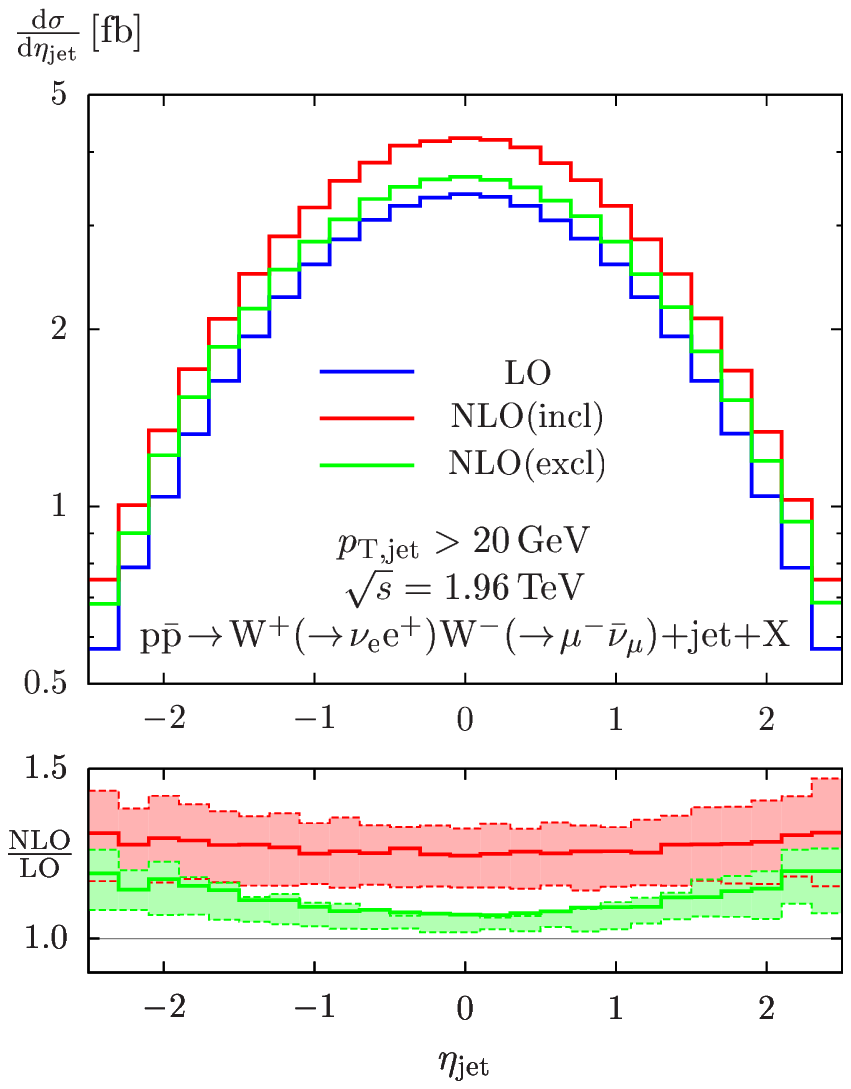}
\includegraphics[bb = 160 400 410 700, scale = .8]{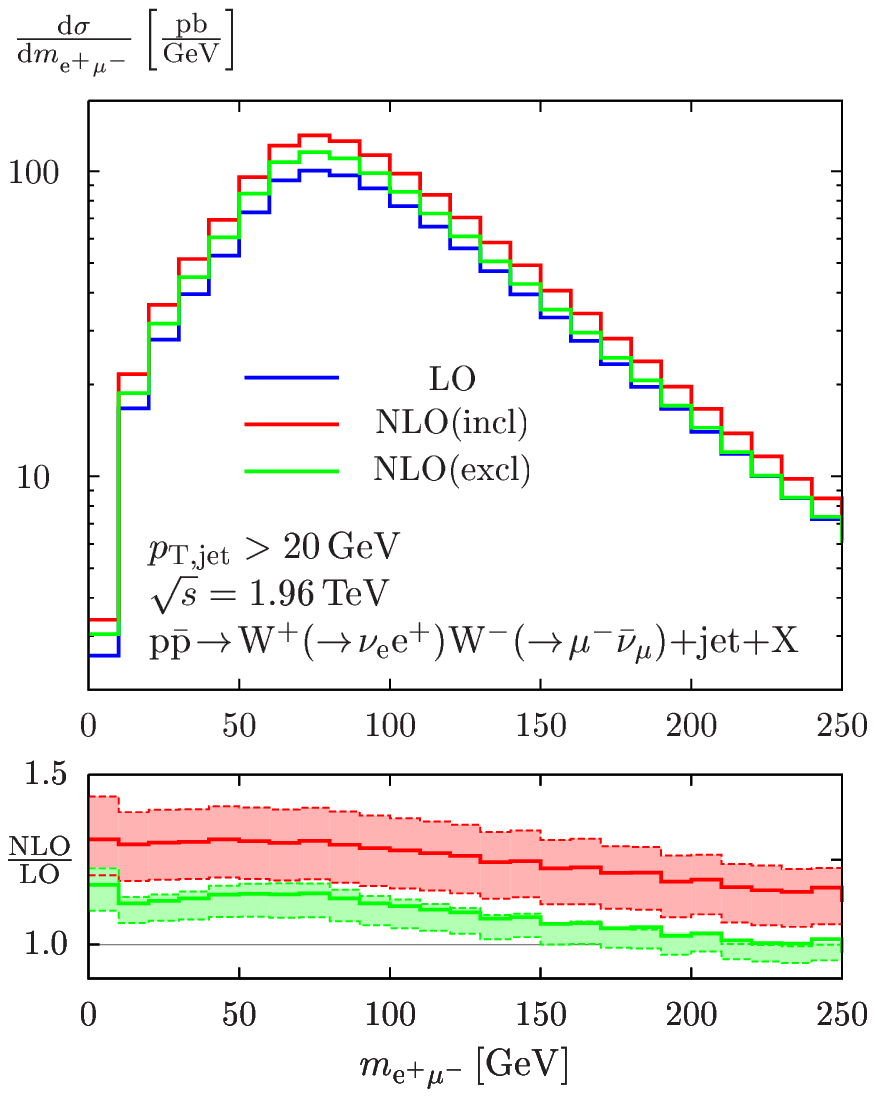}\\
\caption{As in \reffi{fi:nr:decayNLOdistWWaTev}, but for the pseudo-rapidity $\eta$ of the charged decay leptons (upper plots), the pseudo-rapidity of the jet
    (lower left plot), and the invariant mass of the charged leptons (lower
    right plot).}
\label{fi:nr:decayNLOdistWWbTev}
\end{figure}

The distributions of the pseudo-rapidities $\eta$ of the leptons,
which are depicted in \reffi{fi:nr:decayNLOdistWWbTev}, are
not symmetric with respect to \mbox{$\eta=0$} due to the
asymmetric hadronic initial state. 
Instead, the distributions of the lepton and
the antilepton are identical if one of the distributions is mirrored around
the \mbox{$\eta=0$} axes. In the Tevatron setup, the positive beam axes
corresponds to the direction of the proton beam, and the negative axes to the
antiproton beam. Correspondingly, positive
$\eta$ values describe momenta tending in the direction of the proton beam
and vice versa. 

\begin{figure}
\centering
\includegraphics[bb = 200 550 450 750, scale = .8]{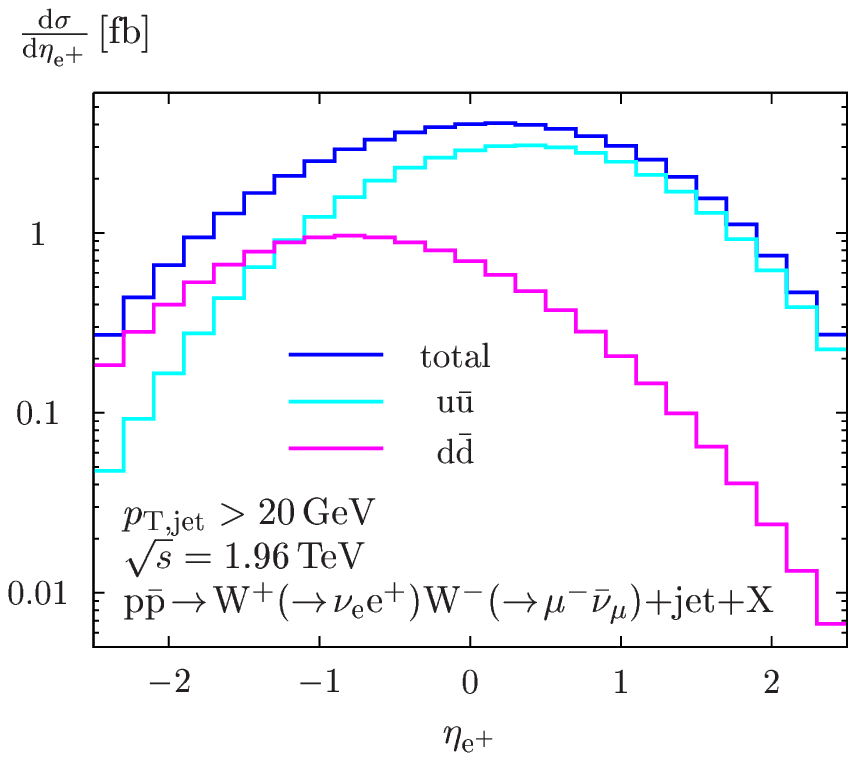}
\includegraphics[bb = 160 550 410 750, scale = .8]{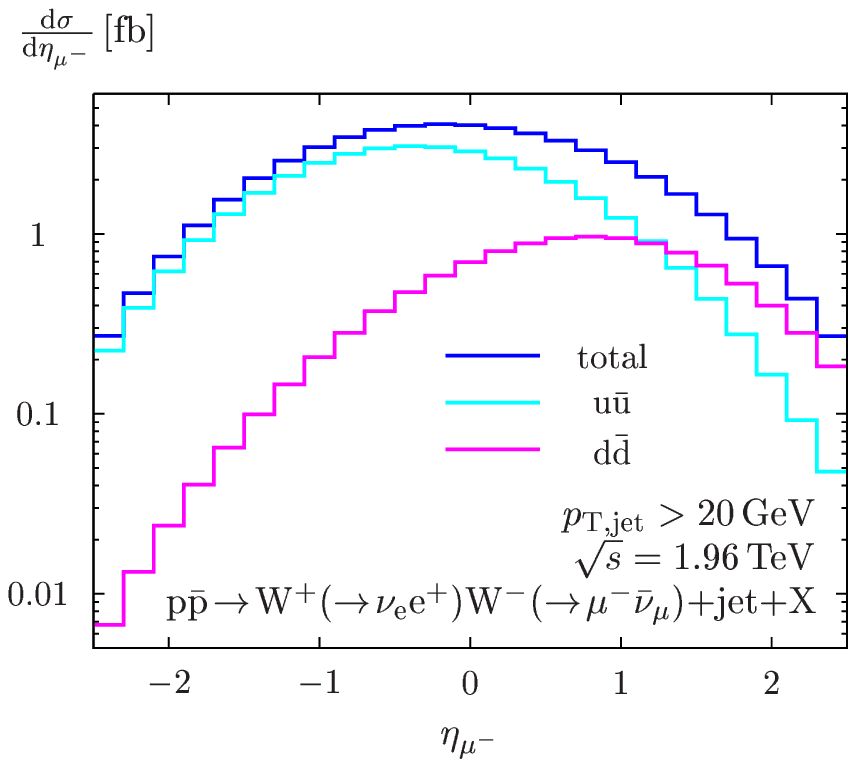}\\
\caption{As in \reffi{fi:nr:decayNLOdistWWbTev}, but only the LO distributions for the pseudo-rapidities $\eta$ of the lepton and the antilepton are depicted. In   addition, the two dominating partonic contributions $\Pu\bPu$ and $\Pd\bPd$ to the hadronic cross section are shown.}
\label{fi:nr:decayNLOdistWWTevextra}
\end{figure}

Considering the non-symmetric pseudo-rapidity distributions, a tendency of the
antilepton to the positive beam direction and, correspondingly, of the lepton
to the negative beam direction is evident. To explain this, the argument of
$t$-channel-like emission of $\PW$ bosons can again be applied. As discussed
in the LHC case, the $\PWp$ boson, and consequently the antilepton, can
only be emitted in this way from an up quark or a down antiquark. 
Since the $\Pq\bPq$
channels dominate at the Tevatron, only these have to be taken into account for
a qualitative discussion. On the average, these subprocesses do not produce
boosted events in a distinguished direction due to the fact that the quark PDFs in the proton are
equal to the respective antiquark PDFs in the antiproton. 
The contributions $\Pu_\Pp(p_1)\bPu_{\bPp}(p_2)$ and $\Pd_\Pp(p_1)\bPd_{\bPp}(p_2)$
dominate over the contributions $\bPu_\Pp(p_1)\Pu_{\bPp}(p_2)$ and $\bPd_\Pp(p_1)\Pd_{\bPp}(p_2)$.
Due to the effect of $t$-channel-like $\PWp$ emission, the first tends to
antilepton emission in the positive beam direction, the latter to the negative
direction, as depicted in \reffi{fi:nr:decayNLOdistWWTevextra}. 
Since the $\Pu\bPu$ channel exceeds the $\Pd\bPd$ channel roughly by a factor
4 in total, a tendency to the positive beam direction results.
The same argumentation holds, mutatis mutandis, for the lepton and the $\PWm$
boson.

The pseudo-rapidity of the jet, which is depicted in the lower left plot of
\reffi{fi:nr:decayNLOdistWWbTev}, is symmetric with respect to $\eta=0$. This
is due to the fact that no distinction can be made between hadronic jets
arising from gluons, quarks, or antiquarks. Therefore, the sum over all
contributions, which are $\mathcal{CP}$ symmetric in pairs, yields a symmetric
distribution. As in the case of LHC, the dependence of the NLO corrections 
on the pseudo-rapidities of the jet and the leptons is moderate. 
Considering the absolute pseudo-rapidity dependence of the cross sections, a
tendency to events that are not strongly boosted in the direction of the beam
axes is observed.  This can be understood from the fact that the dominating
partonic channels with initial states of valence quark and valence
antiquark are not strongly boosted in general. 

The  behaviour of the invariant-mass distribution of the two charged
  leptons, which is depicted in the lower right plot of
  \reffi{fi:nr:decayNLOdistWWbTev}, can be explained analogously to the LHC
  case.

\begin{figure}
\centering
\includegraphics[bb = 200 380 450 700, scale = .8]{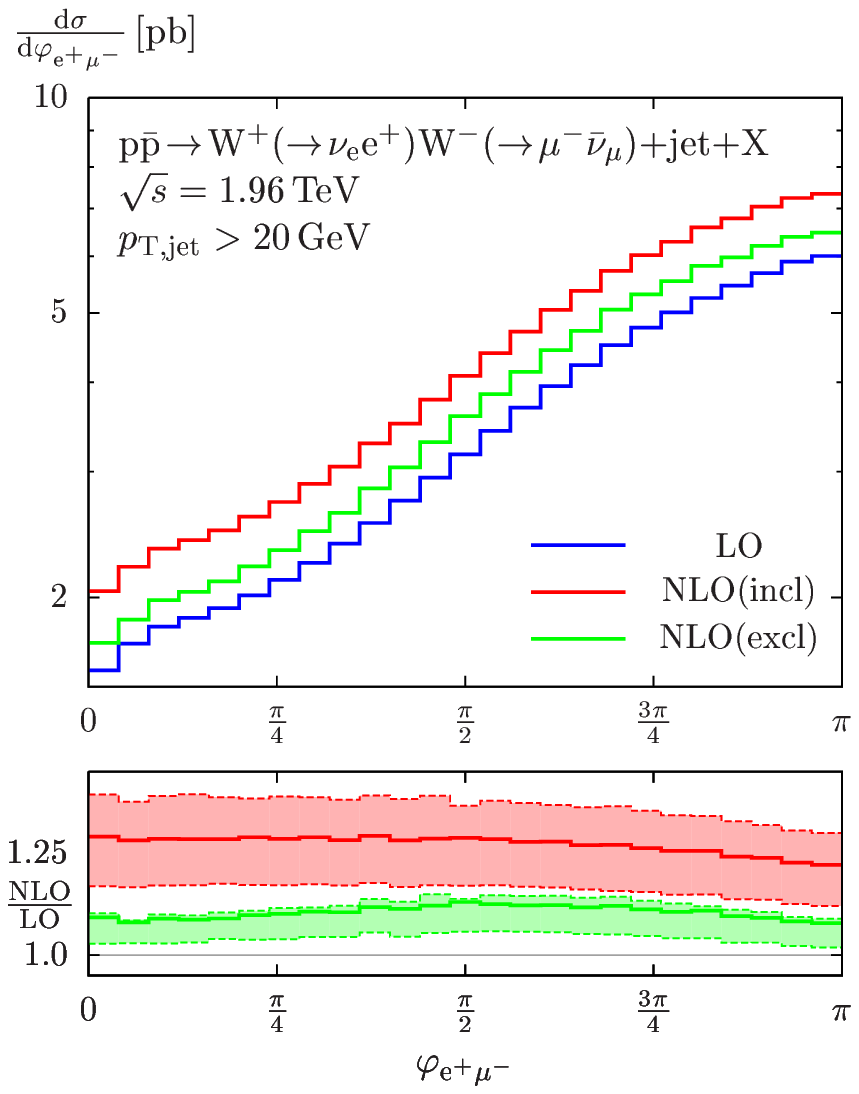}
\includegraphics[bb = 160 380 410 700, scale = .8]{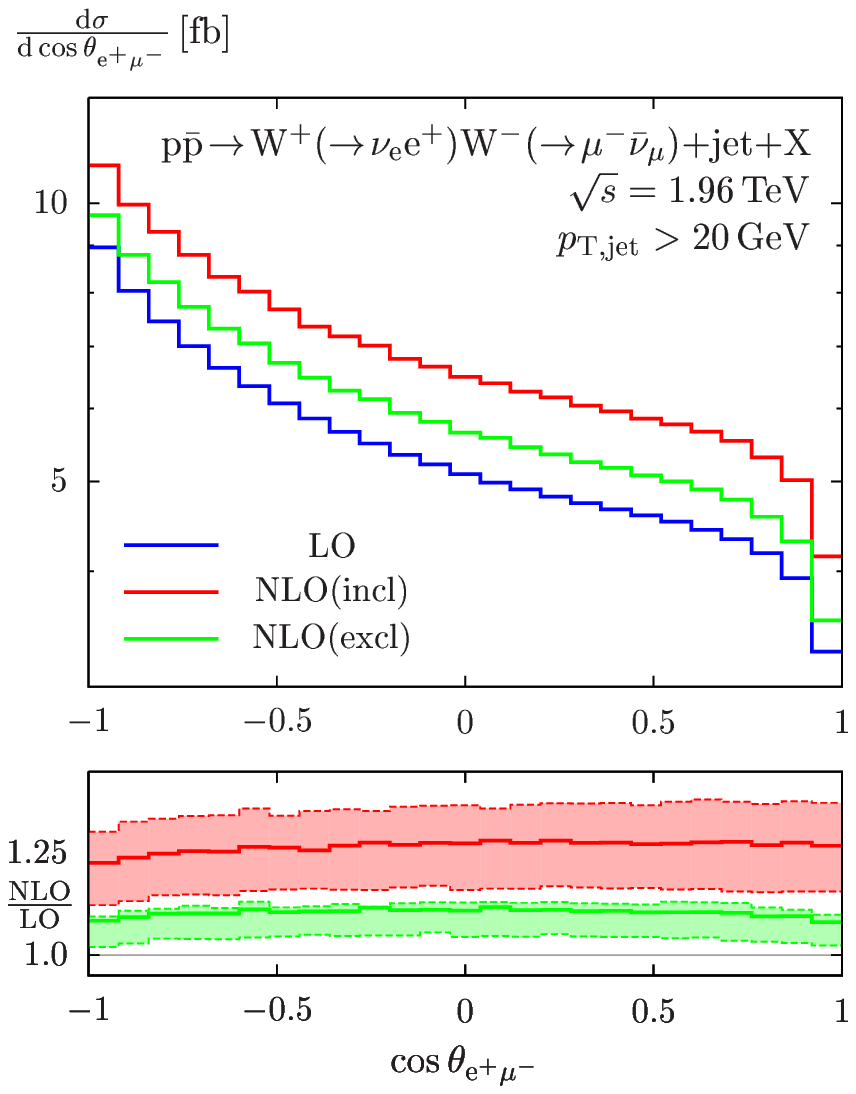}\\[-2em]
\caption{As in \reffi{fi:nr:decayNLOdistWWaTev}, but for the angle $\varphi$ in the
  transverse plane and the cosine of the opening angle $\theta$ between the two leptons.}
\label{fi:nr:decayNLOdistWWcTev}
\end{figure}

Finally, the angle correlations between the two decay leptons are
considered. The angle in the transverse plane is again labelled by $\varphi$, and $\theta$
is the opening angle between the two leptons. The corresponding distributions are depicted in
\reffi{fi:nr:decayNLOdistWWcTev}.
These angle correlations are very important also at Tevatron to
distinguish the background process WW+jet from the signal process
$\PH(\to\PW\PW^\ast)$+jet. Details on this topic are given in the discussion
of the respective LHC distributions.
Also in the Tevatron setup, the two leptons tend to fly into opposite directions in the transverse
plane. In contrast to the situation at the LHC, this tendency is still observed 
in the $\cos\theta$ distribution of the opening angle $\theta$, i.e.\
the effect is not overcompensated by boost effects.
This is again understood from the fact that no tendency to
strongly boosted events arises in the dominant partonic channels at the Tevatron.

\subsection{Benchmark results and comparison to results of other groups}

Independently of our calculation described in \citere{Dittmaier:2007th}, 
two further
groups have performed NLO QCD calculations on WW+jet production. The results
of one of these groups are given in \citere{Campbell:2007ev}, while those of
the second group are not published yet~\cite{Sanguinetti:2008xt}. 
In \citere{Bern:2008ef}, a tuned
comparison of the three calculations is provided for the integrated LO cross
section, which agree again within statistical errors, and of the virtual 
corrections at one specific phase-space point.
There, the renormalized 
matrix elements are subdivided into bosonic and fermionic corrections 
with only the two light generations included in closed quark loops.
The results are given in terms 
of the coefficients to the poles $\frac{1}{\epsilon^2}$ and 
$\frac{1}{\epsilon}$, and the finite part, i.e.\ in a way that is 
independent of the specific method for cancelling
the infrared singularities.%
\footnote{Unfortunately an important detail was not explicitly stated
in \citere{Campbell:2007ev}: In the dipole subtraction method the
IR divergences are usually factorized from $D$-dimenional LO
structures (``conventional dimensional regularization'').
In the comparison shown in \citere{Campbell:2007ev}, 
however, absolute numbers on the corrections to the squared matrix elements
are given where the $D$-dimenional LO structures are evaluated in four 
dimensions (without expanding around $D=4$). This procedure is legitimate,
because the same replacement of $D$-dimensional by four-dimensional LO
structures is done in the IR-divergent part of the real corrections.
The numbers given in \citere{Campbell:2007ev} are, thus, equivalent
to results in 
the ``'t~Hooft--Veltman scheme'' for dimensional regularization where
only particles in loops or in singular splittings are extended to $D$
dimensions.
}
A comparison of results for the third-generation loops has not been 
performed so far. Agreement is achieved between the results of the three
groups at an accuracy level comparable to our internal checks.

\section{Conclusions}
\label{se:concl}

The production of W-boson pairs in association with a hard jet
is an important source for background to Higgs
and new-physics searches both at the Tevatron and at the LHC.
A proper theoretical prediction for this process requires at least 
the inclusion of perturbative QCD corrections at the 
next-to-leading order. 

Continuing previous work, we have described
in detail the calculation of these corrections,
which takes into account leptonic decays of the W~bosons. 
To this end, we employ an improved
narrow-width approximation that treats the $\PW$ bosons
as on-shell particles, but keeps the $\PW$-spin information.
While the naive narrow-width approximation, which neglects the information
on the W~spins, is only good within 5--10\% in differential distributions,
the improved version roughly reaches percent accuracy. This is the result of
a comparison of leading-order predictions in these two approximations 
with results of a calculation fully based on off-shell W~bosons.

The treatment of bottom quarks in the initial or final states deserves
particular attention. Numerically such contributions only play a
significant role if the top-quark propagator present in this
  case becomes resonant. The contributions are thus essentially the
  off-shell continuations of $\PWp\bPt$, $\PWm\Pt$, or $\Pt\bPt$ production
  including the subsequent decay of the top-quarks.
These subprocesses,
however, should not be counted as part of WW+jet production. Therefore,
we have excluded contributions from external bottom quarks. 
The reliability of our procedure has been verified in a comparison
of next-to-leading-order results obtained in two different schemes,
where one is based on four, another on five active quark flavours
in the proton.

Our detailed discussion of numerical results shows that the
QCD corrections stabilize the leading-order prediction for
the WW+jet cross section considerably with respect to a
  variation of the factorization and renormalization scales which we
  identify with each other. At the LHC, this stabilization
of the prediction, however, requires a veto on a second hard jet.
Otherwise the production of final states with WW+2jets, which 
yields a leading-order component of the next-to-leading-order
correction, introduces again a large scale dependence.
As far as the differential distributions are concerned the
  corrections are typically of the order of about 25\%. For a
  remarkable number of distributions the $K$-factor is only mildly
  dependent on the kinematical region. At the LHC the $\eta$
  distributions in the dominant region and also the
  distributions in the angles between the two charged leptons
  have an almost constant
  $K$-factor of about 1.3 (inclusive cross-section definition);
  for the exclusive cross-section definition the corrections are even smaller
  and rather close to 1. The $p_{\mathrm{T}}$ spectra, on
  the other hand, show a much more phase-space-dependent $K$-factor
  with the exclusive cross-section definition showing an even larger
  dependence than the inclusive one. This is not surprising since the
  $p_{\mathrm{T}}$ introduces an additional scale which could introduce
  potentially large logarithms which are badly treated by a constant
  renormalization scale.
  At the Tevatron our findings are similar.  Again the $\eta$ and angular
  distributions receive corrections
  in form of an almost constant $K$-factor of about 1.3 (exclusive
  definition).
  The
  corrections for the exclusive cross-section definition are again
  smaller than for the inclusive definition. In case of the $p_{\mathrm{T}}$
  spectra we observe again a phase-space-dependent $K$-factor.
  We note that the almost constant $K$-factor which holds for a
  remarkable number of distributions has also been observed by
  Campbell, Ellis and Zanderighi~\cite{Campbell:2007ev}.

The QCD corrections to the related processes of
ZZ+jet and WZ+jet production can be obtained in an analogous way
as presented here for WW+jet. The corresponding calculations are
in progress.

\section*{Acknowledgements}
This work is supported in part by the European Community's Marie-Curie
Research Training Network under contract MRTN-CT-2006-035505 ``Tools
and Precision Calculations for Physics Discoveries at Colliders''
and by Deutsche Forschungsgemeinschaft (DFG) through 
SFB-TR~9 ``Computational Theoretical Particle Physics''.
P.U. acknowledges the support of the Initiative and
Networking Fund of the Helmholtz Association, contract HA-101
(``Physics at the Terascale'').

\appendix
\section*{Appendix}
\section{Decaying on-shell gauge bosons}
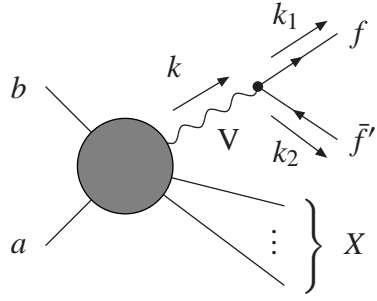
\begin{figure}
\centerline{
\begin{picture}(150,115)
\Line(20,20)(50,50)
\Line(20,80)(50,50)
\Photon(50,50)(100,80){2}{5}
\ArrowLine(100,80)(130,100)
\ArrowLine(130,60)(100,80)
\Vertex(100,80){2}
\Line(50,50)(110,35)
\Line(50,50)(110, 5)
\GCirc(50,50){18}{.5}
\Text( 7,20)[l]{$a$}
\Text( 7,80)[l]{$b$}
\LongArrow(105,91)(125,104)
\LongArrow(105,69)(125, 54)
\LongArrow(68,71)(88,83)
\Text( 66,88)[l]{$k$}
\Text(106,107)[l]{$k_1$}
\Text(106,55)[l]{$k_2$}
\Text( 85,60)[l]{$\PV$}
\Text(135,100)[l]{$f$}
\Text(135,60)[l]{$\bar f'$}
\Text(105,20)[l]{$\vdots \hspace{.5em} 
  \left.\vphantom{\rule{1em}{1.8em}}\right\}\;\;X$}
\end{picture}
}
\caption{Schematic diagram for the process $ab\to \PV X\to f\bar f' X$.}
\label{fig:Vdecay}
\end{figure}

\renewcommand{\theequation}{A.\arabic{equation}}
\label{se:onshelldecays}
In this appendix we describe a simple method to include 
gauge-boson decays with the correct spin correlations
in processes for which the helicity
amplitudes for on-shell gauge bosons are already known.
Specifically we consider the situation illustrated in
\reffi{fig:Vdecay} where the vector boson $\PV$ with momentum $k$ decays into
the pair $f(k_1) \bar f'(k_2)$ of massless fermions with momenta
$k_{1,2}$.
We start with the expression for the cross section 
\beq
\sigma_{ab\to \PV X\to f\bar f' X} = 
\frac{1}{2s_{ab}} \int\rd\Phi_{f\bar f' X} \,
\left|\M_{ab\to \PV X\to f\bar f' X}\right|^2
\label{eq:sigmaab1}
\eeq
for the full reaction $ab\to \PV X\to f\bar f' X$ integrated over its
phase space $\Phi_{f\bar f' X}$ at the CM energy
$\sqrt{s_{ab}}$. Of course, not only diagrams with a $\PV$ resonance,
as shown in \reffi{fig:Vdecay}, contribute to the full amplitude
$\M_{ab\to \PV X\to f\bar f' X}$. We are, however, interested in the
resonant diagrams which factorize according to
\beq
\M^{(\mathrm{res})}_{ab\to \PV X\to f\bar f' X} =
T_\mu(k) \, \frac{-1}{k^2-\MV^2+\ri\MV\GV} \, j^\mu(k_1,k_2),
\eeq
where $T_\mu(k)$ describes the production of an off-shell $\PV$
and $j^\mu(k_1,k_2)$ is the current of the $\PV$ decay.
For small decay widths $\GV$ of $\PV$ the phase-space integral
\refeq{eq:sigmaab1} is dominated by the resonant diagrams with 
momenta near the mass shell of $\PV$. In order to extract the resonant
terms, we factorize the phase space into $\PV$ production and decay
as follows,
\beqar
\lefteqn{\int\rd\Phi_{f\bar f' X} \, 
\frac{1}{|k^2-\MV^2+\ri\MV\GV|^2}
\cdots}
\nn\\
&=& \int\frac{\rd k^2}{2\pi} \,
\frac{1}{|k^2-\MV^2+\ri\MV\GV|^2}
\int\rd\Phi_{\PV X}(k) \,
\int\rd\Phi_{f\bar f'}(k_1,k_2) \, \cdots
\nn\\
&\; \asymp{\GV\to0} \;&
\int\frac{\rd k^2}{2\pi} \,
\frac{\pi}{\MV\GV} \, \de(k^2-\MV^2)
\int\rd\Phi_{\PV X}(k) \,
\int\rd\Phi_{f\bar f'}(k_1,k_2) \, \cdots
\nn\\
&=&
\frac{1}{2\MV\GV} 
\int\rd\Phi_{\PV X}(\hat k) \,
\int\rd\Phi_{f\bar f'}(\hat k_1,\hat k_2) \, \cdots,
\eeqar
where the hats over the momenta indicate that the on-shell condition
$\hat k^2=(\hat k_1+\hat k_2)^2=\MV^2$ is fulfilled.
On resonance we can decompose the contraction $T_\mu j^\mu$
into helicity amplitudes 
$\M_{ab\to \PV X}$ and $\M_{\PV\to f\bar f'}$ for the $\PV$ production and decay
upon inserting the completeness relation for the $\PV$ polarization
vectors $\veps^\mu(\la)$,
\beqar
-T_\mu(\hat k) j^\mu(\hat k_1,\hat k_2) &=&
T_\mu(\hat k) \left(\sum_{\la=0,\pm1} \veps_\PV^\mu(\la)^* \veps_\PV^\nu(\la)
-\frac{\hat k^\mu \hat k^\nu}{\MV^2} \right) j_\nu(\hat k_1,\hat k_2) 
\nn\\
&=&
\sum_{\la=0,\pm1} \M_{ab\to \PV X}(\la) \, \M_{\PV\to f\bar f'}(\la)
\eeqar
where we have used current conservation 
$\hat k^\nu j_\nu(\hat k_1,\hat k_2)=0$ (for massless fermions)
and identified
\beq
\M_{ab\to \PV X}(\la) = T_\mu(\hat k) \veps_\PV^\mu(\la)^*, \qquad
\M_{\PV\to f\bar f'}(\la) = \veps_\PV^\nu(\la) j_\nu(\hat k_1,\hat k_2).
\eeq
In summary, the cross section in resonance approximation reads
\beqar
\sigma_{ab\to \PV X\to f\bar f' X} 
&\; \asymp{\GV\to0} \;&
\frac{1}{2s_{ab}} \int\rd\Phi_{\PV X}(\hat k) \,
\sum_{\la,\la'=0,\pm1} \M_{ab\to \PV X}(\la')^* \M_{ab\to \PV X}(\la) \, 
\nn\\
&& {} \times
\frac{1}{2\MV\GV} 
\int\rd\Phi_{f\bar f'}(\hat k_1,\hat k_2) \, 
\M_{\PV\to f\bar f'}(\la')^* \M_{\PV\to f\bar f'}(\la).
\label{eq:sigmaab2}
\eeqar

If the phase space of the decay fermions is integrated over
completely, rotational invariance implies that 
\beq
\frac{1}{2\MV} 
\int\rd\Phi_{f\bar f'}(\hat k_1,\hat k_2) \, 
\M_{\PV\to f\bar f'}(\la')^* \M_{\PV\to f\bar f'}(\la)
= \Ga_{\PV\to f\bar f'} \, \de_{\la'\la},
\eeq
where $\Ga_{\PV\to f\bar f'}$ is the partial decay width.
Inserting this result into \refeq{eq:sigmaab2}, yields the usual
narrow-width approximation for the cross section,
\beqar
\sigma_{ab\to \PV X\to f\bar f' X} 
&\; \asymp{\GV\to0} \;&
\frac{1}{2s_{ab}} \int\rd\Phi_{\PV X}(\hat k) \,
\sum_{\la=0,\pm1} \left|\M_{ab\to \PV X}(\la)\right|^2 \, 
\mathrm{BR}_{\PV\to f\bar f'}
\nn\\
&=& \sigma_{ab\to \PV X}\, \mathrm{BR}_{\PV\to f\bar f'},
\eeqar
with $\sigma_{ab\to \PV X}$ denoting the production cross section
for unpolarized vector bosons $\PV$ and
$\mathrm{BR}_{\PV\to f\bar f'}=\Ga_{\PV\to f\bar f'}/\GV$
denoting the branching ratio for the decay of $\PV$ into the $f\bar f'$
pair.

\begin{sloppypar}
If the decay fermions are not integrated over the full phase space
$\Phi_{f\bar f'}$, i.e.\ if cuts on the fermions are applied or if
distributions in the fermion kinematics are considered, it is
convenient to introduce the ``decay correlation matrix''
\beq
\De^{\PV\to f\bar f'}_{\la'\la}(\hat k_1,\hat k_2) =
\frac{\Phi_{f\bar f'}}{2\MV\Ga_{\PV\to f\bar f'}} \,
\M_{\PV\to f\bar f'}(\la')^* \M_{\PV\to f\bar f'}(\la),
\label{eq:decaycorrelmat}
\eeq
so that the full cross section takes the form
\beqar
\sigma_{ab\to \PV X\to f\bar f' X} 
&\; \asymp{\GV\to0} \;&
\frac{1}{2s_{ab}} \int\rd\Phi_{\PV X}(\hat k) \,
\sum_{\la,\la'=0,\pm1} \M_{ab\to \PV X}(\la')^* \M_{ab\to \PV X}(\la) \, 
\nn\\
&& {} \times
\mathrm{BR}_{\PV\to f\bar f'}
\int\frac{\rd\Phi_{f\bar f'}(\hat k_1,\hat k_2)}{\Phi_{f\bar f'}} \, 
\De^{\PV\to f\bar f'}_{\la'\la}(\hat k_1,\hat k_2),
\label{eq:sigmaab3}
\eeqar
where $\Phi_{f\bar f'}$ is the volume of the $f\bar f'$ phase space
$\int\rd\Phi_{f\bar f'}(\hat k_1,\hat k_2)$.
The matrix $\De^{\PV\to f\bar f'}$ is widely independent of the production 
process. Only the $\PV$ helicity states entering $\De^{\PV\to f\bar f'}$ 
must be the same as in the production. 
The explicit calculation of $\De^{\PV\to f\bar f'}$ is conveniently
performed in the rest frame of $\PV$.
Defining the momentum and the polarization vectors of $\PV$ 
in the CM frame $\Si$ of $ab$ by%
\footnote{For $\veps_\PV^\mu(\la)$ this choice follows the phase
convention of \citere{Dittmaier:1998nn}.}
\beqar
\hat k^\mu &=& E_\PV (1,\beta_\PV\cos\phi_\PV\sin\theta_\PV,
\beta_\PV\sin\phi_\PV\sin\theta_\PV,\beta_\PV\cos\theta_\PV), \quad
\beta_\PV = \sqrt{1-\MV^2/E_\PV^2},
\nn\\
\veps_\PV^\mu(\pm 1) &=& \frac{\Pe^{\mp\ri\phi_\PV}}{\sqrt{2}}
(0,-\cos\phi_\PV\cos\theta_\PV\pm\ri\sin\phi_\PV,
-\sin\phi_\PV\cos\theta_\PV\mp\ri\cos\phi_\PV,\sin\theta_\PV),
\nn\\
\veps_\PV^\mu(0) &=& \gamma_\PV (\beta_\PV,\cos\phi_\PV\sin\theta_\PV,
\sin\phi_\PV\sin\theta_\PV,\cos\theta_\PV), \quad
\gamma_\PV = 1/\sqrt{1-\beta_\PV^2} = E_\PV/\MV,
\nn\\
\eeqar
we transform them into the rest frame $\tilde\Si$ of $\PV$ with the 
Lorentz transformation matrix 
\beq
\left({\La^\mu}_\nu\right) = 
\pmatrix{\ga_\PV & 0 & 0 & -\beta_\PV\ga_\PV \cr
0 & 1 & 0 & 0 \cr 0 & 0 & 1 & 0 \cr 
-\beta_\PV\ga_\PV & 0 & 0 & \ga_\PV}
\pmatrix{1 & 0 & 0 & 0 \cr
0& \cos\phi_\PV\cos\theta_\PV & \sin\phi_\PV\cos\theta_\PV & -\sin\theta_\PV \cr
0 & -\sin\phi_\PV & \cos\phi_\PV & 0 \cr
0 & \cos\phi_\PV\sin\theta_\PV & \sin\phi_\PV\sin\theta_\PV & \cos\theta_\PV},
\eeq
which is factorized into a rotation in $\Si$ (rotating ${\bf k}$
to the $z$ axis) and a boost along the new $\PV$ direction.
The explicit results for the considered vectors
$\tilde a^\mu = {\La^\mu}_\nu a^\nu$ in $\tilde\Si$ are
\beq
\tilde {\hat k}{}^\mu = (\MV,0,0,0),
\quad
\tilde \veps_\PV^\mu(\pm 1) = \frac{\Pe^{\mp\ri\phi_\PV}}{\sqrt{2}}
(0,-1,\mp\ri,0),
\quad
\tilde \veps_\PV^\mu(0) = (0,0,0,1).
\eeq
In $\tilde\Si$ the decay momenta $\tilde{\hat k}_{1,2}$ are
distributed isotropically and parametrized by
\beq
\tilde{\hat k}{}_1^\mu = \frac{\MV}{2}
(1, \cos\tilde\phi_1\sin\tilde\theta_1,
\sin\tilde\phi_1\sin\tilde\theta_1,\cos\tilde\theta_1), \quad
\tilde{\hat k}{}_2^\mu = \tilde {\hat k}{}^\mu-\tilde{\hat k}{}_1^\mu.
\eeq
Equipped with these kinematical definitions we evaluate the
decay matrix elements 
\beq
\M_{\PV\to f\bar f'}^\si(\la) = e g_{\PV ff'}^\si
\bar u_f(\tilde{\hat k}_1) \tilde{\dsl{\veps}}(\la) \omega_\si
v_{\bar f'}(\tilde{\hat k}_2),
\eeq
where we made the chirality $\si=\pm$ explicit
(which coincides with the sign of the helicity of $f$) and used
an obvious notation for the Dirac spinors $\bar u_f$, $v_{\bar f'}$
and the chirality projectors $\omega_\pm=(1\pm\ga_5)/2$.
The chiral couplings $g_{\PV ff'}^\si$ are defined as in
\refeq{eq:chiralcouplings}.
Ordering the $\la$ values according to $(+,0,-)$, the decay amplitudes
can be written as
\beqar
\M_{\PV\to f\bar f'}^+(\la) &=&
e g_{\PV ff'}^+\MV
\pmatrix{+\frac{1}{\sqrt{2}}\Pe^{\ri(\tilde\phi_1-\phi_\PV)}
(1+\cos\tilde\theta_1),
\sin\tilde\theta_1,
-\frac{1}{\sqrt{2}}\Pe^{-\ri(\tilde\phi_1-\phi_\PV)}
(1-\cos\tilde\theta_1) },
\nn\\
\M_{\PV\to f\bar f'}^-(\la) &=&
e g_{\PV ff'}^-\MV
\pmatrix{-\frac{1}{\sqrt{2}}\Pe^{\ri(\tilde\phi_1-\phi_\PV)}
(1-\cos\tilde\theta_1),
\sin\tilde\theta_1,
+\frac{1}{\sqrt{2}}\Pe^{-\ri(\tilde\phi_1-\phi_\PV)}
(1+\cos\tilde\theta_1) }.
\nn\\
\eeqar
Inserting these amplitudes into 
\refeq{eq:decaycorrelmat} and summing over fermion helicities,
we obtain for the (hermitian) decay correlation matrix
\beqar
\De^{\PV\to f\bar f'}
&=& 3c_{\PV ff'}^+ \pmatrix{
\frac{1}{4}(1+\cos\tilde\theta_1)^2 & * & * \cr
\frac{\Pe^{\ri(\tilde\phi_1-\phi_\PV)}}{2\sqrt{2}}(1+\cos\tilde\theta_1)
\sin\tilde\theta_1 &
\frac{1}{2}\sin^2\tilde\theta_1 & * \cr
-\frac{\Pe^{2\ri(\tilde\phi_1-\phi_\PV)}}{4}\sin^2\tilde\theta_1 &
-\frac{\Pe^{\ri(\tilde\phi_1-\phi_\PV)}}{2\sqrt{2}}(1-\cos\tilde\theta_1)
\sin\tilde\theta_1 & \frac{1}{4}(1-\cos\tilde\theta_1)^2
}
\nn\\
&& {} +3c_{\PV ff'}^- \pmatrix{
\frac{1}{4}(1-\cos\tilde\theta_1)^2 & * & * \cr
-\frac{\Pe^{\ri(\tilde\phi_1-\phi_\PV)}}{2\sqrt{2}}(1-\cos\tilde\theta_1)
\sin\tilde\theta_1 &
\frac{1}{2}\sin^2\tilde\theta_1 & * \cr
-\frac{\Pe^{2\ri(\tilde\phi_1-\phi_\PV)}}{4}\sin^2\tilde\theta_1 &
\frac{\Pe^{\ri(\tilde\phi_1-\phi_\PV)}}{2\sqrt{2}}(1+\cos\tilde\theta_1)
\sin\tilde\theta_1 & \frac{1}{4}(1+\cos\tilde\theta_1)^2
},
\nn\\
\eeqar
where we used the shorthand
\beq
c_{\PV ff'}^\si =
\frac{(g_{\PV ff'}^\si)^2}{(g_{\PV ff'}^+)^2+(g_{\PV ff'}^-)^2},
\qquad
c_{\PV ff'}^+ + c_{\PV ff'}^- = 1.
\eeq
From this result it is obvious that the full integration of the
decay phase space yields
\beq
\int\frac{\rd\Phi_{f\bar f'}(\hat k_1,\hat k_2)}{\Phi_{f\bar f'}} \, 
\De^{\PV\to f\bar f'}(\hat k_1,\hat k_2) =
\int\frac{\rd\tilde\Omega_f}{4\pi} \, 
\De^{\PV\to f\bar f'}(\tilde{\hat k}_1,\tilde{\hat k}_2) = {\bf 1},
\eeq
as it should be for an isotropic decay in the rest frame of $\PV$.
\end{sloppypar}

\section{Benchmark numbers for the virtual corrections}
\label{se:benchmark}
In order to facilitate a comparison to our calculation, 
we provide explicit numbers on the
squared LO amplitude and the corresponding virtual corrections for a
single non-exceptional phase-space point. The set of momenta for
$ab\to \PWp(\to\Pne\Pe^+)\PWm(\to\Pmu^-\Pnmubar) c$ with the explicit partonic reactions 
$\Pq\bPq\to \PWp\PWm\Pg$,
$\Pq\bar \Pg \to \PWp\PWm\Pq$, and
$\Pg\bPq \to \PWp\PWm \bPq$
is chosen as
\begin{eqnarray}
\label{eq:samplepoint}
p_a &=& \scriptstyle (250,0,0,250),
\nn\\
p_b &=& \scriptstyle (250,0,0,-250),
\nn\\
p_{\Pne} &=& \scriptstyle
(49.43764668100422,
 -18.32747340442861,
 -42.74042990560614,
 16.77618190917520),
\nn\\
p_{\Pe^+} &=& \scriptstyle
(175.4405115231198,
 -125.7805594116281,
 -26.45739027367693,
 119.4095073999478),
\nn\\
p_{\Pmu^-} &=& \scriptstyle
(32.78781503058663,
 -29.71117802999691,
 7.333312535152912,
 -11.76899493633845),
\nn\\
p_{\Pnmubar} &=& \scriptstyle
(81.55218598686336,
 45.19373677627630,
 61.86450764413016,
 -27.94759183511103),
\nn\\
p_c &=& \scriptstyle
(160.7818407784260,
 128.6254740697773,
 0,
 -96.46910253767348),
\label{eq:PSpoint}
\end{eqnarray}
with the obvious notation
$p = (p^0,p^1,p^2,p^3)$
and all the components given in GeV. Note that the presented results are calculated using the improved NWA, i.e.\ with the $\PW$~bosons on-shell, but with their spin information kept.

Factoring out the couplings for the Born amplitude, we define
\begin{eqnarray}
  {1\over 4} {1\over {\cal N}_c }
  \sum_{\mbox{\scriptsize spin,colour} } |{\cal M}_\mathrm{LO}|^2
  = e^4\gs^2 a_0,
\end{eqnarray}
where 
\begin{eqnarray}
\label{eq:Aborn}
|{\cal M}_\mathrm{LO}|^2&=&\sum_{\la_+,\la'_+,\la_-,\la'_-=0,\pm1}
\M_{ab\to \PWp\PWm c}(\la'_+,\la'_-)^\ast\M_{ab\to \PWp\PWm c}(\la_+,\la_-)\\\nonumber
&&{}\hspace*{2.3cm}\times\De^{\PWp\to \Pne\Pe^+}_{\la_+'\la_+}(\hat{k}_{\Pne} ,\hat{k}_{\Pe^+})
\De^{\PWm\to \Pmu^-\Pnmubar}_{\la_-'\la_-}(\hat{k}_{\Pmu^-},\hat{k}_{\Pnmubar})
\end{eqnarray}
with the definitions of \refapp{se:onshelldecays}. The factor $ {1/ {\cal N}_c }$ is due to the average over the incoming
colour. For the channel $\Pq\bPq$, $\Pq\Pg$, $\Pg\bPq$ we have ${\cal N}_c = 9, 24, 24$, respectively.

The finite remainder of the virtual amplitudes after renormalization and addition of the $\cal I$ operator defines the relative correction $c_0$ according to  
\begin{eqnarray}
  {1\over 4} {1\over {\cal N}_c }
  \sum_{\mbox{\scriptsize spin,colour} } 
  2\mathrm{Re}\left({\left({\cal
  M}_{\mathrm{virt}}+{\cal M}_{\mathrm{ct}}
+{\cal M}_{\mathrm{{\cal I}-op}}\right)}^\ast{\cal M}_\mathrm{LO}\right) 
= e^4\gs^2 a_0\,c_0,
\end{eqnarray}
where 
${\cal M}_{\mathrm{virt}}$, ${\cal M}_{\mathrm{ct}}$, and
${\cal M}_{\mathrm{{\cal I}-op}}$ are defined in analogy
to \refeq{eq:Aborn}.

The results of our independent calculations at the sample phase-space
point \refeq{eq:samplepoint} are collected in \refta{tab:bornres} 
in terms of $a_0$ and $c_0$, namely for the six different partonic 
insertions for $a$, $b$, and $c$.

\def\SubTitle#1{\hline \multicolumn{2}{|c|}{#1}\\ \hline}
\def\Vone{{ Version 1\ }}
\def\Vtwo{{ Version 2\ }}
\def\Mad{Madgraph}
\def\SubTitle3#1{\hline \multicolumn{3}{|c|}{#1}\\ \hline}
\begin{table}
  \begin{center}
    \begin{tabular}{|c|c|c|}
      \hline
      &$a_0 [\GeV^{-2}]$ &$c_0$\\
      \SubTitle3{$ \Pu\bPu \to \PWp(\to\Pne\Pe^+)\PWm(\to\Pmu^-\Pnmubar) \Pg $} 
      \Vone  & $0.1543423047392003\cdot10^{-4}$ & $-0.01279611285522559$\\
      \Vtwo  & $0.1543423047392000\cdot10^{-4}$ & $-0.01279611284828043$\\
      \SubTitle3{$ \Pd\bPd \to \PWp(\to\Pne\Pe^+)\PWm(\to\Pmu^-\Pnmubar) \Pg $} 
      \Vone  & $0.9381083061358519\cdot10^{-6}$ & $0.02297241368680441$\\
      \Vtwo  & $0.9381083061358457\cdot10^{-6}$ & $0.02297241372542423$\\
      \SubTitle3{$\Pu\Pg\to \PWp(\to\Pne\Pe^+)\PWm(\to\Pmu^-\Pnmubar)\Pu$}
      \Vone  & $0.2766898219323333\cdot10^{-5}$ & $-0.04800046567722612$\\
      \Vtwo  & $0.2766898219323326\cdot10^{-5}$ & $-0.04800046570175944$\\
      \SubTitle3{$\Pd\Pg\to \PWp(\to\Pne\Pe^+)\PWm(\to\Pmu^-\Pnmubar)\Pd$}
      \Vone  & $0.1995963102379381\cdot10^{-6}$ & $-0.08921487131748859$\\
      \Vtwo  & $0.1995963102379386\cdot10^{-6}$ & $-0.08921487141443678$\\
      \SubTitle3{$\Pg\bPu\to \PWp(\to\Pne\Pe^+)\PWm(\to\Pmu^-\Pnmubar) \bPu$}
      \Vone  & $0.1351044519951966\cdot10^{-4}$ & $-0.2945621757876244$\\
      \Vtwo  & $0.1351044519951967\cdot10^{-4}$ & $-0.2945621758243954$\\
      \SubTitle3{$\Pg\bPd\to \PWp(\to\Pne\Pe^+)\PWm(\to\Pmu^-\Pnmubar) \bPd$}
      \Vone  & $0.4409954556585700\cdot10^{-5}$ & $-0.08497157067204103$\\
      \Vtwo  & $0.4409954556585692\cdot10^{-5}$ & $-0.08497157060895164$\\
      \hline
   \end{tabular}
    \caption{Colour and spin averaged LO matrix elements squared and virtual corrections.}
    \label{tab:bornres}
  \end{center}
\end{table}

For the LO amplitudes we find an agreement of at least 14 digits---as expected from a calculation using 64bit double precision. For the finite remainder of the one-loop corrections we find an agreement of at least 8 digits.

\section{Tables for histograms}
\label{se:tableshistograms}
In this appendix we present the  tables corresponding to the differential distributions 
presented in \refse{se:nr:decayNLOdist}. For each distribution, we list all predictions for the 
LO cross section and for both definitions of NLO cross sections---namely NLO~(excl) and NLO~(incl)---with
the scale choice $\mu=\mu_{\mathrm{ren}}=\mu_{\mathrm{fact}}=\MW$. 
The given errors result from the Monte Carlo integration. The bin is
specified by its central value. The bin width---which we chose
constant for the entire histogram---is obtained from the distance of
two neighboring bin positions. 
\begin{table}
\begin{center}
\rotatebox{90}{
\small
\begin{tabular}{|r|r|r|r|}
\hline
&&&\\[-2ex]
$p_{\mathrm{T,jet}}[\GeV]$ & $\frac{d\sigma_{\mathrm{LO}}}{dp_{\mathrm{T,jet}}}\Bigl[\frac{\pba}{\GeV}\Bigr]$ & $\frac{d\sigma_{\mathrm{NLO,excl}}}{dp_{\mathrm{T,jet}}}\Bigl[\frac{\pba}{\GeV}\Bigr]$ & $\frac{d\sigma_{\mathrm{NLO,incl}}}{dp_{\mathrm{T,jet}}}\Bigl[\frac{\pba}{\GeV}\Bigr]$\\[1.5ex]
\hline
$         55 $ & $        1.47878 \,(26)$  & $         1.7903 \,(23)$  & $         1.8600 \,(22)$ \\
$         65 $ & $        1.19041 \,(23)$  & $         1.3622 \,(21)$  & $         1.5178 \,(20)$ \\
$         75 $ & $        0.98096 \,(21)$  & $         1.0595 \,(20)$  & $         1.2619 \,(19)$ \\
$         85 $ & $        0.82324 \,(19)$  & $         0.8511 \,(20)$  & $         1.0706 \,(20)$ \\
$         95 $ & $        0.70022 \,(18)$  & $         0.6875 \,(19)$  & $         0.9150 \,(19)$ \\
$        105 $ & $        0.60034 \,(16)$  & $         0.5648 \,(18)$  & $         0.7878 \,(16)$ \\
\hline
\end{tabular}
}
\rotatebox{90}{
\small
\begin{tabular}{|r|r|r|r|}
\hline
&&&\\[-2ex]
$p_{\mathrm{T,miss}}[\GeV]$ & $\frac{d\sigma_{\mathrm{LO}}}{dp_{\mathrm{T,miss}}}\Bigl[\frac{\pba}{\GeV}\Bigr]$ & $\frac{d\sigma_{\mathrm{NLO,excl}}}{dp_{\mathrm{T,miss}}}\Bigl[\frac{\pba}{\GeV}\Bigr]$ & $\frac{d\sigma_{\mathrm{NLO,incl}}}{dp_{\mathrm{T,miss}}}\Bigl[\frac{\pba}{\GeV}\Bigr]$\\[1.5ex]
\hline
$       27.5 $ & $        1.10787 \,(34)$  & $         1.1683 \,(35)$  & $         1.5232 \,(31)$ \\
$       32.5 $ & $        1.21978 \,(36)$  & $         1.2943 \,(38)$  & $         1.6798 \,(31)$ \\
$       37.5 $ & $        1.28976 \,(37)$  & $         1.3484 \,(33)$  & $         1.7642 \,(28)$ \\
$       42.5 $ & $        1.32096 \,(37)$  & $         1.3665 \,(37)$  & $         1.7964 \,(31)$ \\
$       47.5 $ & $        1.31712 \,(37)$  & $         1.3687 \,(38)$  & $         1.8037 \,(27)$ \\
$       52.5 $ & $        1.28158 \,(36)$  & $         1.3103 \,(34)$  & $         1.7481 \,(27)$ \\
\hline
\end{tabular}
}
\end{center}
\label{ta:dist.lhc.ptjet.ptmiss}
\caption{Distribution of $p_{\mathrm{T,jet}}$ and $p_{\mathrm{T,miss}}$ at the LHC with $\mu=\muren=\mufact=\MW$.}
\end{table}

\begin{table}
\begin{center}
\rotatebox{90}{
\small
\begin{tabular}{|r|r|r|r|}
\hline
&&&\\[-2ex]
$p_{\mathrm{T,e^+}}[\GeV]$ & $\frac{d\sigma_{\mathrm{LO}}}{dp_{\mathrm{T,e^+}}}\Bigl[\frac{\pba}{\GeV}\Bigr]$ & $\frac{d\sigma_{\mathrm{NLO,excl}}}{dp_{\mathrm{T,e^+}}}\Bigl[\frac{\pba}{\GeV}\Bigr]$ & $\frac{d\sigma_{\mathrm{NLO,incl}}}{dp_{\mathrm{T,e^+}}}\Bigl[\frac{\pba}{\GeV}\Bigr]$\\[1.5ex]
\hline
$       27.5 $ & $        2.21854 \,(59)$  & $         2.1958 \,(33)$  & $         3.0506 \,(27)$ \\
$       32.5 $ & $        2.08923 \,(52)$  & $         2.0766 \,(38)$  & $         2.8827 \,(28)$ \\
$       37.5 $ & $        1.89828 \,(46)$  & $         1.8785 \,(32)$  & $         2.6195 \,(30)$ \\
$       42.5 $ & $        1.69226 \,(41)$  & $         1.6640 \,(27)$  & $         2.3308 \,(25)$ \\
$       47.5 $ & $        1.49414 \,(37)$  & $         1.4507 \,(26)$  & $         2.0407 \,(23)$ \\
$       52.5 $ & $        1.31341 \,(34)$  & $         1.2568 \,(24)$  & $         1.7847 \,(22)$ \\
\hline
\end{tabular}
}
\rotatebox{90}{
\small
\begin{tabular}{|r|r|r|r|}
\hline
&&&\\[-2ex]
$p_{\mathrm{T,\mu^-}}[\GeV]$ & $\frac{d\sigma_{\mathrm{LO}}}{dp_{\mathrm{T,\mu^-}}}\Bigl[\frac{\pba}{\GeV}\Bigr]$ & $\frac{d\sigma_{\mathrm{NLO,excl}}}{dp_{\mathrm{T,\mu^-}}}\Bigl[\frac{\pba}{\GeV}\Bigr]$ & $\frac{d\sigma_{\mathrm{NLO,incl}}}{dp_{\mathrm{T,\mu^-}}}\Bigl[\frac{\pba}{\GeV}\Bigr]$\\[1.5ex]
\hline
$       27.5 $ & $        1.75438 \,(39)$  & $         1.7529 \,(22)$  & $         2.3634 \,(20)$ \\
$       32.5 $ & $        1.72719 \,(38)$  & $         1.7365 \,(37)$  & $         2.3444 \,(34)$ \\
$       37.5 $ & $        1.63647 \,(37)$  & $         1.6451 \,(33)$  & $         2.2317 \,(31)$ \\
$       42.5 $ & $        1.51599 \,(36)$  & $         1.5200 \,(22)$  & $         2.0711 \,(20)$ \\
$       47.5 $ & $        1.38745 \,(35)$  & $         1.3796 \,(26)$  & $         1.8914 \,(21)$ \\
$       52.5 $ & $        1.25864 \,(34)$  & $         1.2391 \,(24)$  & $         1.7103 \,(22)$ \\
\hline
\end{tabular}
}
\end{center}
\label{ta:dist.lhc.ptep.ptmum}
\caption{Distribution of $p_{\mathrm{T,e^+}}$ and $p_{\mathrm{T,\mu^-}}$ at the LHC with $\mu=\muren=\mufact=\MW$.}
\end{table}

\begin{table}
\begin{center}
\rotatebox{90}{
\small
\begin{tabular}{|r|r|r|r|}
\hline
&&&\\[-2ex]
$\eta_{\mathrm{e^+}}$ & $\frac{d\sigma_{\mathrm{LO}}}{d\eta_{\mathrm{e^+}}}[\pba]$ & $\frac{d\sigma_{\mathrm{NLO,excl}}}{d\eta_{\mathrm{e^+}}}[\pba]$ & $\frac{d\sigma_{\mathrm{NLO,incl}}}{d\eta_{\mathrm{e^+}}}[\pba]$\\[1.5ex]
\hline
$       -2.4 $ & $        13.6439 \,(77)$  & $         12.462 \,(66)$  & $         17.890 \,(69)$ \\
$       -2.2 $ & $        15.4666 \,(77)$  & $         14.058 \,(74)$  & $         20.255 \,(64)$ \\
$         -2 $ & $        17.1745 \,(76)$  & $         15.362 \,(55)$  & $         22.547 \,(42)$ \\
$       -1.8 $ & $        18.7412 \,(74)$  & $         16.866 \,(62)$  & $         24.688 \,(43)$ \\
$       -1.6 $ & $        20.1160 \,(71)$  & $         17.900 \,(69)$  & $         26.546 \,(45)$ \\
$       -1.4 $ & $        21.3037 \,(69)$  & $         18.923 \,(73)$  & $         28.009 \,(46)$ \\
\hline
\end{tabular}
}
\rotatebox{90}{
\small
\begin{tabular}{|r|r|r|r|}
\hline
&&&\\[-2ex]
$\eta_{\mathrm{\mu^-}}$ & $\frac{d\sigma_{\mathrm{LO}}}{d\eta_{\mathrm{\mu^-}}}[\pba]$ & $\frac{d\sigma_{\mathrm{NLO,excl}}}{d\eta_{\mathrm{\mu^-}}}[\pba]$ & $\frac{d\sigma_{\mathrm{NLO,incl}}}{d\eta_{\mathrm{\mu^-}}}[\pba]$\\[1.5ex]
\hline
$       -2.4 $ & $        11.8284 \,(58)$  & $         10.997 \,(56)$  & $         15.461 \,(42)$ \\
$       -2.2 $ & $        13.7789 \,(61)$  & $         12.593 \,(65)$  & $         18.033 \,(73)$ \\
$         -2 $ & $        15.7736 \,(63)$  & $         14.326 \,(85)$  & $         20.574 \,(48)$ \\
$       -1.8 $ & $        17.7086 \,(66)$  & $         15.974 \,(63)$  & $         23.258 \,(51)$ \\
$       -1.6 $ & $        19.5103 \,(68)$  & $         17.584 \,(61)$  & $         25.653 \,(47)$ \\
$       -1.4 $ & $        21.1445 \,(70)$  & $         18.862 \,(81)$  & $         27.866 \,(51)$ \\
\hline
\end{tabular}
}
\end{center}
\label{ta:dist.lhc.etaep.etamum}
\caption{Distribution of $\eta_{\mathrm{e^+}}$ and $\eta_{\mathrm{\mu^-}}$ at the LHC with $\mu=\muren=\mufact=\MW$.}
\end{table}

\begin{table}
\begin{center}
\rotatebox{90}{
\small
\begin{tabular}{|r|r|r|r|}
\hline
&&&\\[-2ex]
$\eta_{\mathrm{jet}}$ & $\frac{d\sigma_{\mathrm{LO}}}{d\eta_{\mathrm{jet}}}[\pba]$ & $\frac{d\sigma_{\mathrm{NLO,excl}}}{d\eta_{\mathrm{jet}}}[\pba]$ & $\frac{d\sigma_{\mathrm{NLO,incl}}}{d\eta_{\mathrm{jet}}}[\pba]$\\[1.5ex]
\hline
$      -4.32 $ & $        0.15890 \,(34)$  & $         0.2762 \,(36)$  & $         0.3316 \,(37)$ \\
$      -3.96 $ & $        0.50253 \,(69)$  & $         0.7129 \,(77)$  & $         0.8899 \,(75)$ \\
$       -3.6 $ & $         1.1812 \,(11)$  & $          1.463 \,(16)$  & $          1.891 \,(13)$ \\
$      -3.24 $ & $         2.3948 \,(16)$  & $          2.654 \,(15)$  & $          3.574 \,(14)$ \\
$      -2.88 $ & $         4.3033 \,(22)$  & $          4.382 \,(22)$  & $          6.085 \,(22)$ \\
$      -2.52 $ & $         6.9684 \,(29)$  & $          6.676 \,(29)$  & $          9.489 \,(26)$ \\
\hline
\end{tabular}
}
\rotatebox{90}{
\small
\begin{tabular}{|r|r|r|r|}
\hline
&&&\\[-2ex]
$m_{\mathrm{e^+\mu^-}}[\GeV]$ & $\frac{d\sigma_{\mathrm{LO}}}{dm_{\mathrm{e^+\mu^-}}}\Bigl[\frac{\pba}{\GeV}\Bigr]$ & $\frac{d\sigma_{\mathrm{NLO,excl}}}{dm_{\mathrm{e^+\mu^-}}}\Bigl[\frac{\pba}{\GeV}\Bigr]$ & $\frac{d\sigma_{\mathrm{NLO,incl}}}{dm_{\mathrm{e^+\mu^-}}}\Bigl[\frac{\pba}{\GeV}\Bigr]$\\[1.5ex]
\hline
$          5 $ & $       0.013441 \,(30)$  & $        0.01412 \,(22)$  & $        0.01878 \,(20)$ \\
$         15 $ & $       0.139490 \,(90)$  & $        0.13426 \,(44)$  & $        0.18829 \,(40)$ \\
$         25 $ & $        0.24176 \,(12)$  & $        0.22708 \,(55)$  & $        0.32658 \,(51)$ \\
$         35 $ & $        0.32454 \,(14)$  & $        0.30439 \,(67)$  & $        0.43912 \,(58)$ \\
$         45 $ & $        0.40439 \,(15)$  & $        0.38108 \,(76)$  & $        0.54827 \,(68)$ \\
$         55 $ & $        0.50351 \,(17)$  & $        0.48153 \,(84)$  & $        0.68487 \,(77)$ \\
\hline
\end{tabular}
}
\end{center}
\label{ta:dist.lhc.etajet.mepmum}
\caption{Distribution of $\eta_{\mathrm{jet}}$ and $m_{\mathrm{e^+\mu^-}}$ at the LHC with $\mu=\muren=\mufact=\MW$.}
\end{table}

\begin{table}
\begin{center}
\rotatebox{90}{
\small
\begin{tabular}{|r|r|r|r|}
\hline
&&&\\[-2ex]
$\cos\theta_{\mathrm{e^+\mu^-}}$ & $\frac{d\sigma_{\mathrm{LO}}}{d\cos\theta_{\mathrm{e^+\mu^-}}}[\pba]$ & $\frac{d\sigma_{\mathrm{NLO,excl}}}{d\cos\theta_{\mathrm{e^+\mu^-}}}[\pba]$ & $\frac{d\sigma_{\mathrm{NLO,incl}}}{d\cos\theta_{\mathrm{e^+\mu^-}}}[\pba]$\\[1.5ex]
\hline
$      -0.96 $ & $         49.745 \,(16)$  & $          48.56 \,(13)$  & $         66.057 \,(93)$ \\
$      -0.88 $ & $         43.269 \,(15)$  & $          41.75 \,(16)$  & $          58.03 \,(14)$ \\
$       -0.8 $ & $         40.279 \,(14)$  & $          38.59 \,(14)$  & $          54.24 \,(10)$ \\
$      -0.72 $ & $         38.446 \,(14)$  & $          36.85 \,(23)$  & $          51.78 \,(14)$ \\
$      -0.64 $ & $         37.272 \,(13)$  & $          35.21 \,(20)$  & $          50.40 \,(17)$ \\
$      -0.56 $ & $         36.580 \,(13)$  & $          34.52 \,(16)$  & $          49.37 \,(14)$ \\
\hline
\end{tabular}
}
\rotatebox{90}{
\small
\begin{tabular}{|r|r|r|r|}
\hline
&&&\\[-2ex]
$\varphi_{\mathrm{e^+\mu^-}}$ & $\frac{d\sigma_{\mathrm{LO}}}{d\varphi_{\mathrm{e^+\mu^-}}}[\pba]$ & $\frac{d\sigma_{\mathrm{NLO,excl}}}{d\varphi_{\mathrm{e^+\mu^-}}}[\pba]$ & $\frac{d\sigma_{\mathrm{NLO,incl}}}{d\varphi_{\mathrm{e^+\mu^-}}}[\pba]$\\[1.5ex]
\hline
$ 0.062831855 $ & $        21.1649 \,(87)$  & $         17.523 \,(65)$  & $         28.681 \,(51)$ \\
$ 0.18849556 $ & $        23.3692 \,(90)$  & $         19.185 \,(66)$  & $         31.436 \,(52)$ \\
$ 0.31415927 $ & $        24.5298 \,(93)$  & $         20.322 \,(80)$  & $         33.167 \,(58)$ \\
$ 0.43982297 $ & $        24.6116 \,(92)$  & $         20.398 \,(67)$  & $         33.105 \,(55)$ \\
$ 0.56548668 $ & $        24.7017 \,(94)$  & $         20.686 \,(72)$  & $         33.341 \,(59)$ \\
$ 0.69115039 $ & $        24.8908 \,(93)$  & $         21.067 \,(83)$  & $         33.678 \,(58)$ \\
\hline
\end{tabular}
}
\end{center}
\label{ta:dist.lhc.cthepmum.phiepmum}
\caption{Distribution of $\cos\theta_{\mathrm{e^+\mu^-}}$ and $\varphi_{\mathrm{e^+\mu^-}}$ at the LHC with $\mu=\muren=\mufact=\MW$.}
\end{table}

\begin{table}
\begin{center}
\rotatebox{90}{
\small
\begin{tabular}{|r|r|r|r|}
\hline
&&&\\[-2ex]
$p_{\mathrm{T,jet}}[\GeV]$ & $\frac{d\sigma_{\mathrm{LO}}}{dp_{\mathrm{T,jet}}}\Bigl[\frac{\pba}{\GeV}\Bigr]$ & $\frac{d\sigma_{\mathrm{NLO,excl}}}{dp_{\mathrm{T,jet}}}\Bigl[\frac{\pba}{\GeV}\Bigr]$ & $\frac{d\sigma_{\mathrm{NLO,incl}}}{dp_{\mathrm{T,jet}}}\Bigl[\frac{\pba}{\GeV}\Bigr]$\\[1.5ex]
\hline
$         25 $ & $        0.38977 \,(12)$  & $        0.49226 \,(78)$  & $        0.51143 \,(76)$ \\
$         35 $ & $       0.214392 \,(77)$  & $        0.25162 \,(56)$  & $        0.28034 \,(60)$ \\
$         45 $ & $       0.133866 \,(56)$  & $        0.14689 \,(38)$  & $        0.17183 \,(41)$ \\
$         55 $ & $       0.089710 \,(43)$  & $        0.09317 \,(33)$  & $        0.11298 \,(29)$ \\
$         65 $ & $       0.063002 \,(34)$  & $        0.06205 \,(22)$  & $        0.07802 \,(20)$ \\
$         75 $ & $       0.045775 \,(28)$  & $        0.04275 \,(21)$  & $        0.05562 \,(18)$ \\
\hline
\end{tabular}
}
\rotatebox{90}{
\small
\begin{tabular}{|r|r|r|r|}
\hline
&&&\\[-2ex]
$p_{\mathrm{T,miss}}[\GeV]$ & $\frac{d\sigma_{\mathrm{LO}}}{dp_{\mathrm{T,miss}}}\Bigl[\frac{\pba}{\GeV}\Bigr]$ & $\frac{d\sigma_{\mathrm{NLO,excl}}}{dp_{\mathrm{T,miss}}}\Bigl[\frac{\pba}{\GeV}\Bigr]$ & $\frac{d\sigma_{\mathrm{NLO,incl}}}{dp_{\mathrm{T,miss}}}\Bigl[\frac{\pba}{\GeV}\Bigr]$\\[1.5ex]
\hline
$       27.5 $ & $        0.22486 \,(12)$  & $        0.26071 \,(41)$  & $        0.29070 \,(40)$ \\
$       32.5 $ & $        0.22916 \,(12)$  & $        0.26346 \,(38)$  & $        0.29475 \,(38)$ \\
$       37.5 $ & $        0.22730 \,(12)$  & $        0.25967 \,(42)$  & $        0.29130 \,(41)$ \\
$       42.5 $ & $        0.21985 \,(11)$  & $        0.25002 \,(38)$  & $        0.28133 \,(36)$ \\
$       47.5 $ & $        0.20813 \,(11)$  & $        0.23658 \,(42)$  & $        0.26677 \,(42)$ \\
$       52.5 $ & $        0.19202 \,(10)$  & $        0.21661 \,(33)$  & $        0.24525 \,(33)$ \\
\hline
\end{tabular}
}
\end{center}
\label{ta:dist.tev.ptjet.ptmiss}
\caption{Distribution of $p_{\mathrm{T,jet}}$ and $p_{\mathrm{T,miss}}$ at the Tevatron with $\mu=\muren=\mufact=\MW$.}
\end{table}

\begin{table}
\begin{center}
\rotatebox{90}{
\small
\begin{tabular}{|r|r|r|r|}
\hline
&&&\\[-2ex]
$p_{\mathrm{T,e^+}}[\GeV]$ & $\frac{d\sigma_{\mathrm{LO}}}{dp_{\mathrm{T,e^+}}}\Bigl[\frac{\pba}{\GeV}\Bigr]$ & $\frac{d\sigma_{\mathrm{NLO,excl}}}{dp_{\mathrm{T,e^+}}}\Bigl[\frac{\pba}{\GeV}\Bigr]$ & $\frac{d\sigma_{\mathrm{NLO,incl}}}{dp_{\mathrm{T,e^+}}}\Bigl[\frac{\pba}{\GeV}\Bigr]$\\[1.5ex]
\hline
$       27.5 $ & $        0.30969 \,(16)$  & $        0.35540 \,(53)$  & $        0.40097 \,(53)$ \\
$       32.5 $ & $        0.30476 \,(14)$  & $        0.35176 \,(47)$  & $        0.39624 \,(46)$ \\
$       37.5 $ & $        0.27514 \,(12)$  & $        0.31748 \,(41)$  & $        0.35780 \,(41)$ \\
$       42.5 $ & $        0.23547 \,(11)$  & $        0.27055 \,(38)$  & $        0.30585 \,(37)$ \\
$       47.5 $ & $       0.195342 \,(96)$  & $        0.22209 \,(32)$  & $        0.25213 \,(33)$ \\
$       52.5 $ & $       0.159304 \,(85)$  & $        0.17864 \,(32)$  & $        0.20402 \,(31)$ \\
\hline
\end{tabular}
}
\rotatebox{90}{
\small
\begin{tabular}{|r|r|r|r|}
\hline
&&&\\[-2ex]
$p_{\mathrm{T,\mu^-}}[\GeV]$ & $\frac{d\sigma_{\mathrm{LO}}}{dp_{\mathrm{T,\mu^-}}}\Bigl[\frac{\pba}{\GeV}\Bigr]$ & $\frac{d\sigma_{\mathrm{NLO,excl}}}{dp_{\mathrm{T,\mu^-}}}\Bigl[\frac{\pba}{\GeV}\Bigr]$ & $\frac{d\sigma_{\mathrm{NLO,incl}}}{dp_{\mathrm{T,\mu^-}}}\Bigl[\frac{\pba}{\GeV}\Bigr]$\\[1.5ex]
\hline
$       27.5 $ & $        0.30973 \,(16)$  & $        0.35406 \,(48)$  & $        0.39974 \,(48)$ \\
$       32.5 $ & $        0.30491 \,(14)$  & $        0.35174 \,(44)$  & $        0.39591 \,(45)$ \\
$       37.5 $ & $        0.27515 \,(13)$  & $        0.31884 \,(49)$  & $        0.35935 \,(49)$ \\
$       42.5 $ & $        0.23554 \,(11)$  & $        0.26988 \,(49)$  & $        0.30501 \,(47)$ \\
$       47.5 $ & $       0.195422 \,(96)$  & $        0.22188 \,(45)$  & $        0.25183 \,(43)$ \\
$       52.5 $ & $       0.159132 \,(84)$  & $        0.17840 \,(26)$  & $        0.20373 \,(26)$ \\
\hline
\end{tabular}
}
\end{center}
\label{ta:dist.tev.ptep.ptmum}
\caption{Distribution of $p_{\mathrm{T,e^+}}$ and $p_{\mathrm{T,\mu^-}}$ at the Tevatron with $\mu=\muren=\mufact=\MW$.}
\end{table}

\begin{table}
\begin{center}
\rotatebox{90}{
\small
\begin{tabular}{|r|r|r|r|}
\hline
&&&\\[-2ex]
$\eta_{\mathrm{e^+}}$ & $\frac{d\sigma_{\mathrm{LO}}}{d\eta_{\mathrm{e^+}}}[\pba]$ & $\frac{d\sigma_{\mathrm{NLO,excl}}}{d\eta_{\mathrm{e^+}}}[\pba]$ & $\frac{d\sigma_{\mathrm{NLO,incl}}}{d\eta_{\mathrm{e^+}}}[\pba]$\\[1.5ex]
\hline
$       -2.4 $ & $        0.27124 \,(43)$  & $         0.3139 \,(20)$  & $         0.3443 \,(19)$ \\
$       -2.2 $ & $        0.43757 \,(56)$  & $         0.5046 \,(25)$  & $         0.5550 \,(25)$ \\
$         -2 $ & $        0.66082 \,(71)$  & $         0.7610 \,(37)$  & $         0.8416 \,(33)$ \\
$       -1.8 $ & $        0.94463 \,(87)$  & $         1.0770 \,(39)$  & $         1.1947 \,(39)$ \\
$       -1.6 $ & $         1.2805 \,(10)$  & $         1.4555 \,(44)$  & $         1.6258 \,(44)$ \\
$       -1.4 $ & $         1.6640 \,(12)$  & $         1.8749 \,(53)$  & $         2.1066 \,(53)$ \\
\hline
\end{tabular}
}
\rotatebox{90}{
\small
\begin{tabular}{|r|r|r|r|}
\hline
&&&\\[-2ex]
$\eta_{\mathrm{\mu^-}}$ & $\frac{d\sigma_{\mathrm{LO}}}{d\eta_{\mathrm{\mu^-}}}[\pba]$ & $\frac{d\sigma_{\mathrm{NLO,excl}}}{d\eta_{\mathrm{\mu^-}}}[\pba]$ & $\frac{d\sigma_{\mathrm{NLO,incl}}}{d\eta_{\mathrm{\mu^-}}}[\pba]$\\[1.5ex]
\hline
$       -2.4 $ & $        0.27168 \,(42)$  & $         0.3040 \,(16)$  & $         0.3409 \,(16)$ \\
$       -2.2 $ & $        0.46828 \,(63)$  & $         0.5228 \,(60)$  & $         0.5895 \,(62)$ \\
$         -2 $ & $        0.74839 \,(87)$  & $         0.8256 \,(40)$  & $         0.9343 \,(40)$ \\
$       -1.8 $ & $         1.1136 \,(11)$  & $         1.2392 \,(57)$  & $         1.4061 \,(57)$ \\
$       -1.6 $ & $         1.5503 \,(15)$  & $         1.6979 \,(98)$  & $          1.948 \,(10)$ \\
$       -1.4 $ & $         2.0437 \,(17)$  & $         2.2547 \,(85)$  & $         2.5763 \,(89)$ \\
\hline
\end{tabular}
}
\end{center}
\label{ta:dist.tev.etaep.etamum}
\caption{Distribution of $\eta_{\mathrm{e^+}}$ and $\eta_{\mathrm{\mu^-}}$ at the Tevatron with $\mu=\muren=\mufact=\MW$.}
\end{table}

\begin{table}
\begin{center}
\rotatebox{90}{
\small
\begin{tabular}{|r|r|r|r|}
\hline
&&&\\[-2ex]
$\eta_{\mathrm{jet}}$ & $\frac{d\sigma_{\mathrm{LO}}}{d\eta_{\mathrm{jet}}}[\pba]$ & $\frac{d\sigma_{\mathrm{NLO,excl}}}{d\eta_{\mathrm{jet}}}[\pba]$ & $\frac{d\sigma_{\mathrm{NLO,incl}}}{d\eta_{\mathrm{jet}}}[\pba]$\\[1.5ex]
\hline
$       -2.4 $ & $        0.57315 \,(68)$  & $         0.6833 \,(78)$  & $         0.7509 \,(86)$ \\
$       -2.2 $ & $        0.78791 \,(83)$  & $          0.901 \,(11)$  & $          1.005 \,(11)$ \\
$         -2 $ & $         1.0390 \,(10)$  & $          1.221 \,(16)$  & $          1.346 \,(16)$ \\
$       -1.8 $ & $         1.3276 \,(12)$  & $          1.533 \,(19)$  & $          1.711 \,(13)$ \\
$       -1.6 $ & $         1.6330 \,(13)$  & $          1.866 \,(18)$  & $          2.083 \,(13)$ \\
$       -1.4 $ & $         1.9453 \,(15)$  & $          2.166 \,(14)$  & $          2.483 \,(15)$ \\
\hline
\end{tabular}
}
\rotatebox{90}{
\small
\begin{tabular}{|r|r|r|r|}
\hline
&&&\\[-2ex]
$m_{\mathrm{e^+\mu^-}}[\GeV]$ & $\frac{d\sigma_{\mathrm{LO}}}{dm_{\mathrm{e^+\mu^-}}}\Bigl[\frac{\pba}{\GeV}\Bigr]$ & $\frac{d\sigma_{\mathrm{NLO,excl}}}{dm_{\mathrm{e^+\mu^-}}}\Bigl[\frac{\pba}{\GeV}\Bigr]$ & $\frac{d\sigma_{\mathrm{NLO,incl}}}{dm_{\mathrm{e^+\mu^-}}}\Bigl[\frac{\pba}{\GeV}\Bigr]$\\[1.5ex]
\hline
$          5 $ & $       0.002587 \,(10)$  & $       0.003042 \,(64)$  & $       0.003389 \,(59)$ \\
$         15 $ & $       0.016720 \,(24)$  & $       0.018743 \,(67)$  & $       0.021646 \,(66)$ \\
$         25 $ & $       0.028051 \,(32)$  & $       0.031663 \,(92)$  & $       0.036463 \,(91)$ \\
$         35 $ & $       0.039593 \,(38)$  & $        0.04497 \,(12)$  & $        0.05156 \,(12)$ \\
$         45 $ & $       0.052892 \,(45)$  & $        0.06066 \,(14)$  & $        0.06926 \,(14)$ \\
$         55 $ & $       0.073373 \,(54)$  & $        0.08432 \,(17)$  & $        0.09572 \,(17)$ \\
\hline
\end{tabular}
}
\end{center}
\label{ta:dist.tev.etajet.mepmum}
\caption{Distribution of $\eta_{\mathrm{jet}}$ and $m_{\mathrm{e^+\mu^-}}$ at the Tevatron with $\mu=\muren=\mufact=\MW$.}
\end{table}

\begin{table}
\begin{center}
\rotatebox{90}{
\small
\begin{tabular}{|r|r|r|r|}
\hline
&&&\\[-2ex]
$\cos\theta_{\mathrm{e^+\mu^-}}$ & $\frac{d\sigma_{\mathrm{LO}}}{d\cos\theta_{\mathrm{e^+\mu^-}}}[\pba]$ & $\frac{d\sigma_{\mathrm{NLO,excl}}}{d\cos\theta_{\mathrm{e^+\mu^-}}}[\pba]$ & $\frac{d\sigma_{\mathrm{NLO,incl}}}{d\cos\theta_{\mathrm{e^+\mu^-}}}[\pba]$\\[1.5ex]
\hline
$      -0.96 $ & $         8.9561 \,(59)$  & $          9.712 \,(22)$  & $         10.986 \,(21)$ \\
$      -0.88 $ & $         8.0391 \,(55)$  & $          8.789 \,(26)$  & $          9.964 \,(23)$ \\
$       -0.8 $ & $         7.4532 \,(53)$  & $          8.214 \,(24)$  & $          9.309 \,(27)$ \\
$      -0.72 $ & $         7.0061 \,(50)$  & $          7.726 \,(29)$  & $          8.794 \,(26)$ \\
$      -0.64 $ & $         6.6385 \,(49)$  & $          7.319 \,(24)$  & $          8.324 \,(23)$ \\
$      -0.56 $ & $         6.3402 \,(47)$  & $          7.054 \,(26)$  & $          8.024 \,(29)$ \\
\hline
\end{tabular}
}
\rotatebox{90}{
\small
\begin{tabular}{|r|r|r|r|}
\hline
&&&\\[-2ex]
$\varphi_{\mathrm{e^+\mu^-}}$ & $\frac{d\sigma_{\mathrm{LO}}}{d\varphi_{\mathrm{e^+\mu^-}}}[\pba]$ & $\frac{d\sigma_{\mathrm{NLO,excl}}}{d\varphi_{\mathrm{e^+\mu^-}}}[\pba]$ & $\frac{d\sigma_{\mathrm{NLO,incl}}}{d\varphi_{\mathrm{e^+\mu^-}}}[\pba]$\\[1.5ex]
\hline
$ 0.062831855 $ & $         1.5811 \,(19)$  & $         1.7284 \,(86)$  & $         2.0408 \,(86)$ \\
$ 0.18849556 $ & $         1.7229 \,(20)$  & $         1.8614 \,(80)$  & $         2.2075 \,(80)$ \\
$ 0.31415927 $ & $         1.8202 \,(20)$  & $         1.9833 \,(98)$  & $          2.340 \,(10)$ \\
$ 0.43982297 $ & $         1.8724 \,(21)$  & $          2.036 \,(14)$  & $          2.405 \,(13)$ \\
$ 0.56548668 $ & $         1.9321 \,(21)$  & $         2.1060 \,(73)$  & $         2.4815 \,(72)$ \\
$ 0.69115039 $ & $         2.0126 \,(21)$  & $         2.2103 \,(85)$  & $         2.5944 \,(84)$ \\
\hline
\end{tabular}
}
\end{center}
\label{ta:dist.tev.cthepmum.phiepmum}
\caption{Distribution of $\cos\theta_{\mathrm{e^+\mu^-}}$ and $\varphi_{\mathrm{e^+\mu^-}}$ at the Tevatron with $\mu=\muren=\mufact=\MW$.}
\end{table}

\end{document}